\documentclass[doublespace,english]{revtex4-1}
\usepackage[T1]{fontenc}
\pdfoutput=1
\usepackage{array}
\usepackage{booktabs}
\usepackage{multirow}

\usepackage{amsmath, amsthm, amssymb, amsfonts, enumerate}
\usepackage{setspace}
\usepackage{natbib, bbm}
\usepackage{graphicx}
\usepackage{color}

\providecommand{\tabularnewline}{\\}

\usepackage{babel}

\usepackage{subfig}

\begin{document}
\title{A Flux-Balanced Fluid Model for Collisional Plasma Edge Turbulence: Model Derivation and Basic Physical Features}

\author{Andrew J. Majda}
\author{Di Qi}

\affiliation{Department of Mathematics, Courant Institute
of Mathematical Sciences, New York University, New York, NY 10012}
\affiliation{Center for Atmosphere and Ocean Science, Courant Institute of Mathematical Sciences, New York University, New York, NY 10012}

\author{Antoine J. Cerfon}
\affiliation{Department of Mathematics, Courant Institute
of Mathematical Sciences, New York University, New York, NY 10012}


\begin{abstract}
We propose a new reduced fluid model for the study of the drift wave -- zonal flow dynamics in magnetically confined plasmas. Our model can be viewed as an extension of the classic Hasegawa-Wakatani (HW) model, and is based on an improved treatment of the electron dynamics parallel to the field lines, to guarantee a balanced electron flux on the magnetic surfaces. Our flux-balanced HW (bHW) model contains the same drift-wave instability as previous HW models, but unlike these models, it converges exactly to the modified Hasegawa-Mima model in the collisionless limit. We rely on direct numerical simulations to illustrate some of the key features of the bHW model, such as the enhanced variability in the turbulent fluctuations, and the existence of stronger and more turbulent zonal jets than the jets observed in other HW models, especially for high plasma resistivity. Our simulations also highlight the crucial role of the feedback of the third-order statistical moments in achieving a statistical equilibrium with strong zonal structures. Finally, we investigate the changes in the observed dynamics when more general dissipation effects are included, and in particular when we include the reduced model for ion Landau damping originally proposed by Wakatani and Hasegawa.
\end{abstract}

\maketitle

\section{Introduction}

In toroidal magnetically confined plasmas, the level of cross field heat and particle transport in the edge region is in large part set by drift-wave turbulence driven by temperature and density gradients \cite{hortonreview1999}. This turbulence can itself generate zonal flows \cite{diamond2005zonal}, which are known to mediate turbulent transport by shearing the turbulent eddies and absorbing some of the drift wave turbulence energy \cite{hammett1993,waltz1994,Beer1994,rogers2000,diamond2005zonal,itohzonalphysics2006,connaughton2011}. Understanding this process is a critical step towards the goals of minimizing heat transport in magnetically confined plasmas and designing compact, economically viable fusion devices.

A wide variety of models have been used to study the drift wave turbulence -- zonal flow dynamics in toroidal plasmas, with a varying degree of physics fidelity \cite{hasegawa1978pseudo,hasegawa1983plasma,hammett1993,waltz1994,rogers2000,numata2007bifurcation,watanabe2007,staebler2016,ashourvan2016}; interested readers may find a longer list of relevant references in \cite{diamond2005zonal}. In the present work, we propose a new fluid model which continues the rich tradition of studies based on reduced fluid models \cite{hasegawa1979,wakatani1984collisional,hasegawawakataniPRL1987,dewar2007zonal,numata2007bifurcation,pushkarev2013,stoltzfus2013,parker2014,parker2016,ashourvan2016} initiated by the pioneering works of Hasegawa and Mima \cite{hasegawa1978pseudo} and Hasegawa and Wakatani \cite{hasegawa1983plasma}.    The approximations made to derive many of the reduced models, including ours, are often only satisfied for parameter regimes which do not correspond to the values measured in the edge of magnetic confinement fusion plasma experiments. We therefore do not expect numerical results obtained with reduced fluid models to be in quantitative agreement with measurements. However, because of their relative simplicity, reduced models have played an important role in characterizing the mechanisms responsible for the formation and dynamics of zonal flows and continue to be important tools to provide more insights into these important phenomena. In order to put our new model in context, we propose a very brief review of the main reduced models considered in the plasma physics literature.

The Hasegawa-Mima (HM) \cite{hasegawa1978pseudo} model is a one-field partial differential equation (PDE) which is the simplest known model containing the drift wave turbulence -- zonal flow feedback loop mechanism. It was recognized in the 1990s \cite{dorland1993gyrofluid,hammett1993} that in the context of magnetic confinement fusion, it is more physically relevant to consider a corrected form of the HM equations, now referred to as the modified Hasegawa-Mima (mHM) \cite{dewar2007zonal}, in which the magnetic surface averaged electron density response is subtracted from the original HM (oHM) equation \cite{dorland1993gyrofluid}. The mHM model has the desirable feature of being Galilean invariant under boosts in the poloidal direction \cite{dewar2007zonal}, and is known to lead to a stronger generation of zonal flows. This modification is also at the heart of the new model we propose.

A limitation of the HM models is that they do not contain a natural instability, so that turbulent forcing must be added externally in order to study the turbulence -- zonal flow dynamics. In contrast, Hasegawa and Wakatani \cite{hasegawa1983plasma} have shown that a generalized version of the HM models which includes electron ion friction in the parallel direction naturally contains a drift instability due to the finite plasma resistivity, and thus drift wave turbulence induced transport. This is the original Hasegawa-Wakatani (oHW) model, a system of coupled PDEs for the ion vorticity (which is the Laplacian of the electrostatic potential) and the ion density. However, the oHW model has shortcomings too. Just like the oHM, it is not Galilean invariant. Furthermore, zonal flows are not generated in the oHW model \cite{numata2007bifurcation}, unless one makes the adiabaticity parameter wave number dependent \cite{pushkarev2013}. Numata \textit{et al.} have proposed a modified Hasegawa-Wakatani (mHW) model obtained by subtracting the zonal components from the resistive coupling term. The mHW model addresses the issues of the oHW mentioned above: its drift wave turbulence can lead to the strong generation of zonal flows, and it can be shown that the model has the desired Galilean invariance. Even if so, the mHW has the weakness that it does not converge to either the oHM model or the mHM model in the limit of zero resistivity and zero dissipation, as we will demonstrate numerically in this article. 

We propose a generalized Hasegawa-Wakatani model with an improved treatment of the electron response along the magnetic field lines. Our model addresses the aforementioned limitation of the mHW model by solving for the mHM potential vorticity instead of the vorticity, and reduces to the mHM model in the adiabatic, nondissipative limit, as desired, while maintaining the Galilean invariance of the mHW model. We call our model the \textit{flux-balanced Hasegawa-Wakatani model} (bHW) to emphasize the fact that in this model, the surface-averaged electron density is constant in time when the electrons are adiabatic, as one would physically expect \cite{dorland1993gyrofluid,hammett1993,dewar2007zonal}. Our simple modification to the mHW model leads to major differences in the observed dynamics. Most significantly, the generation of zonal flows is enhanced and the turbulent fluctuations about the zonal mean state are increased. In our new Hasegawa-Wakatani model, we have also chosen to generalize the form of dissipative effects introduced in the oHW model to represent a wider variety of physical processes. As we acknowledged previously, we do not expect quantitative agreement between our fluid model and measurements from the edge of magnetic confinement fusion plasma experiments, but we hope to improve our understanding of the interplay between competing physical effects and of the implications of making simplifying fluid assumptions to model subtle kinetic effects. As an illustration, in this article we make the observation that the simplified linear Landau damping term of the oHW model \cite{hasegawa1983plasma} which we also included in our model, acting mostly on the largest scale modes, may not always act as a pure dissipative effect and can effectively increase the variability in the flow fluctuations. To help the reader throughout the article, we provide a summary of the similarities and differences between the flux-balanced Hasegawa-Wakatani model and the modified Hasegawa-Wakatani model in Table \ref{tab:Comparison-of-models}. 

The structure of this paper is as follows. We first briefly review the classical Hasegawa-Mima and Hasegawa-Wakatani models in Section \ref{sec:historical_models}, in order to provide a historical context for our modifications to the Hasegawa-Wakatani model. In Section \ref{sec:gHW}, we present our new Hasegawa-Wakatani model and derive its conserved quantities. In Section \ref{sec:Features-of-BHW} and \ref{sec:Direct-Numerical-Simulations}, we study the main differences between the mHW model and our model using numerical simulations; Section \ref{sec:Features-of-BHW} focuses on statistical considerations while Section \ref{sec:Direct-Numerical-Simulations} illustrates our conclusions and observations from Section \ref{sec:Features-of-BHW} with snapshots of results of direct numerical simulations, and explores the role of the different dissipation terms. We summarize our work in Section \ref{sec:conclusion}, and present the main linear stability properties of our new balanced model in Appendix \ref{sec:Linear-Instability}.

\begin{table}
\begin{tabular}{>{\raggedright}p{0.15\paperwidth}>{\centering}p{0.3\paperwidth}>{\centering}p{0.3\paperwidth}}
\toprule 
 & \textbf{Flux-balanced Hasegawa-Wakatani (bHW) Model} & \multicolumn{1}{>{\centering}p{0.3\paperwidth}}{\textbf{Modified Hasegawa-Wakatani (mHW) Model}}\tabularnewline
\midrule
\midrule 
Zonal jet structure (Figure \ref{fig:Snapshots}, \ref{fig:Time-series-of-jets}) & Stronger and more turbulent zonal jets are generated & Jets are less persistent, but more steady when present, with weaker fluctuations\tabularnewline
\midrule 
Hydrodynamic limit $\alpha\rightarrow0$ (Figure \ref{fig:Total-statistical-energy}, \ref{fig:Second-moment}, and \ref{fig:Snapshots}) & The bHW model maintains zonal jet structures, with a reduced particle
flux $\overline{\tilde{u}\tilde{n}}$ & \multicolumn{1}{>{\centering}p{0.3\paperwidth}}{The mHW model reduces to fully homogeneous turbulence with strong
particle flux}\tabularnewline
& Stronger variability in the entire variance spectrum
and especially in zonal modes with wavenumber $k_{y}=0$ & Weaker variability in the variance spectrum and no anisotropy in the
zonal direction\tabularnewline
\midrule 
Adiabatic limit $\alpha\rightarrow\infty$ (Figure \ref{fig:Total-statistical-energy}, \ref{fig:Comparison-decay}) & The bHW model converges uniformly to the mHM model with little small-scale fluctuation & \multicolumn{1}{>{\centering}p{0.3\paperwidth}}{The mHW model saturates in a final state different from the mHM final state, with intermittent small-scale vortices}\tabularnewline
\midrule 
\multirow{2}{0.15\paperwidth}{Third-order moment statistics (Figure \ref{fig:Second-moment})} & \multicolumn{2}{>{\centering}p{0.6\paperwidth}}{The mHW and bHW models both have important third-order
moment feedbacks to the statistical equations for the mean and variance.}\tabularnewline
\cmidrule{2-3} 
 & Stronger third-order moment feedback, especially along zonal
modes  & \multicolumn{1}{>{\centering}p{0.3\paperwidth}}{Weaker zonal feedback in the third-order moments}\tabularnewline
\bottomrule
\end{tabular}

\caption{Summary of the similarities and differences between the mHW model and the bHW model observed in the results of direct numerical simulations.\label{tab:Comparison-of-models}}
\end{table}

\section{The Hasegawa-Mima and Hasegawa-Wakatani Models}\label{sec:historical_models}

In this section, we review the central features of the classical
Hasegawa-Mima (HM) and Hasegawa-Wakatani (HW) models, as well as their modified versions \cite{dorland1993gyrofluid,numata2007bifurcation}, known to
excite more realistic and stronger drift-wave -- zonal flow dynamics. This short review is meant to provide the motivation for our new model. The presentation is purposefully brief as clear, longer and more detailed presentations are readily available in the literature \cite{hortonreview1999,diamond2005zonal,dewar2007zonal,numata2007bifurcation}

\subsection{Slab geometry on a two-dimensional rectangular domain}

\begin{figure}
\begin{raggedright}
\centering\includegraphics[height=6.5cm]{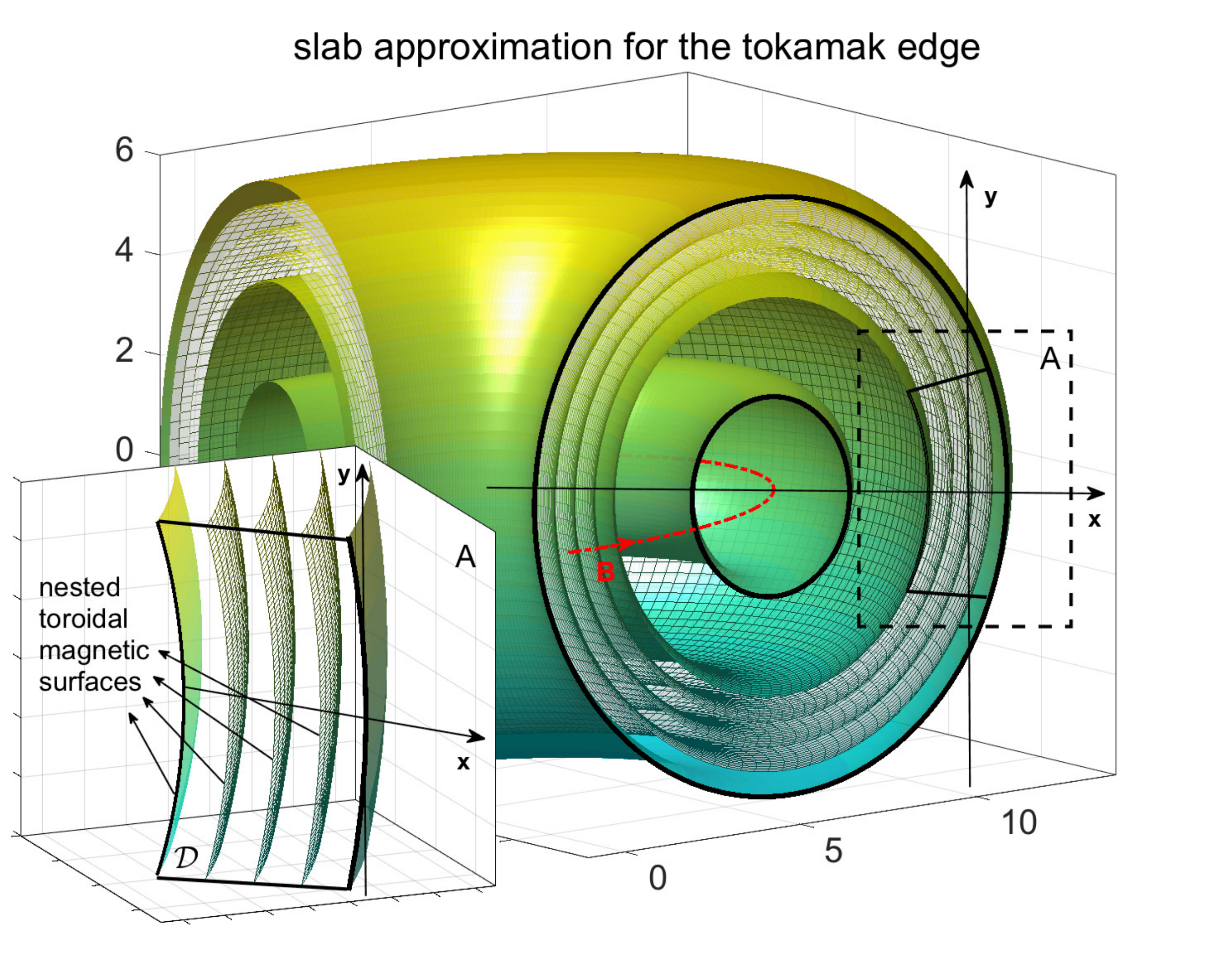}
\par\end{raggedright}

\caption{Nested toroidal flux surfaces in a tokamak geometry. The subplot (A) in the bottom left corner gives an illustration of the slab approximation for the plasma edge, in which the curved flux surfaces are flattened into planes, and shows the coordinate system used in this article. \label{fig:Illustration-of-domain}}
\end{figure}

For simplicity, all the models we will discuss in this article are considered for a shearless slab geometry \cite{balescutransport,dewar2007zonal}, in which the toroidal magnetic surfaces are imagined to be flattened into planes parallel to the $y$ and $z$ axes, as shown in Figure \ref{fig:Illustration-of-domain}, where $(x,y,z)$ is the Cartesian coordinate system used to describe the geometry, $x$ representing the radial distance, which can be viewed as a flux surface label, and $y$ and $z$ playing the roles of the poloidal and toroidal angles respectively. The magnetic field is assumed to be solely in the $z$ direction, $\mathbf{B}=B_{0}\nabla z$, and $B_{0}$ is constant and uniform. The equilibrium density depends on the radial variable, $n_{0}(x)$, and we will treat the density profile within the framework of the standard ``local approximation'' \cite{RicciRogers}, in which $n_{0}'(x)/n_{0}(x)=\mbox{constant}$ \cite{wakatani1984collisional,numata2007bifurcation}. The electron temperature is uniform throughout the plasma, and the ion temperature is assumed to be small compared to the electron temperature: $T_{i}/T_{e}\ll 1$. In all the models discussed in this article, the physical quantities will be assumed to be uniform in the $z$-direction, corresponding to the fact that in strongly magnetized plasmas, the dynamics is highly anisotropic, with much slower variations of the physical quantities along the magnetic field than across the magnetic field. The problem is therefore reduced to a two-dimensional problem in which all quantities only depend on $x$, $y$, and the time $t$, and the computational domain $\Omega$ is a rectangle whose sides have lengths $L_{x}$ and $L_{y}$. Finally, the boundary conditions on the perturbed quantities ($n$, $\varphi$), which are the quantities the models solve for, are periodic in both $x$ and $y$ \cite{wakatani1984collisional}. 



\subsection{The Hasegawa-Mima models\label{sub:The-Hasegawa-Mima-model}}

The Hasegawa-Mima models, both original and modified, are equations for the ion vorticity obtained by combining the ion continuity equation and the ion momentum equation, together with the assumption of an adiabatic electron response to close the system of equations \cite{hasegawa1978pseudo,dewar2007zonal}. The only difference between the oHM and mHM models lies in the treatment of that adiabatic electron response. More than a decade after the introduction of the oHM model, it was indeed recognized that one should subtract the zonal mean of the electrostatic potential in the equation relating the adiabatic electron density and the electrostatic potential in order to prevent unphysical net radial transport of electrons \cite{Dorland1993,dorland1993gyrofluid,dewar2007zonal,pushkarev2013}.

The oHM and mHM equations can be formulated under the same framework
by defining a switch parameter $j$ with $j=0$ for the oHM model and $j=1$ for
the mHM model, which accounts for the different treatment of the adiabatic electrons. The unified equation is \cite{hasegawa1979,hortonreview1999}
\begin{equation}
\frac{\partial q}{\partial t}+J\left(\varphi,q\right)-\kappa\frac{\partial\tilde{\varphi}}{\partial y}=0,\quad q=\nabla^{2}\varphi-\left(\tilde{\varphi}+\delta_{j0}\overline{\varphi}\right).\label{eq:HM_nondim}
\end{equation}
In Eq. (\ref{eq:HM_nondim}), $J\left(\varphi,q\right)=\partial_{x}\varphi\partial_{y}q-\partial_{y}\varphi\partial_{x}q$ is 
the Jacobian associated with the advection term $\mathbf{v}_{E}\cdot\nabla q$, where $\mathbf{v}_{E}$ is the $\mathbf{E}\times\mathbf{B}$ velocity, $t$ is the time normalized to the ion cyclotron frequency $\omega_{ci}=eB_{0}/m$, and $x$ and $y$ are normalized in terms of the hybrid ion thermal Larmor radius $\rho_{s}=\omega_{\mathrm{ci}}^{-1}\left(T_{e}/m_{i}\right)^{1/2}=\sqrt{m_{i}T_{e}}/eB_{0}$, $\delta_{j0}$ is the Kronecker delta, which is equal to 1 if $j=0$ and 0 otherwise, $q=\nabla^2\varphi-(\tilde{\varphi}+\delta_{j0}\overline{\varphi})$ is the \emph{potential vorticity}, with $\varphi=e\phi/T_{e}$ the normalized electrostatic potential, where $e$ is the charge of the electron and $\phi$ the electrostatic potential, and $\kappa=-\mathrm{d}\ln n_{0}/\mathrm{d}x$. A bar over a quantity $f$ represents the zonally-averaged mean of that quantity, which only depends on $x$, and a tilde represents the fluctuation component of $f$, obtained by removing the mean from $f$:
\[
\overline{f}\left(x\right)=\frac{1}{L_{y}}\int f\left(x,y\right)dy,\quad\tilde{f}=f-\overline{f},
\]
where $L_{y}$ is length of the domain in the $y$ direction.

The mHM modification to the oHM model is simple, only appearing in the definition of the potential vorticity $q$. Yet it has important physical implications. First, unlike the oHM model, the mHM model is Galilean invariant under boosts in the poloidal direction \cite{dewar2007zonal}, which is a desired property for our shearless slab geometry. Second, it can be shown, using linear theory, that in the absence of mean flow, the drift wave dispersion relation is identical in both models, and given by $\omega=k_{y}\kappa/(1+k^2)$, where $k_{y}$ is the $y$ component of the wavevector $\mathbf{k}$, and $k$ the magnitude of $\mathbf{k}$. However, in the presence of a constant and uniform background mean flow in the $y$ direction, $\mathbf{v}_{E}=\bar{v}\hat{y}$, the dispersion relation is modified in different ways in the two models \cite{dewar2007zonal}:
\[
\begin{aligned}\mathrm{oHM:\quad} & \omega=\frac{k_{y}\kappa}{1+k^{2}}+\frac{k^{2}}{1+k^{2}}k_{y}\bar{v},\\
\mathrm{mHM:}\quad & \omega=\frac{k_{y}\kappa}{1+k^{2}}+k_{y}\bar{v}.
\end{aligned}
\]
We see that at small scales, $k\gg1$, the dispersion relations of the oHM and mHM models agree, and are given by $\omega=\omega_{*}+k_{y}\bar{v}$, where $\omega_{*}=\kappa k_{y}/(1+k^2)$ is the drift wave frequency. However, at larger scales, i.e. scales comparable to $\rho_{s}$ corresponding to $k\sim 1$, the dispersion relations differ. In the mHM model, the mean flow leads to a simple Doppler shift in the dispersion relation, as expected, but in the oHM model the Doppler shift is reduced by  the factor $k^{2}/(1+k^{2})$. 

It is clear from the dispersion relations above that in the absence of mean flow, or in the presence of a steady, uniform mean flow in the poloidal direction, the Hasegawa-Mima models do not have instabilities. This does not remain true if we assume the presence of a radially varying zonal mean flow. As we will show in an accompanying article more focused on the analysis of direct numerical simulations \cite{qmc2018}, a drift instability can grow in these conditions, and break up the zonal jet structure. Here too, the oHM and mHM models differ, and we find that the zonal jets are more likely to be broken up by this instability in the oHM model than in the mHM model. This provides a hint for the most important physical difference between the oHM and mHM models from the point of view of the present article as well as applications to magnetic confinement fusion, namely the well-known result that the proper treatment of the electron adiabatic response in the mHM model leads to much stronger zonal jet structures when a random forcing term is added to the equation in order to mimick an instability leading to turbulent behavior \cite{hammett1993,dewar2007zonal,chandre2014}. This crucial observation provides the motivation for the mHW model and our new bHW model, which will be the focus of the remainder of this article. 

\subsection{The Hasegawa-Wakatani models}

The Hasegawa-Mima models capture essential features of the drift wave -- zonal flow dynamics when a forcing term is included in the equation, but does not include any internal drift instability to drive the drift wave -- zonal flow feedback loop in the absence of forcing. The \emph{ Hasegawa-Wakatani models} \cite{wakatani1984collisional,numata2007bifurcation} address this limitation by including electron-ion friction, which relaxes the slaving relation between the electron density and the electrostatic potential, leading to a drift wave instability \cite{stoltzfus2013}. Because the one-to-one correspondence between density and potential is lost, the HW models are two-field models for the ion vorticity and the ion density fluctuation. As for the HM models, one can make the distinction between the original HW model (oHW) \cite{wakatani1984collisional} and the modified HW model (mHW) \cite{numata2007bifurcation}, which differ in their treatment of the parallel current. The mHW model accounts for the fact that zonal modes, with wavenumber $k_{y}=0$, do not contribute to the parallel current, while the oHW does not. 
As for the HM models, we can write the two HW models in
a unified form, with the switch parameter $j$ such that $j=0$ for the oHW model, and $j=1$ for the mHW model:

\addtocounter{equation}{0}\begin{subequations}\label{plasma}
\begin{eqnarray}
\frac{\partial\zeta}{\partial t}+J\left(\varphi,\zeta\right) & = & \alpha\left[\left(\tilde{\varphi}-\tilde{n}\right)+\delta_{j0}\left(\overline{\varphi}-\overline{n}\right)\right]+\mu\Delta\zeta\label{eq:plasma1}\\
\frac{\partial n}{\partial t}+J\left(\varphi,n\right)+\kappa\frac{\partial\tilde{\varphi}}{\partial y} & = & \alpha\left[\left(\tilde{\varphi}-\tilde{n}\right)+\delta_{j0}\left(\overline{\varphi}-\overline{n}\right)\right]\label{eq:plasma2}
\end{eqnarray}
\end{subequations} where $\zeta=\Delta\varphi$ is the ion
vorticity and $n=n_{1}/n_{0}$ is the relative density fluctuation, with $N=n_{0}+n_{1}$ the total ion density, and where all the quantities have been normalized in the same way as for the HM models. The term $\mu\Delta\zeta$ is an approximate model for collisional ion viscosity perpendicular to the magnetic field, where $\mu=\tilde{\mu}/(\rho_{s}^2\omega_{ci})$ with $\tilde{\mu}=3T_{i}\nu_{ii}/(10 m_{i}\omega_{ci}^2)$ the kinematic ion viscosity coefficient, $T_{i}$ the ion temperature, $\nu_{ii}$ the ion-ion collision frequency, and $m_{i}$ the ion mass. We note that strictly speaking, Eqs. (\ref{eq:plasma1})--(\ref{eq:plasma2}) are not the mHW model, in the sense that the mHW model has different dissipation terms than the oHW model. Specifically, in the mHW model, dissipation for the ion vorticity is written as $D\Delta^2\zeta$ and dissipation for the ion density is written as $D\Delta^2 n$, with $D$ the same unspecified constant in both equations. We do not dwell further on this distinction, since these terms are not given any physical justification in \cite{numata2007bifurcation}, and seem to only be included in order to guarantee numerical stability. 

The parameter $\alpha\equiv T_{e}k_{z}^{2}/n_{0}e^{2}\eta\omega_{\mathrm{ci}}$, where $\eta$ is the resistivity parallel to the field line, is often referred to as the adiabaticity parameter \cite{camargo1995,pushkarev2013,numata2007bifurcation}. In the collisionless limit, $\alpha\rightarrow\infty$, the electrons have an adiabatic response along the field lines, whereas for small $\alpha$ electron-ion friction decouples the density and electrostatic potential. In order to make the mathematical connection with the HM models in the asymptotic collisionless limit $\alpha\rightarrow\infty$ and $\mu\rightarrow 0$, we can derive an equation for the potential vorticity $q=\nabla^{2}\varphi-n$ valid for both models, by assuming that the oHW and mHW models both only include the viscous dissipation $\mu \Delta\zeta$ as dissipation mechanism (these dissipative terms will not matter in the end since we will take the limit $\mu\rightarrow 0$). Subtracting Eq. (\ref{eq:plasma2}) from Eq. (\ref{eq:plasma1}), we find

\begin{equation}
\frac{\partial q}{\partial t}+J\left(\varphi,q\right)-\kappa\frac{\partial\tilde{\varphi}}{\partial y}=\mu\Delta \zeta,\quad q=\nabla^{2}\varphi-n.\label{eq:HW_vort}
\end{equation}
We see that in the collisionless limit $\mu\rightarrow 0$, Eq. (\ref{eq:HW_vort}) has the same form as the equivalent equation (\ref{eq:HM_nondim}) for the HM models, with the HM potential vorticities $\nabla^{2}\varphi-(\tilde{\varphi}+\delta_{j0}\overline{\varphi})$ replaced with the HW potential vorticity $q=\nabla^{2}\varphi-n$. In the collisionless limit, $\alpha\rightarrow\infty$, Eqs. (\ref{eq:plasma1})--(\ref{eq:plasma2}) give $n=\varphi$ in the oHW model, so the oHW potential vorticity becomes $q=\nabla^{2}\varphi-\varphi$, and the oHW model coincides with the oHM model as desired. 

However, in the mHW model, $\alpha\rightarrow\infty$ leads to $\tilde{n}=\tilde{\varphi}$, leaving $\overline{n}$ unspecified. The potential vorticity converges
to $q\rightarrow\nabla^{2}\varphi-\tilde{\varphi}-\overline{n}$ and the mHW equations (\ref{eq:plasma1})--(\ref{eq:plasma2}) do in general not coincide with the mHM model in the zero resistivity limit. Physically, the mHW model does not guarantee the absence of a net radial transport of electrons when the electrons are adiabatic. Throughout this article we will return to this important conclusion, which is confirmed by our numerical simulations, as it is a key difference between the mHW model and the new bHW model we introduce in the next section. Even if the mHW model does not converge to the mHM model in the collisionless limit, it has been shown that the improved treatment of the parallel current as compared to the oHW model leads to stronger zonal mean structures than in the oHW model \cite{numata2007bifurcation}, just as the mHM model does compared to the oHM model.

Appendix \ref{sec:Linear-Instability} contains a detailed discussion of the linear stability of resistive drift waves in the HW models. At this point, it suffices to say that for $\alpha\neq 0$, $\kappa\neq 0$ and $\mu=0$, all the modes are unstable and the growth rates are smaller for small-scale modes, $k\gg 1$.

\section{A Generalized Flux-Balanced Hasegawa-Wakatani Model}\label{sec:gHW}

We are now ready to introduce the new bHW model, which addresses a major shortcoming of the mHW model, namely the fact that it does not converge to the mHM model in the collisionless limit. In this new model, we also consider generalized dissipation effects, which allow us to study separately the roles of the different dissipation terms, which are often included for the mere sake of numerical stability, and whose simple forms are rarely justified physically.

\subsection{The flux-balanced Hasegawa-Wakatani model}\label{subsec:fbpres}
Let us start with the dissipation terms. Starting from the mHW model in Eq. (\ref{plasma}), we relax the \emph{ad hoc} assumption that the viscosity coefficients are identical for the ion vorticity and the density fluctuation. Furthermore, we include the model Landau damping term considered in the original Hasegawa-Wakatani article when the drift wave frequency $\omega_{*}$ is comparable to $k_{z}v_{T_{i}}$, where $v_{T_{i}}$ is the ion thermal velocity \cite{wakatani1984collisional}. The equations become
\addtocounter{equation}{0}\begin{subequations}\label{bHM_inter}
\begin{eqnarray}
\frac{\partial\zeta}{\partial t}+J\left(\varphi,\zeta\right)\quad\quad\quad\;\: & =&\alpha\left(\tilde{\varphi}-\tilde{n}\right)\;+\mu\Delta\zeta+C\omega_{*}\varphi,\label{eq:bHM_inter1}\\
\frac{\partial n}{\partial t}+J\left(\varphi,n\right)+\kappa\frac{\partial\tilde{\varphi}}{\partial y} & =&\alpha\left(\tilde{\varphi}-\tilde{n}\right)\;+D\Delta n.\label{eq:bHM_inter2}
\end{eqnarray}
\end{subequations}
Here, $\mu\Delta\zeta$ is the same ion collisional viscosity term as in (\ref{eq:plasma1}), and $D\Delta n$ can be viewed as a dissipation term due to the collisional diffusion of the electrons perpendicular to the magnetic field \cite{camargo1995}. Lastly, $C\omega_{*}\varphi$, with $C=T_{i}/(\omega_{c_{i}}T_{e})$, is the simple model for ion Landau damping introduced in \cite{wakatani1984collisional}. Observe that Hasegawa and Wakatani suggest a cutoff for this term, which they set to zero when $\omega_{*}$ is large. For simplicity, we do not impose such a cutoff, noticing that the absence of cutoff will only be noticeable when $k_{x}$ is small and $k_{y}$ is moderate, since $\omega_{*}$ is small for large $k$, as well as for small $k_{y}$. Furthermore, even if $\omega_{*}$ can be larger than $k_{z}v_{T_{i}}$ when $k_{x}$ is small and $k_{y}$ moderate, the ion viscosity term, $\mu\Delta\zeta$, is still likely to dominate in these circumstances, since it contains a term proportional to $k_{y}^4$.

We would like to stress the fact that to the best of our knowledge, the forms of our dissipation terms $\mu\Delta\zeta$, $D\Delta n$ and $C\varphi$ cannot be derived rigorously from kinetic theory for magnetic confinement plasmas. However, our choice to consider distinct coefficients $\mu$ and $D$ and to introduce different forms of dissipation allows us to analyze the relative importance of these effects, and determine which effects would need to be modeled more accurately in a higher fidelity model. For example, as we will discuss in more detail in Section \ref{sec:Direct-Numerical-Simulations} in the light of our numerical results, and in our linear stability analysis in Appendix \ref{sec:Linear-Instability}, we find that Landau damping as modeled in Eq. (\ref{bHM_inter}) has a stabilizing effect on the largest scales and at the same time increases the variability of the small-scale fluctuations. 

We now turn to the key modification to the mHW model we make in the bHW model. As we did in Section \ref{sec:historical_models}, if we subtract Eq. (\ref{eq:bHM_inter2}) from Eq. (\ref{eq:bHM_inter1}) the terms with the adiabaticity parameter cancel, and we obtain an equation for the potential vorticity in the mHW model $q^{\mathrm{m}}\equiv\zeta-n=\Delta\varphi-n$ which reads
\begin{equation}
\frac{\partial q}{\partial t}+J\left(\varphi,q\right)-\kappa\frac{\partial\tilde{\varphi}}{\partial y}=\left[\left(\mu-D\right)\Delta^{2}+C\omega_{*}\right]\varphi+D\Delta q.\label{eq:pv_incomplete}
\end{equation}

As discussed in Section \ref{sec:historical_models}, the issue with the equation above is that in the limit $\mu,D,C\rightarrow 0$ and $\alpha\rightarrow\infty$ it violates the balanced electron response on the magnetic surfaces in the mHM sense, due to the inclusion of a non-zero zonal mean state $\overline{n}$ in the potential vorticity. We propose a simple modification to the mHW model to address this limitation, namely to replace the original potential vorticity $q^{\mathrm{m}}$ with the corrected form $q^{\mathrm{b}}=\zeta-\tilde{n}$. The resulting flux-balanced equation for the new potential vorticity $q^{\mathrm{b}}$ does not depend on the zonal mean density $\overline{n}$ explicitly. The immediate and desired implication of this modification is that in the collisionless limit, $\alpha\rightarrow\infty$ gives the slaving relation $\tilde{n}\rightarrow\tilde{\varphi}$ from Eq. (\ref{eq:bHM_inter2}), and the potential vorticity $q^{\mathrm{b}}$
converges to the mHM potential vorticity $q\rightarrow\nabla^{2}\varphi-\tilde{\varphi}$ unlike the mHW potential vorticity. As a result, in the adiabatic limit, and in the absence of the dissipative terms, Eq. (\ref{eq:pv_incomplete}) when expressed for $q^{\mathrm{b}}$ is identical to Eq. (\ref{eq:HM_nondim}) of the mHM model.

We now have all the elements to introduce the \emph{flux-balanced Hasegawa-Wakatani (bHW) equations} for the balanced potential vorticity $q^{\mathrm{b}}=\nabla^{2}\varphi-\tilde{n}$
and the relative particle density $n=\overline{n}+\tilde{n}$ as \addtocounter{equation}{0}\begin{subequations}\label{plasma_balanced}
\begin{eqnarray}
\frac{\partial q^{\mathrm{b}}}{\partial t}+J\left(\varphi,q^{\mathrm{b}}\right)-\kappa\frac{\partial\tilde{\varphi}}{\partial y} & = & \left[\left(\mu-D\right)\Delta^{2}+C\omega_{*}\right]\varphi\;+D\Delta q^{\mathrm{b}},\qquad q^{\mathrm{b}}=\Delta \varphi-\tilde{n}\label{eq:plasma_balc1}\\
\frac{\partial n}{\partial t}+J\left(\varphi,n\right)+\kappa\frac{\partial\tilde{\varphi}}{\partial y} & = & \alpha\left(\tilde{\varphi}-\tilde{n}\right)\qquad\qquad\quad+D\Delta n,\label{eq:plasma_balc2}
\end{eqnarray}
\end{subequations}
The potential vorticity equation (\ref{eq:plasma_balc1}) has a form analogous to the vorticity equation in the mHM model, Eq. (\ref{eq:HM_nondim}), but unlike the mHM model, it contains a resistive drift instability through the coupling with the density fluctuation $\tilde{n}$. The equation for the relative particle density fluctuation (\ref{eq:plasma_balc2}) is the same as in the mHW equation (\ref{eq:plasma2}), except for the dissipation term $D\Delta n$.


\subsection{Comparison with the modified Hasegawa-Wakatani model}

In this section and the next, we present some elementary properties of our model. We begin with a direct comparison with the mHW model. A good starting point is to write the equations for the balanced potential vorticity $q^{\mathrm{b}}=\Delta\varphi-\tilde{n}$ according to the mHW model and to the bHW model side by side

\addtocounter{equation}{0}\begin{subequations}\label{eq:comp_BMHW}
\begin{eqnarray}
\mathrm{mHW:}&\quad\frac{\partial q^{\mathrm{b}}}{\partial t}+J\left(\varphi,q^{\mathrm{b}}\right)+\frac{\partial\left(\overline{\tilde{u}\tilde{n}}\right)}{\partial x}+\left(\frac{\partial\overline{n}}{\partial x}-\kappa\right)\frac{\partial\tilde{\varphi}}{\partial y}\;& =\mathrm{dissip.}\label{eq:comp1}\\
\mathrm{bHW:}&\quad\frac{\partial q^{\mathrm{b}}}{\partial t}+J\left(\varphi,q^{\mathrm{b}}\right)\qquad\qquad\qquad\quad\quad-\kappa\frac{\partial\tilde{\varphi}}{\partial y}\;& =\mathrm{dissip.}\label{eq:comp2}
\end{eqnarray}
\end{subequations}
with $\tilde{u}=-\partial\tilde{\varphi}/\partial y$ the velocity fluctuation
along the $x$ direction and
\[
\frac{\partial\left(\overline{\tilde{u}\tilde{f}}\right)}{\partial x}\equiv\overline{J}\left(\tilde{\varphi},\tilde{f}\right)=\frac{1}{L_{y}}\int\left(\frac{\partial\tilde{\varphi}}{\partial x}\frac{\partial\tilde{f}}{\partial y}-\frac{\partial\tilde{\varphi}}{\partial y}\frac{\partial\tilde{f}}{\partial x}\right)dy,
\]
with $\tilde{f}=\tilde{n}$ as in the equation above, or $\tilde{f}=\tilde{q}$ as we will have below. The two differences in the mHW model are: 1) the zonal density gradient feedback term $(\partial_x\overline{n})\partial\tilde{\varphi}/\partial y$, which modifies the original background density gradient profile $-\kappa=\mathrm{d}(\ln n_{0})/\mathrm{d}x$;
and 2) the additional eddy flux feedback, $\partial_x\left(\overline{\tilde{u}\tilde{n}}\right)$,
due to the transport of the particle density along the $x$ direction.
Both terms are zonal mean density feedbacks which can act on the largest scales (to change the zonal jet structure), and can also modify the structures
in the small-scale fluctuations (to change the particle flux). It
is important to observe that although the zonal mean density $\overline{n}$ does not appear in Eq. (\ref{eq:comp2}), $q^{\mathrm{b}}$ depends on $\overline{n}$ in the bHW model through the dependence of $\tilde{n}$ on $\overline{n}$, which we will write explicitly shortly. For the same reason, the vorticity $\zeta=\nabla^2\varphi$ depends on $\overline{n}$ in the mHW model.

Let us now turn to the equations for the zonal mean quantities $\overline{q}^{b}\left(x\right)$
and $\overline{n}\left(x\right)$. If, in order to highlight similarities between the mHW and bHW models, we modify the \emph{ad hoc} dissipation terms in the mHW model in such a way that the dissipation in the density equation is $D\Delta n$ in both models, we obtain the following equations for the mean balanced potential vorticity $\overline{q}^{\mathrm{b}}=\partial^2\overline{\varphi}/\partial x^2$, 
\addtocounter{equation}{0}\begin{subequations}\label{eq:plasma_mean-bhw-potvor}
\begin{eqnarray}
\mathrm{mHW:}&\quad\frac{\partial\overline{q}^{\mathrm{b}}}{\partial t}+\frac{\partial\left(\overline{\tilde{u}\tilde{q}^{\mathrm{b}}}\right)}{\partial x}+\frac{\partial\left(\overline{\tilde{u}\tilde{n}}\right)}{\partial x}&=D\frac{\partial^2\overline{q}^{\mathrm{b}}}{\partial x^{2}}\label{eq:meanqmHW}\\
\mathrm{bHW:}&\quad\frac{\partial\overline{q}^{\mathrm{b}}}{\partial t}+\frac{\partial\left(\overline{\tilde{u}\tilde{q}^{\mathrm{b}}}\right)}{\partial x}\quad\;\quad\quad&=\left[\left(\mu-D\right)\frac{\partial^4}{\partial x^{4}}+C\omega_{*}\right]\overline{\varphi}+D\frac{\partial^2\overline{q}^{\mathrm{b}}}{\partial x^{2}},\label{eq:meanqbHW}
\end{eqnarray}
\end{subequations}
and an equation for the zonal mean density $\overline{n}$ which is identical in both models:
\begin{equation}
\frac{\partial\overline{n}}{\partial t}+\frac{\partial\left(\overline{\tilde{u}\tilde{n}}\right)}{\partial x}  =D\frac{\partial ^2\overline{n}}{\partial x^2}.
\label{eq:plasma_mean-bhw-density}
\end{equation}
Equations (\ref{eq:plasma_mean-bhw-potvor}) and (\ref{eq:plasma_mean-bhw-density}) highlight the feedback of the fluctuations on the zonal mean states, through the nonlinear coupling terms $\partial_x\left(\overline{\tilde{u}\tilde{q}^{\mathrm{b}}}\right)$ and $\partial_x\left(\overline{\tilde{u}\tilde{n}}\right)$. In the mHW model, the density fluctuation feedback $\partial_x\left(\overline{\tilde{u}\tilde{n}}\right)$ is also present in the equation for the mean balanced potential vorticity (\ref{eq:meanqmHW}), just like it is in Eq. (\ref{eq:comp1}). Considering the steady-state version of Eq. (\ref{eq:plasma_mean-bhw-density}), we see that a strong particle flux $\overline{\tilde{u}\tilde{n}}$ can lead to a large zonal mean density gradient $\partial\overline{n}/\partial x$ which can then become a significant contribution in the term $(\partial_x\overline{n}-\kappa)\partial\tilde{\varphi}/\partial y$ in Eq. (\ref{eq:comp1}), and effectively alter the background density gradient in the mHW model. 

A summary of the similarities and differences between the bHW and mHW models is given in the table in Appendix \ref{sec:comparison-table}, including the equations for the total energy $E$ and enstrophy $W$ which we will discuss in the next section. Once more, in order to draw the appropriate parallels between the two models, we modified the form of the \emph{ad hoc} dissipation originally proposed in the mHW model \cite{numata2007bifurcation} in such a way as to make the dissipation term agree in the density equation of both models.



\subsection{Conservation laws for the flux-balanced Hasegawa-Wakatani model}\label{subsec:cons}

For the analysis of the fundamental properties of a new model and the verification of numerical codes, it is always helpful to identify dynamical invariants. We define the total energy $E$ and potential enstrophy $W$ \cite{numata2007bifurcation} for the bHW model as
\begin{equation}
\begin{aligned}E=\overline{E}+\tilde{E} & =\frac{1}{2}\int_{\Omega}\left(\left|\overline{\varphi}_{x}\right|^{2}+\overline{n}^{2}\right)dV+\frac{1}{2}\int_{\Omega}\left(\left|\nabla\tilde{\varphi}\right|^{2}+\tilde{n}^{2}\right)dV,\\
W=\overline{W}+\tilde{W} & =\frac{1}{2}\int_{\Omega}\overline{q^{\mathrm{b}}}^{2}dV+\frac{1}{2}\int_{\Omega}\widetilde{q^{\mathrm{b}}}^{2}dV,
\end{aligned}
\label{eq:energy_plasma}
\end{equation}
where the integrals are over the computational domain $\Omega$, which is a periodic box with sides $L_{x}$ and $L_{y}$, and where we have separated $\overline{E},\overline{W}$, the energy and enstrophy of the zonal mean state with $\overline{q^{\mathrm{b}}}=\partial^2\overline{\varphi}/\partial x^2$,
from $\tilde{E},\tilde{W}$, the energy and enstrophy of the fluctuations
about the zonal mean state, with $\widetilde{q^{\mathrm{b}}}=\Delta\tilde{\varphi}-\tilde{n}$. Note that the enstrophy $W$ in the bHW model is defined in terms of the balanced potential
vorticity $q^{\mathrm{b}}=\nabla^{2}\varphi-\tilde{n}$. It is easy to verify that the nonlinear terms $J(\varphi,q^{\mathrm{b}})$ and $J(\varphi,n)$ in (\ref{plasma_balanced}) conserve both the energy and the balanced enstrophy, and we obtain the following dynamical equations for the total energy and potential enstrophy
\begin{equation}
\begin{aligned}\frac{dE}{dt} & =\kappa\int_{\Omega}\overline{\tilde{u}\tilde{n}}\left(1+\kappa^{-1}\overline{v}\right)dV-\alpha\int_{\Omega}\left(\tilde{n}-\tilde{\varphi}\right)^{2}dV-D_{E},\\
\frac{dW}{dt} & =\kappa\int_{\Omega}\overline{\tilde{u}\tilde{n}}dV-D_{W},
\end{aligned}
\label{eq:dyn_ene}
\end{equation}
where $\overline{v}=\frac{\partial\overline{\varphi}}{\partial x}$ and $D_{E}$ and $D_{W}$ come from the dissipation terms in the bHW equations:
\[
\begin{aligned}D_{E} & =\int_{\Omega}\left(C\omega_{*}\varphi^{2}+\mu\left|\Delta\varphi\right|^{2}+D\left|\nabla n\right|^{2}\right)dV,\\
D_{W} & =\int_{\Omega}\left[\left(D-\mu\right)\Delta\tilde{\varphi}\Delta\tilde{n}+C\omega_{*}\tilde{n}\tilde{\varphi}\right]dV+\int_{\Omega}\left[C\omega_{*}\left|\nabla\varphi\right|^{2}+\mu\left|\nabla q^{\mathrm{b}}\right|^{2}+\left(D-\mu\right)\left|\nabla\tilde{n}\right|^{2}\right]dV.
\end{aligned}
\]
Comparing the energy and enstrophy equations (\ref{eq:dyn_ene}) with the analogous equations for $E$ and $W$ in the mHW model, as given in the table above, we notice the additional term $\int_{\Omega}\overline{v}\overline{\tilde{u}\tilde{n}}dV$ in the energy equation for the bHW model. This contribution to the total energy from the fluctuations originates from the eddy diffusivity term $\left(\overline{\tilde{u}\tilde{n}}\right)_{x}$
in the equation for the vorticity in the bHW model, which is absent in the mHW model. This additional term, in which we recognize the advection by the mean velocity $\overline{v}$, represents the zonal flow transport of the particle flux $\overline{\tilde{u}\tilde{n}}$. 

As in the mHW model, the resistive coupling of $\tilde{n}$ and $\tilde{\varphi}$ through the adiabaticity $\alpha(\tilde{\varphi}-\tilde{n})$ gives a negative-definite term in the energy equation and acts as an energy sink. That term does not enter in the evolution equation for the enstrophy $W$. The dissipation for
the total energy $D_{E}$ is always positive definite. This is not true for the dissipation $D_{W}$ of the potential enstrophy, whose terms may not all always be positive. Specifically, the first integral on the right
hand side contains interactions between the fluctuating potential
and density $\Delta\tilde{\varphi}\Delta\tilde{n},\tilde{n}\tilde{\varphi}$
due to ion Landau damping and the fact that the damping coefficients $\mu$ and $D$ are not equal; the last term in the second integral, $\left(D-\mu\right)\left|\nabla\tilde{n}\right|^{2}$, is always negative if $D<\mu$ and can thus act as a source of potential enstrophy instead of a sink. In Section \ref{sub:The-effects-inhomo}, we illustrate these different dissipation effects on the flow field and energy spectra through direct numerical simulations.

To close this section, we propose a more mathematically rigorous proof of the convergence of the bHW model to the mHM model in the adiabatic limit $\alpha\rightarrow\infty$. We have already observed that formally, the limit $\alpha\rightarrow\infty$ implies that $\tilde{\varphi}=\tilde{n}$, meaning that the bHW potential vorticity is equal to the mHM potential vorticity, and that the models are unified in this limit. Here, we rely on the energy and enstrophy equations (\ref{eq:dyn_ene}) to find a more precise description of the relation between the bHW
model and the mHM model in the adiabatic limit. For simplicity, we assume that $\mu=D$ and $C=0$, so that the dissipation terms in the
energy/enstrophy equations take the simple form $D_{W}=D\int\left|\nabla q^{\mathrm{b}}\right|^{2},$ and $D_{E}=D\int\left(\left|\Delta\varphi\right|^{2}+\left|\nabla n\right|^{2}\right)$. We then consider states in statistical equilibrium, i.e. such that the time derivatives for the statistical
expectations $\langle E\rangle$ and $\langle W\rangle$ vanish. Here, $\langle\cdot\rangle$ represents an ensemble average, which could in principle be estimated with a Monte-Carlo approach, using a large number of numerical simulations. When statistical equilibrium has been reached, the enstrophy
equation in (\ref{eq:dyn_ene}) gives us the following estimate for the total particle flux,
\[
\Gamma^{\mathrm{eq}}\equiv\langle\int_{\Omega}\overline{\tilde{u}\tilde{n}}dV\rangle_{eq}=\kappa^{-1}D\langle\int_{\Omega}\left|\nabla q^{\mathrm{b}}\right|^{2}dV\rangle_{eq}>0,
\]
where $\langle\cdot\rangle_{\mathrm{eq}}$ can be viewed as the time average along the stationary trajectory of the functional $f$ by making the usual ergodic hypothesis. This relation shows that there always exists a statistically positive particle flux towards the boundary of the domain, and that the total
particle flux $\Gamma^{\mathrm{eq}}$ depends on the ratio $D/\kappa$.

Turning to the equilibrium statistical equation for the total energy,
we obtain the following estimate from the balance between the flux
and dissipation terms 
\[
\langle\int_{\Omega}\left(\tilde{n}-\tilde{\varphi}\right)^{2}dV\rangle_{\mathrm{eq}}=\alpha^{-1}\langle\left(\kappa\int_{\Omega}\overline{\tilde{u}\tilde{n}}\left(1+\kappa^{-1}\overline{v}\right)dV-D_{E}\right)\rangle_{\mathrm{eq}}\leq\alpha^{-1}\langle\left(VD_{W}-D_{E}\right)\rangle_{\mathrm{eq}}.
\]
To derive the inequality above, we used the identity $\kappa\Gamma^{\mathrm{eq}}=\langle D_{W}\rangle_{\mathrm{eq}}$ and assumed that the zonal mean flow can be bounded as $V=\max_{\Omega}\left|1+\kappa^{-1}\overline{v}\right|$.
The statistical expectation on the right hand side of the inequality
is positive and finite in the equilibrium state. Therefore, the expectation $\langle\left(\tilde{n}-\tilde{\varphi}\right)^{2}\rangle_{\mathrm{eq}}$
vanishes as $\alpha\rightarrow\infty$ since the right-hand side of the inequality goes to zero. This demonstrates that in the adiabatic limit, the fluctuations $\tilde{n}$ and $\tilde{\varphi}$ approach the Boltzmann relation $\tilde{n}=\tilde{\varphi}$ in the mean square sense under expectation.
Note however that there may still exist a non-zero zonal structure in the
density $\overline{n}\left(x\right)$ in this limit, in addition to the fluctuation $\tilde{n}=\tilde{\varphi}$.

\section{Statistical analysis of the properties of the Flux-Balanced Hasegawa-Wakatani Model\label{sec:Features-of-BHW}}

For the remainder of this article, we turn to direct numerical simulations to highlight characteristic features of the bHW model, and compare them with the mHW model. We solve the equations on a doubly periodic square domain of size $L$ along each side, so that the smallest wavenumber is $2\pi/L$, which is also the spacing $\Delta k$ between any two wavenumbers. We write the quantities $\left(\varphi,n,\zeta\right)$ we solve for 
as the following Fourier series
\[
\varphi=\sum_{\mathbf{k}}\hat{\varphi}_{\mathbf{k}}\left(t\right)e^{i\mathbf{\tilde{k}\cdot x}},\quad n=\sum_{\mathbf{k}}\hat{n}_{\mathbf{k}}\left(t\right)e^{i\mathbf{\tilde{k}\cdot x}},\quad\zeta=\sum_{\mathbf{k}}-\tilde{k}^{2}\hat{\varphi}_{\mathbf{k}}\left(t\right)e^{i\mathbf{\tilde{k}\cdot x}},
\]
where the spatial variables $\mathbf{x}=\left(x,y\right)\in\left[-L/2,L/2\right]\times\left[-L/2,L/2\right]$ and the corresponding wavenumbers are
\[
\mathbf{\tilde{k}}=\frac{2\pi}{L}\mathbf{k},\quad\mathbf{k}=\left(k_{x},k_{y}\right)\in[-\frac{N}{2}+1,\frac{N}{2}]\times[-\frac{N}{2}+1,\frac{N}{2}].
\]
where $N$ is the number of modes we keep in our simulations. For all the results shown in this article, we used $L=40$ and $N=256$. We know from our linear stability analysis (see Appendix \ref{sec:Linear-Instability}) that the larger the domain size $L$, the more unstable modes we will have inside the circle of strong instability. We solve the equations using a pseudo-spectral approach, in which the nonlinear terms are calculated in real space instead of Fourier space. To stabilize the truncated numerical system, hyperviscosities $\nu\Delta^{s}q^{\mathrm{b}}$ and $\nu\Delta^{s}n$ are added in the vorticity and density equations respectively, with $\nu=7\times 10^{-21}$ and order $s=8$. We rely on a fourth-order
explicit-implicit Runge-Kutta scheme for time stepping, where the stiff hyperviscosity operator $\nu\Delta^{8}$ is the only term treated implicitly.
In agreement with the difference in the formulations of the mHW and bHW models, we numerically solve for the unknowns $\left(\zeta,n\right)$ when we consider the mHW model, and solve for the balanced potential vorticity $q^{\mathrm{b}}=\nabla^{2}\varphi-\tilde{n}$ and $n$ when we consider the bHW model.

We keep the background density gradient fixed at $\kappa=0.5$ and will vary the value of $\alpha$ in the range $[0.01,10]$. In this section, the dissipation coefficients $\mu$ and $D$ are set to $\mu=5\times10^{-4},D=5\times10^{-4}$, and the ion Landau damping term is turned off by setting $C=0$. In contrast, in section \ref{sec:Direct-Numerical-Simulations} we will consider different values for the parameters $\mu,D$, and $C$ to analyze the role of dissipation effects more closely. The parameters we used for all our numerical simulations in this article are summarized in Table \ref{tab:Basic-model-parameter}. 

\begin{table}
\begin{tabular}{cc}
\toprule 
domain size $L$ & 40\tabularnewline
\midrule 
spatial resolution $N$ & 256\tabularnewline
\midrule 
time step $\Delta t$ & $1\times10^{-2}$\tabularnewline
\midrule 
mean density gradient $\kappa$ & 0.5\tabularnewline
\midrule 
adiabatic particle resistivity $\alpha$ & 0.01 -- 10\tabularnewline
\midrule 
kinetic ion viscosity $\mu$ & $5\times10^{-4},20\times10^{-4}$\tabularnewline
\midrule 
cross-field diffusion $D$ & $5,20\times10^{-4}$\tabularnewline
\midrule 
ion Landau damping $C$ & 0, 0.01\tabularnewline
\midrule 
hyperviscosity $\nu$  & $7\times10^{-21}$\tabularnewline
\midrule 
order of hyperviscosity $s$ & 8\tabularnewline
\bottomrule
\end{tabular}

\caption{Model parameters and their chosen values for our numerical simulations.\label{tab:Basic-model-parameter}}
\end{table}

\subsection{Statistical comparison between the balanced and modified Hasegawa-Wakatani models\label{sub:Comparison-in-transition}} 

\subsubsection{Transition from the turbulence dominated regime to the zonal flow dominated regime}

Previous work has shown that the randomness of the turbulent flow depends sensitively on the parameters $\kappa$ and $\alpha$ \cite{numata2007bifurcation}. This can be readily understood from the linear stability analysis we present in Appendix A, since $\kappa/\alpha$ determines the size of the region of instability in $k$-space. Hence, if $\kappa$ is held fixed, a smaller value of $\alpha$, corresponding to higher plasma resistivity, leads to a larger region of instability, and a more energetic and turbulent vorticity field. When the value of $\alpha$ is increased, the vorticity field is regularized into dominant anisotropic zonal jets. This is the phenomenon we now study, comparing the bHW and mHW models.

Because of the turbulent nature of the flows, it is more appropriate to adopt a statistical viewpoint, and instead of looking at physical quantities at a given instant in time, we take time-averages of the solution once the stationary state is reached. We will focus on the total kinetic energy and the relative enstrophy for both models, given by
\[
\int_{\Omega}\left|\nabla\varphi\right|^{2}dV=\sum_{\mathbf{k}}\tilde{k}^{2}\left|\hat{\varphi}_{k}\right|^{2},\quad\int_{\Omega}\zeta^{2}dV=\sum_{\mathbf{k}}\tilde{k}^{4}\left|\hat{\varphi}_{k}\right|^{2}.
\]
The relative enstrophy gives more weight to the small scale modes, while the kinetic energy puts more emphasis on the large scale structures. We decompose $\varphi$ and $\zeta$ into their time averages $\langle\varphi\rangle_{\mathrm{eq}}$, $\langle\zeta\rangle_{\mathrm{eq}}$ and their time fluctuating parts $\varphi'$, $\zeta'$ according to $\varphi =\langle\varphi\rangle_{\mathrm{eq}}+\varphi'$, $\zeta=\langle\zeta\rangle_{\mathrm{eq}}+\zeta'$, and compute the energy and enstrophy in the time-averaged mean
\[
\int_{\Omega}\left|\langle \nabla\varphi\rangle_{\mathrm{eq}}\right|^{2}dV\;,\qquad\int_{\Omega}\langle\zeta\rangle_{\mathrm{eq}}^{2}dV
\]
and the time-averaged kinetic energy and enstrophy of the variances
\[
\int_{\Omega}\langle|\nabla\varphi'|^2\rangle_{\mathrm{eq}} dV=\int_{\Omega}\langle\left|\nabla\varphi-\langle\nabla\varphi\rangle_{\mathrm{eq}}\right|^2\rangle dV\;,\qquad\int_{\Omega}\langle\zeta'^2\rangle_{\mathrm{eq}} dV=\int_{\Omega}\langle(\zeta-\langle\zeta\rangle_{\mathrm{eq}})^2\rangle_{\mathrm{eq}} dV
\]

\begin{figure}
\subfloat[Kinetic energy $\int_{\Omega}|\nabla\varphi|^{2}dV$ of the variance and of the mean]{\includegraphics[scale=0.34]{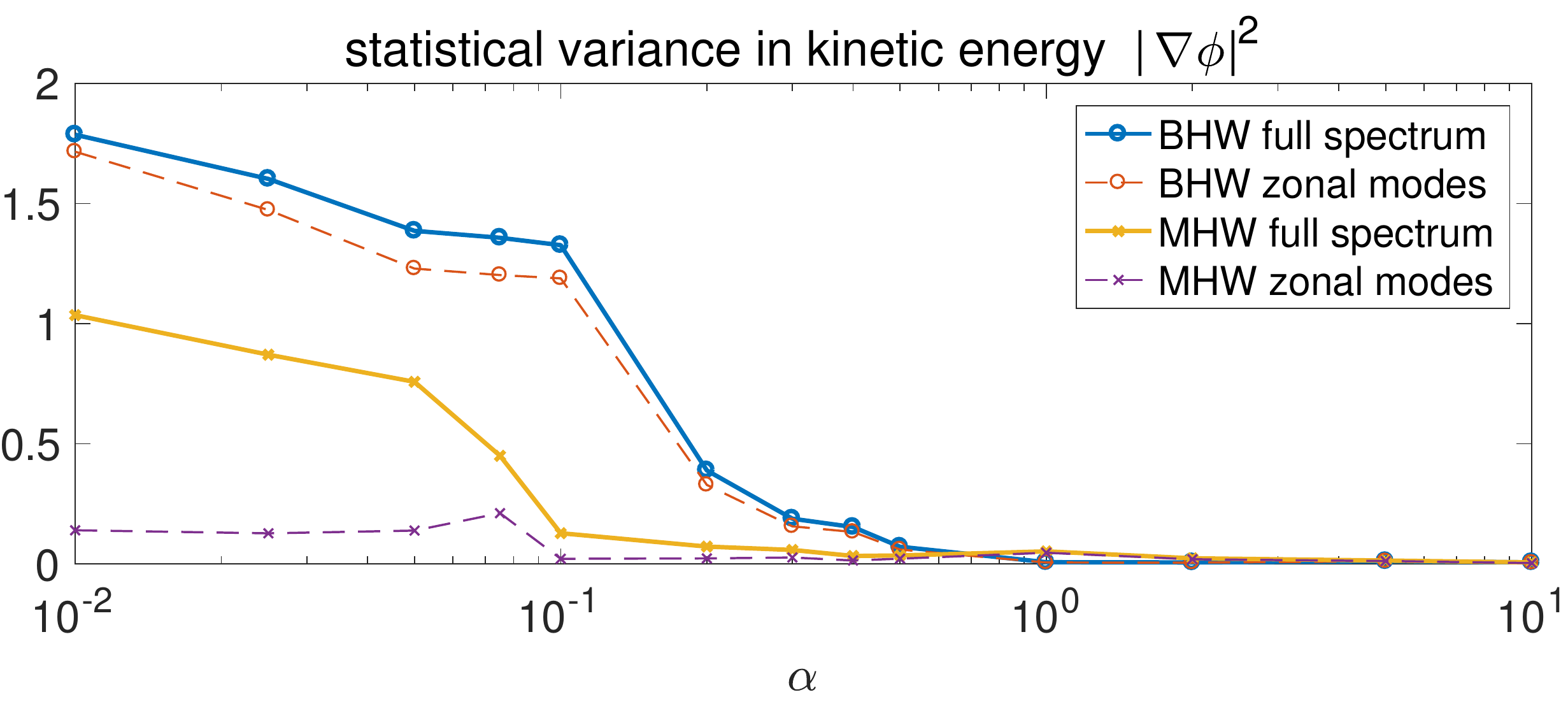}\includegraphics[scale=0.34]{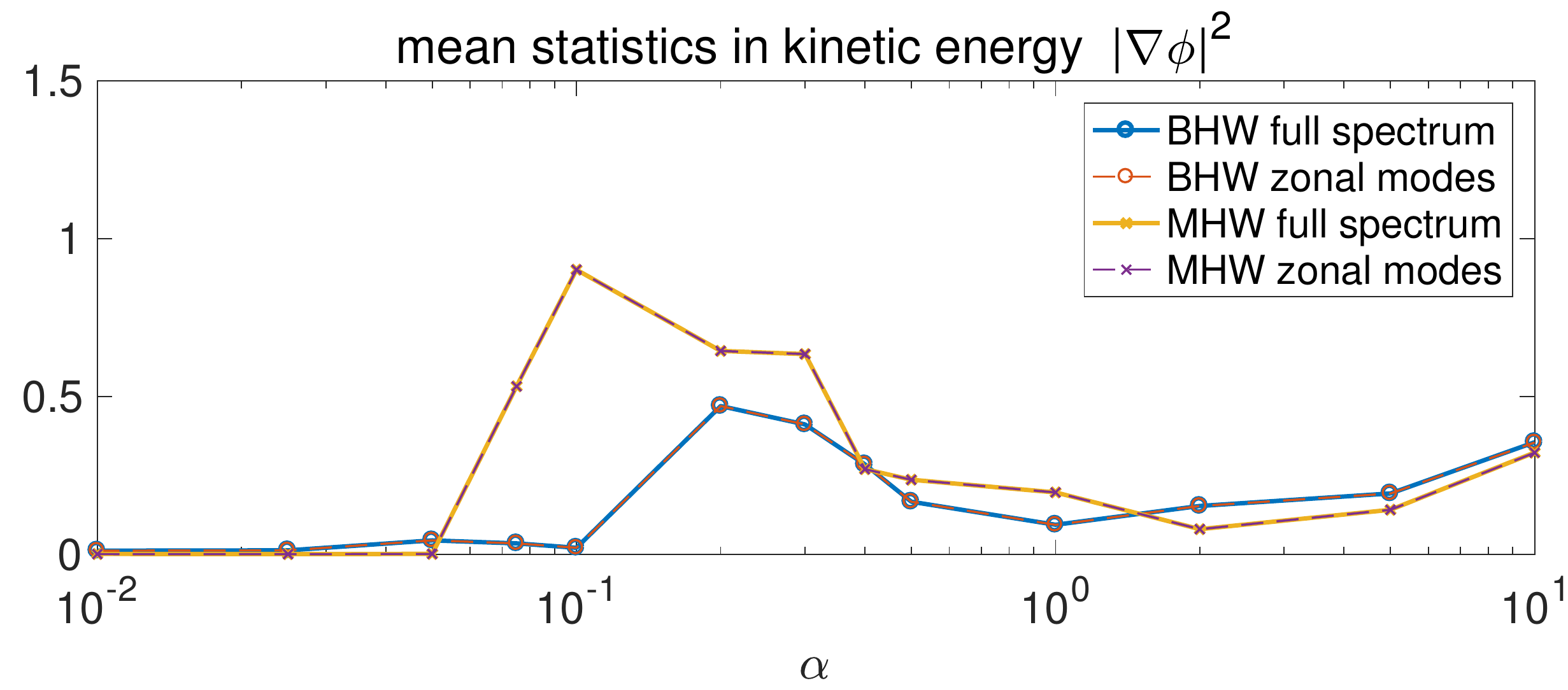}
}

\subfloat[Relative enstrophy $\int_{\Omega}\zeta^{2}dV$ of the variance and of the mean ]{\includegraphics[scale=0.34]{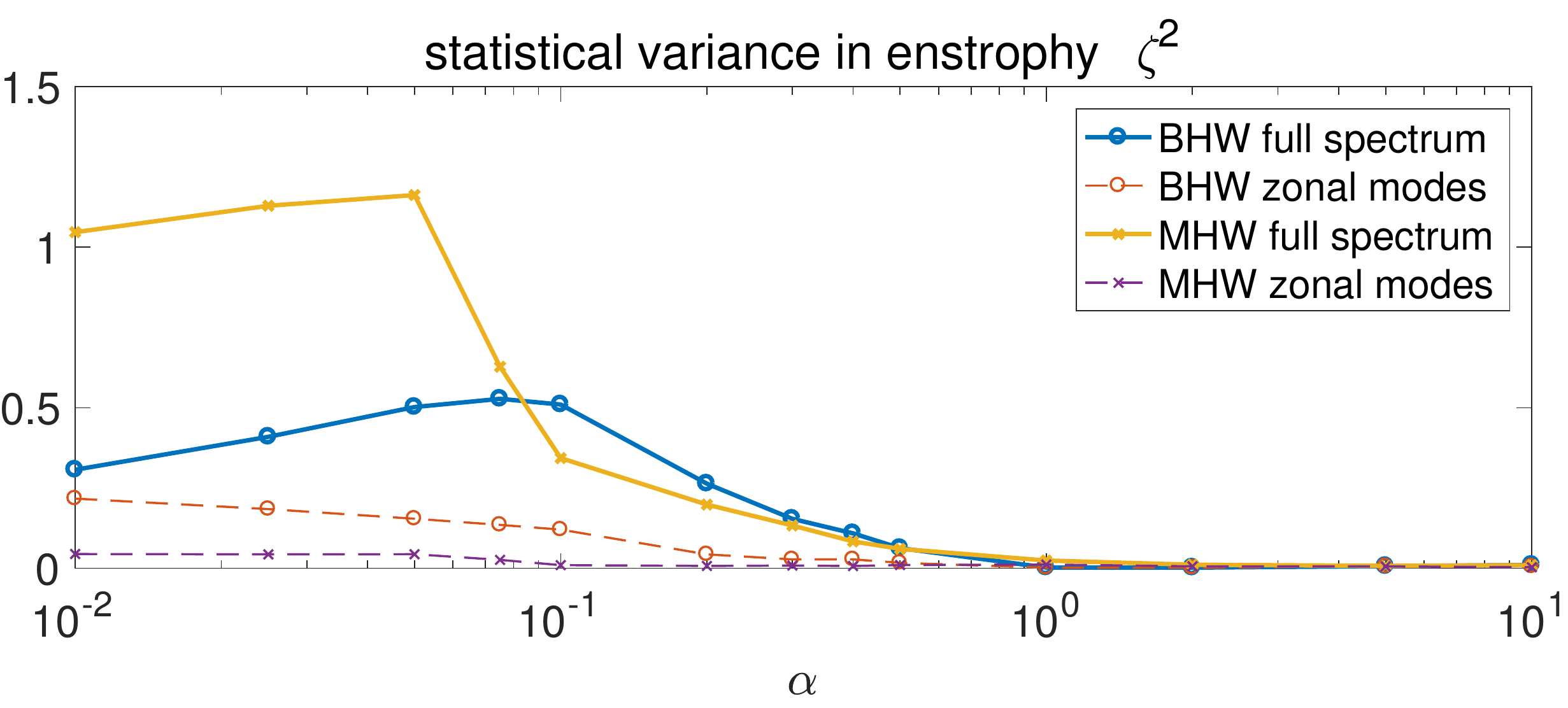}\includegraphics[scale=0.34]{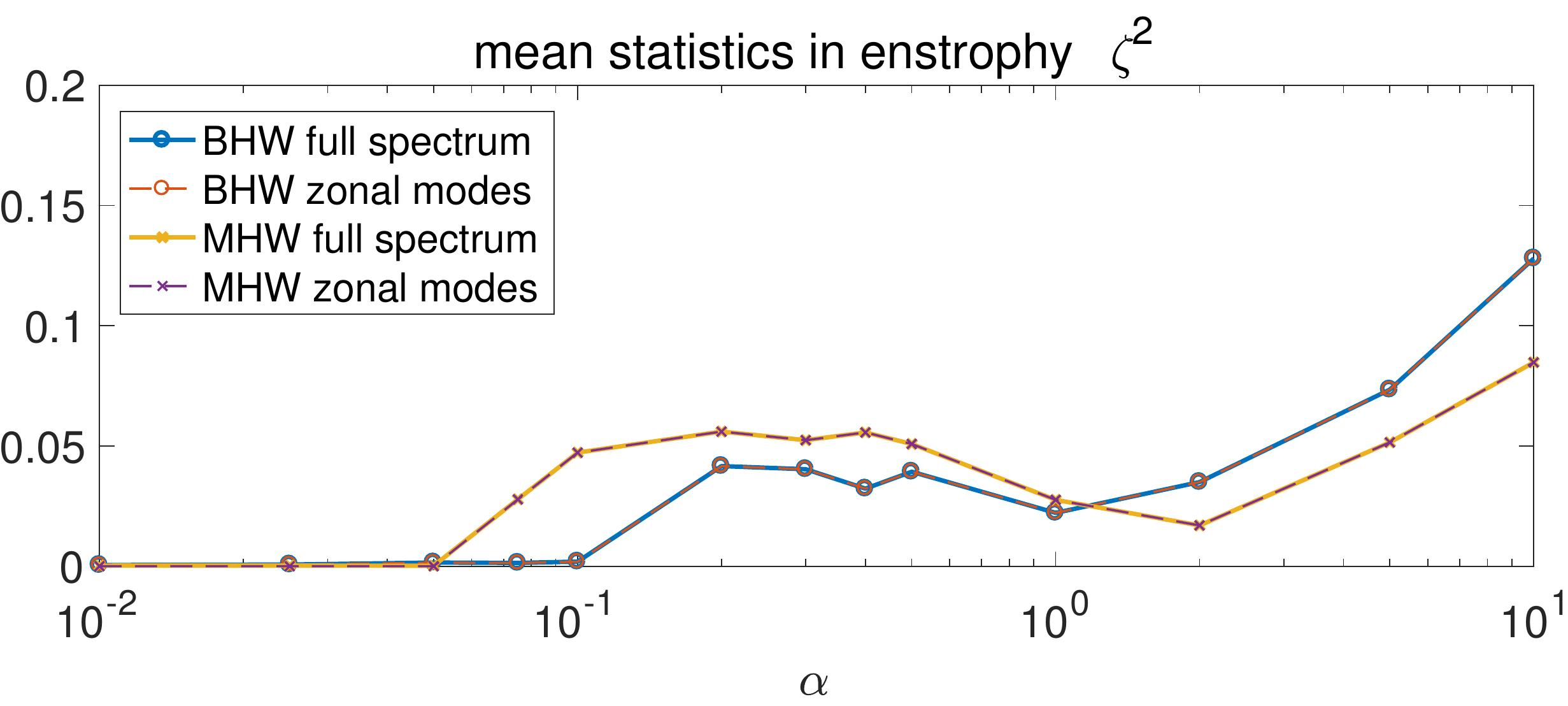}

}
\caption{Comparison of the statistical energy of the variance $\int_{\Omega}\left\langle |f-\left\langle f\right \rangle_{\mathrm{eq}}|^{2}\right\rangle dV$ and of the mean $\int_{\Omega}|\left\langle f\right\rangle_{\mathrm{eq}}|^{2}dV$ for both
the bHW and mHW models and varying values of $\alpha$, with $f=\nabla\varphi$ (corresponding to the kinetic energy) and $f=\zeta$ (corresponding to the enstrophy). The dashed lines show the energy contained in the zonal modes $k_{y}=0$. The other parameters are kept fixed, with $\kappa=0.5$, $\mu=D=5\times10^{-4}$.\label{fig:Total-statistical-energy} }
\end{figure}

The results are shown in Figure \ref{fig:Total-statistical-energy} for simulations of the bHW equations and the mHW equations, with $\alpha$ varying in the interval $\left[10^{-2},10\right]$.
Since one of the main reasons to study the reduced HM and HW fluids models is to better understand the drift wave -- zonal flow interplay, we separately plot the kinetic energy and enstrophy contained in zonal modes, i.e. with $k_{y}=0$, and the total kinetic energy and enstrophy obtained by summing over all the resolved modes. We first focus on the plots for the variances, in the left-hand column, to observe that in both models the kinetic energy and enstrophy of the variances decrease as $\alpha$ increases. This is a clear signature of the transition from a strongly turbulent regime to a zonal flow dominated regime, as we expected. For $\alpha$ small, the flow is strongly fluctuating with large and small scale modes both containing considerable amount of energy. As $\alpha$ increases,
stronger zonal jets develop and the variances in kinetic energy and enstrophy decrease. In the regime $\alpha\gg 1$, the system tends to the HM equations which do not have an internal instability, and the variances reach minimum values. 

Still looking at this first column, we highlight differences between the bHW and mHW models. The bHW model always contains a large variance for the kinetic energy in comparison to the mHW model, which has weaker variability. Furthermore, much of the variance in the bHW model is in zonal modes, unlike the mHW model. Our improved treatment of the electron dynamics parallel to the field lines enhances the zonal flow variability as the system transitions to the turbulence dominated regime when $\alpha$ decreases. The plot for the variance of the relative enstrophy, which is a physical quantity which gives more weight to smaller scale structures, helps to explain the different dynamics in the two models. Indeed, we notice that the mHW model has a much larger variance in enstrophy as $\alpha<0.1$ than the bHW model. This is due to the generation of many small and strong vortices, as Figure \ref{fig:Snapshots} in Section \ref{sec:Direct-Numerical-Simulations} illustrates explicitly with a contour plot of the vorticity field. Even if so, the zonal mode variance in the relative
enstrophy remains larger in the bHW model. This is a central point of this article: even for very small values of $\alpha$ the variability of the turbulent fluctuations is in zonal modes in the bHW model, while the mHW equations describe fully homogeneous turbulence in the small $\alpha$ limit.

We now consider the kinetic energy and enstrophy in the mean, in the second column of Figure \ref{fig:Total-statistical-energy}. The kinetic energy and enstrophy in zonal modes almost overlap with the total kinetic energy and enstrophy, proving that the mean state is always dominantly in the zonal direction. We can distinguish three different regimes depending on the value of $\alpha$. Starting at around $\alpha\sim 0.1$, we notice a significant jump in the value of the kinetic energy and the enstrophy in the mean as compared to situations with $\alpha<0.1$, indicating the presence of stronger zonal jets. A close comparison between the plots for the kinetic energy and enstrophy shows a large kinetic energy and a moderate enstrophy for $\alpha\in\left(0.1,0.5\right)$, which is a signature of the presence of one or very few zonal jets. For $\alpha>0.5$, the kinetic energy is a bit lower and fairly constant as a function of $\alpha$, whereas the enstrophy increases. This indicates the presence of multiple zonal jets. These conclusions are confirmed by the snapshots of the vorticity field for each of the $\alpha$ regimes discussed here, which are shown in Figure \ref{fig:Snapshots} in the next Section. 

\subsubsection{Statistical energy equations and higher-order moment feedbacks}

Besides the kinetic energy and enstrophy in the statistical equilibrium state discussed in the previous section, it is also important to investigate the dynamical evolution of the statistical quantities, in particular to better understand the nonlinear energy exchange processes in the model and the model sensitivity to various perturbations. While in many situations we
are mostly interested in the statistics of the two lowest order moments, that is, the statistical mean and variance, the higher-order feedbacks to these low order moments play a key role in the dynamics and should be accurately treated, as has been highlighted and illustrated in \cite{majda2016introduction,majda2018strategies,qi2016low} for general systems with quadratic nonlinearity. In this section, we highlight these mechanisms for the bHW and mHW models.

We consider here the statistical average $\left\langle \tilde{k}^{2}|\hat{\varphi}_{\mathbf{k}}|^{2}\right\rangle$ of the kinetic energy in each spectral mode $\hat{\varphi}_{\mathbf{k}}$ of the electrostatic potential. To derive the dynamical equations for $\left\langle \tilde{k}^{2}|\hat{\varphi}_{\mathbf{k}}|^{2}\right\rangle$, we first project the equations for the vorticity in the mHW and bHW models onto each individual spectral mode, to obtain equations for each $\hat{\varphi}_{k}$. We then multiply by $\hat{\varphi}_{k}^{*}$ on both sides of the equations
and take the statistical ensemble average (see \cite{majda2018strategies}
for details) to obtain the desired equations for $\left\langle \tilde{k}^{2}|\hat{\varphi}_{\mathbf{k}}|^{2}\right\rangle$. For the bHW model, the equations are
\begin{equation}
\begin{aligned}\frac{d}{dt}\frac{1}{2}\left\langle \tilde{k}^{2}|\hat{\varphi}_{\mathbf{k}}|^{2}\right\rangle +Q^{\mathrm{b}}_{F,k}  & =-\alpha\left(\left\langle |\hat{\varphi}_{k}|^{2}\right\rangle -\left\langle \hat{\varphi}_{k}^{*}\hat{n}_{k}\right\rangle \right)-D\tilde{k}^{4}\left\langle |\hat{\varphi}_{k}|^{2}\right\rangle +c.c.,\quad k_{y}\neq0,\\
\frac{d}{dt}\frac{1}{2}\left\langle \tilde{k}^{2}|\hat{\varphi}_{\mathbf{k}}|^{2}\right\rangle +Q^{\mathrm{b}}_{F,k}  & =-D\tilde{k}^{4}\left\langle |\hat{\varphi}_{k}|^{2}\right\rangle+c.c.,\quad k_{y}=0.
\end{aligned}
\label{eq:dyn_stat_BHW}
\end{equation}
Here, $c.c.$ stands for the complex conjugate completion. Importantly, the nonlinear interaction term in the vorticity equation produces the higher-order feedback $Q^{\mathrm{b}}_{F,k}=-\left\langle \hat{\varphi}_{k}^{*}J\left(\varphi,\Delta\varphi+\overline{n}\right)_{k}\right\rangle+c.c.$ for the non-zonal modes $k_y\neq0$; and $Q^{\mathrm{b}}_{F,k}=-\left\langle \hat{\varphi}_{k}^{*}J\left(\varphi,\Delta\varphi-\tilde{n}\right)_{k}\right\rangle+c.c.$ for the zonal modes $k_y=0$.
For the mHW model, the equations are
\begin{equation}
\begin{aligned}\frac{d}{dt}\frac{1}{2}\left\langle \tilde{k}^{2}\left|\hat{\varphi}_{\mathbf{k}}\right|^{2}\right\rangle +Q^{\mathrm{m}}_{F,k}  & =-\alpha\left(\left\langle \left|\hat{\varphi}_{k}\right|^{2}\right\rangle -\left\langle \hat{\varphi}_{k}^{*}\hat{n}_{k}\right\rangle \right)-D\tilde{k}^{4}\left\langle \left|\hat{\varphi}_{k}\right|^{2}\right\rangle +c.c.,\quad k_{y}\neq0,\\
\frac{d}{dt}\frac{1}{2}\left\langle \tilde{k}^{2}\left|\hat{\varphi}_{\mathbf{k}}\right|^{2}\right\rangle +Q^{\mathrm{m}}_{F,k}  & =-D\tilde{k}^{4}\left\langle \left|\hat{\varphi}_{k}\right|^{2}\right\rangle +c.c.,\quad k_{y}=0,
\end{aligned}
\label{eq:dyn_stat_MHW}
\end{equation} 
with the higher-order feedback $Q^{\mathrm{m}}_{F,k}=-\left\langle \hat{\varphi}_{k}^{*}J\left(\varphi,\Delta\varphi\right)_{k}\right\rangle +c.c.$
The removal of the zonal mean density $\overline{n}$ in the term involving the adiabaticity parameter $\alpha$ in both models, as well as in the definition of the potential vorticity in the bHW model, are responsible for the different form the equations take for the zonal modes. 
We finally note the presence of the cross-covariance term $\left\langle \hat{\varphi}_{k}\hat{n}_{k}^{*}\right\rangle $ in both models, representing the interactions between the potential and density. 

As emphasized in \cite{majda2016introduction,majda2018strategies},
the third-order moments on the left hand side of Eqs. (\ref{eq:dyn_stat_BHW}) and (\ref{eq:dyn_stat_MHW}) play the critical role of mediating the growing unstable modes and driving the system to the final equilibrium. As we know from the linear stability analysis, the linear operators on the right hand sides of the statistical equations contain positive eigenvalues, corresponding to the linearly unstable directions. If the third order moments on the left hand sides are not included, the internal instability will lead to fast growth in kinetic energy among the unstable modes and fast decay in the other over-damped modes. When the third-order moments are kept, they transfer the growing energy from the unstable subspace to the stable subspace, where the energy eventually gets dissipated. 

We thus see the importance of a rigorous analysis of the contributions from the third-order moments. Unfortunately, it is usually expensive to compute these moments from direct numerical simulations since it requires the inclusion of all the triad modes across the entire spectrum \cite{majda2018strategies}. On the other hand, the third-order moment feedbacks in statistical equilibrium are much easier to compute. Indeed, in statistical equilibrium the time derivatives on the left hand sides of Eqs. (\ref{eq:dyn_stat_BHW}) and (\ref{eq:dyn_stat_MHW}) vanish, so the third-order moments can be expressed in terms of the first and second order moments appearing on the right-hand sides. Specifically, for the bHW model, we can write
\[
\begin{aligned}Q^{\mathrm{b}}_{F,k,\mathrm{eq}} & =\alpha\left(\left\langle |\hat{\varphi}_{k}|^{2}\right\rangle _{\mathrm{eq}}-\left\langle \hat{\varphi}_{k}^{*}\hat{n}_{k}\right\rangle _{\mathrm{eq}}\right)+D\tilde{k}^{4}\left\langle |\hat{\varphi}_{k}|^{2}\right\rangle _{\mathrm{eq}}+c.c.,\quad k_{y}\ne0,\\
 & =D\tilde{k}^{4}\left\langle |\hat{\varphi}_{k}|^{2}\right\rangle _{\mathrm{eq}}+c.c.,\qquad\qquad\quad k_{y}=0.
\end{aligned}
\]
The equilibrium moments, $\left\langle |\hat{\varphi}_{k}|^{2}\right\rangle _{\mathrm{eq}},\left\langle \hat{\varphi}_{k}^{*}\hat{n}_{k}\right\rangle _{\mathrm{eq}}$ can be readily calculated from the equilibrium mean and variance by averaging along the stationary time trajectory. The equilibrium third-order moments for the mHW model can be computed in an analogous way.

\begin{figure}
\subfloat[bHW model]{\includegraphics[scale=0.29]{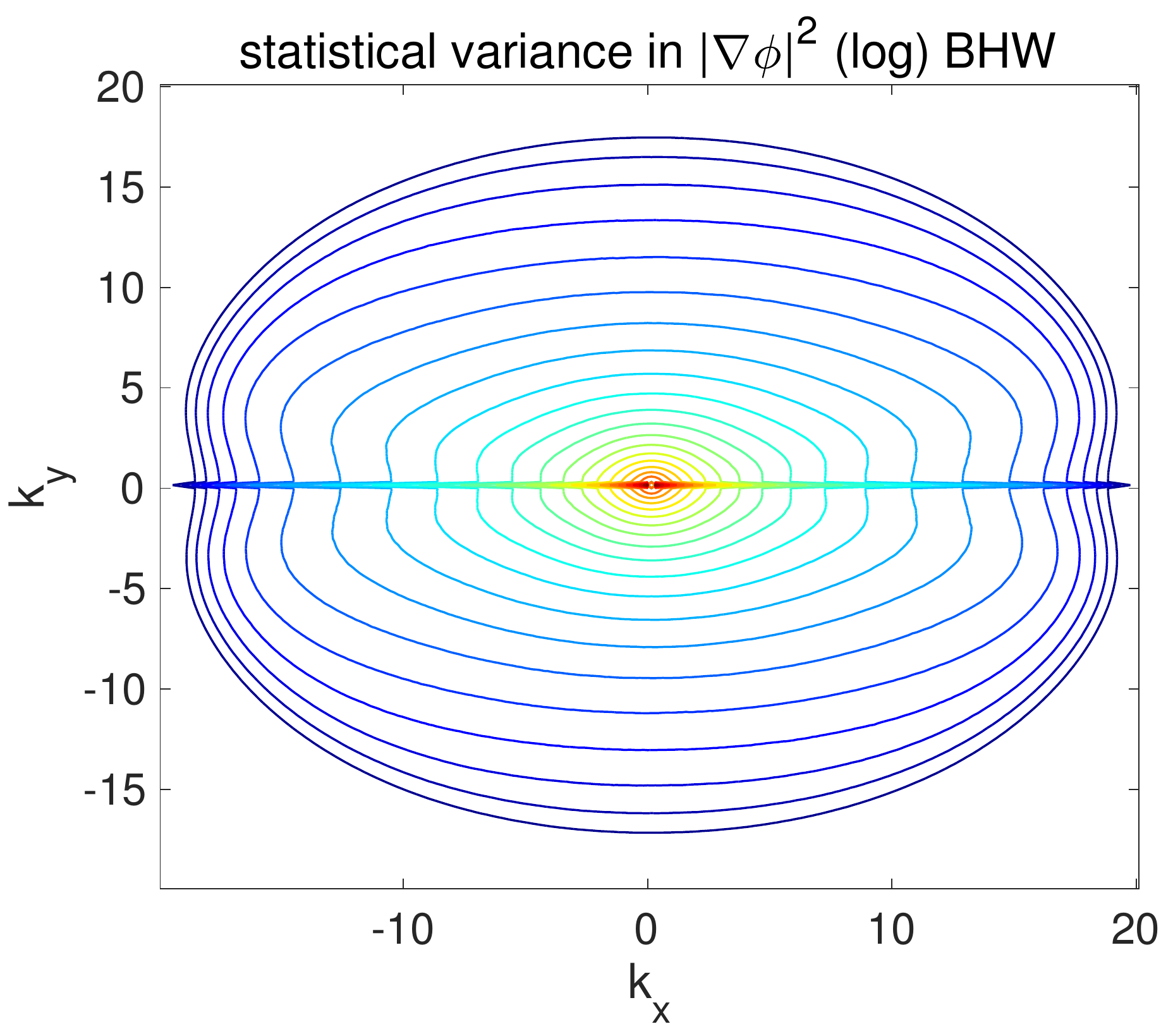}\includegraphics[scale=0.29]{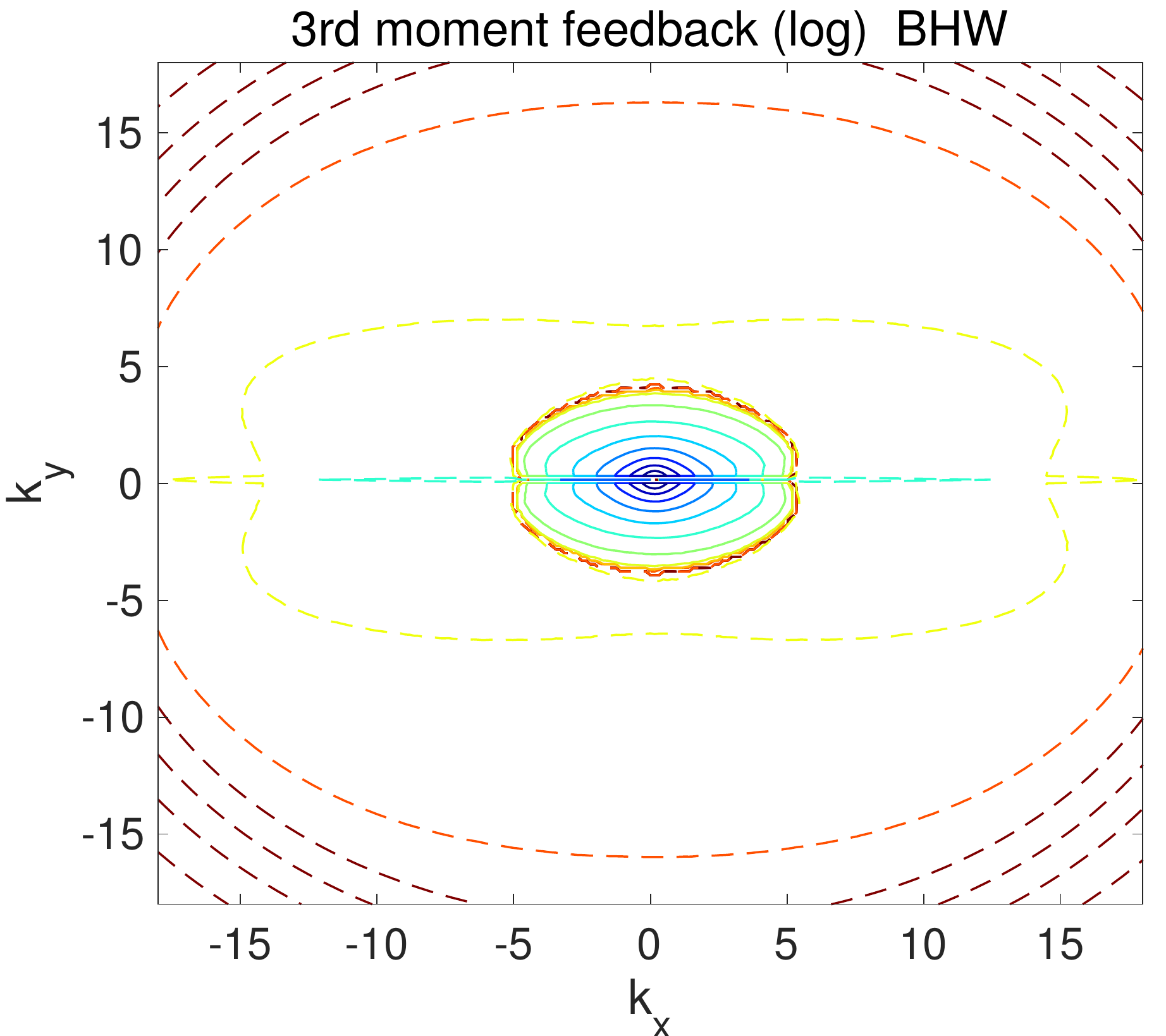}\includegraphics[scale=0.29]{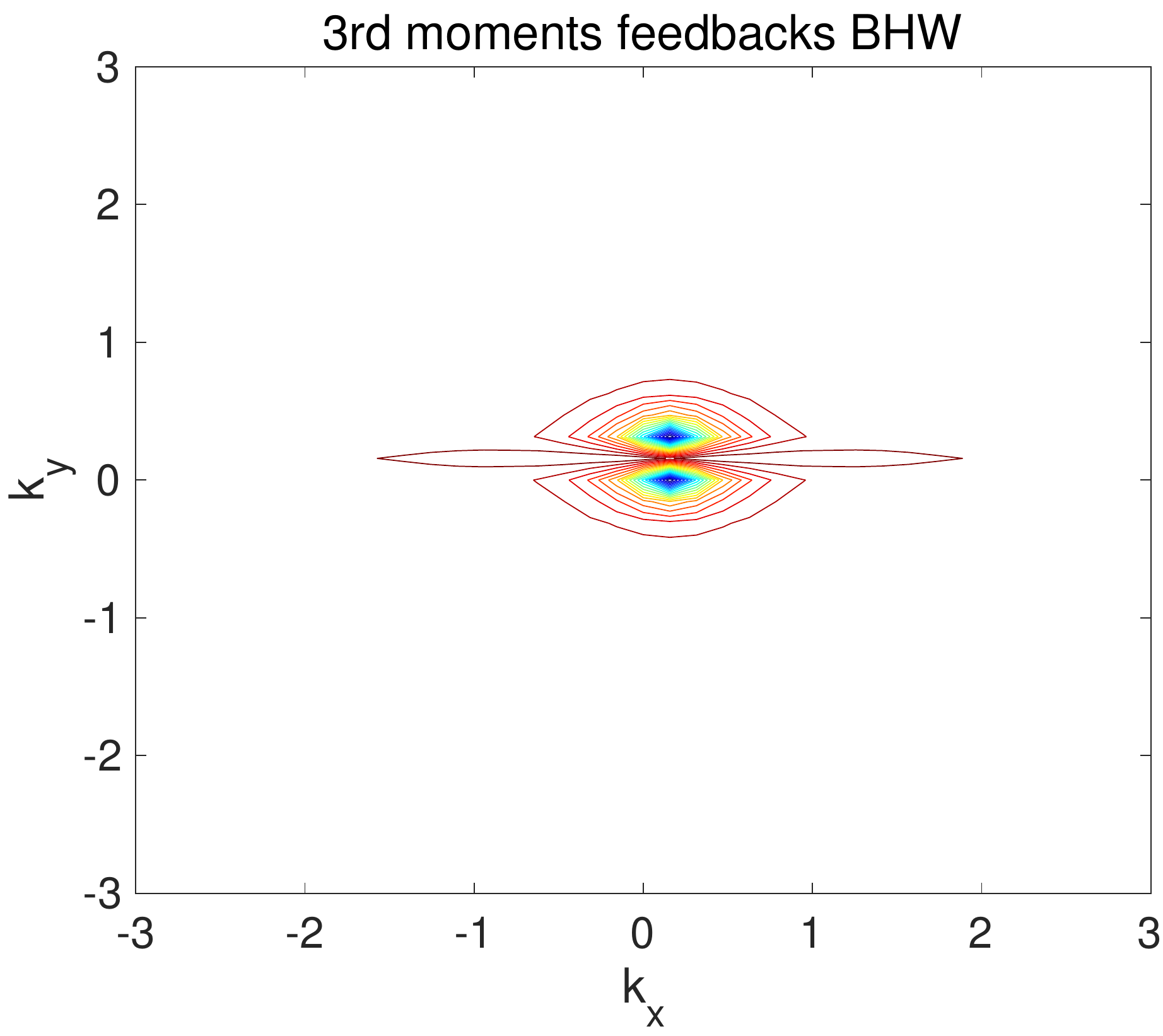}

}

\subfloat[mHW model]{\includegraphics[scale=0.29]{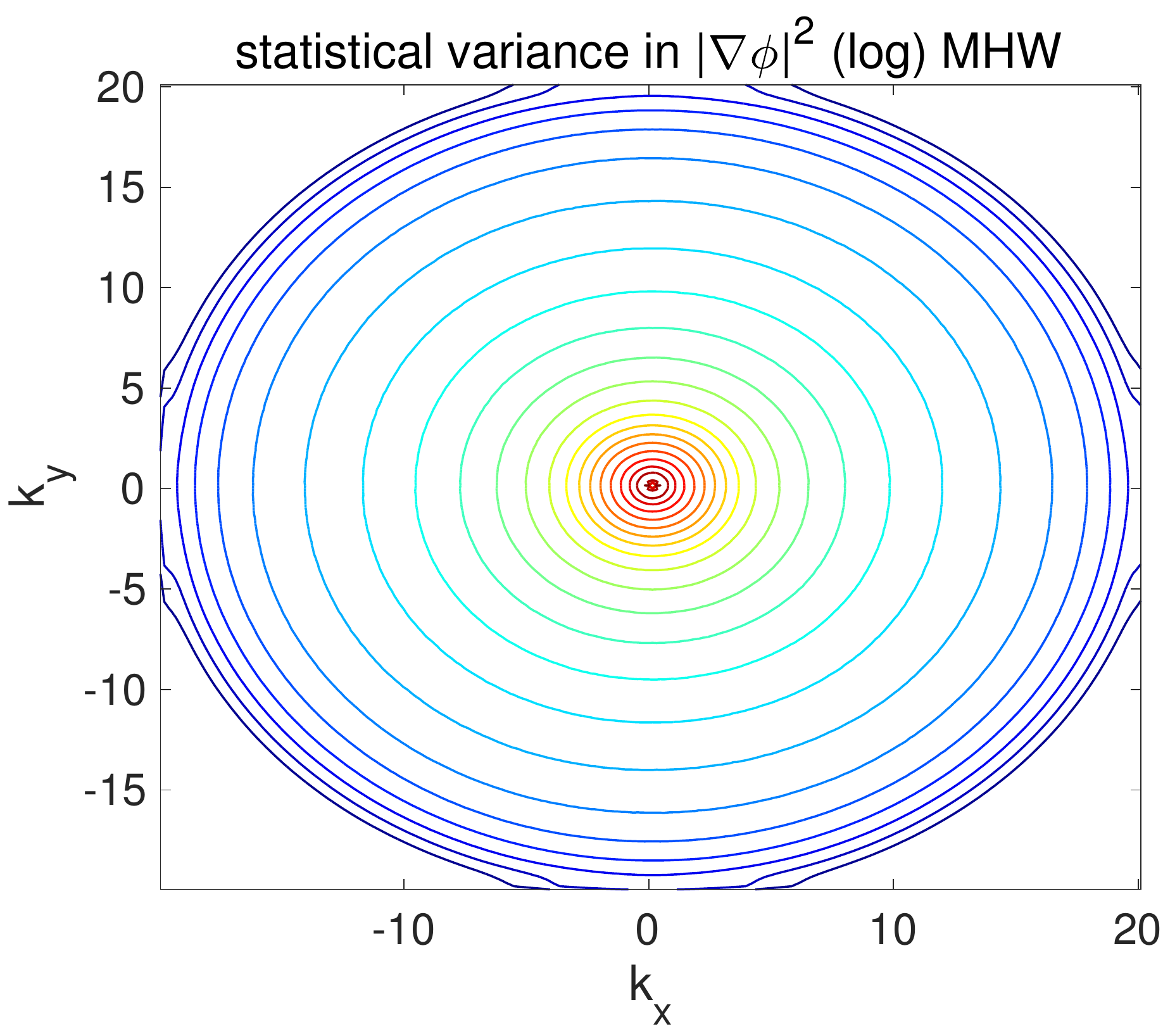}\includegraphics[scale=0.29]{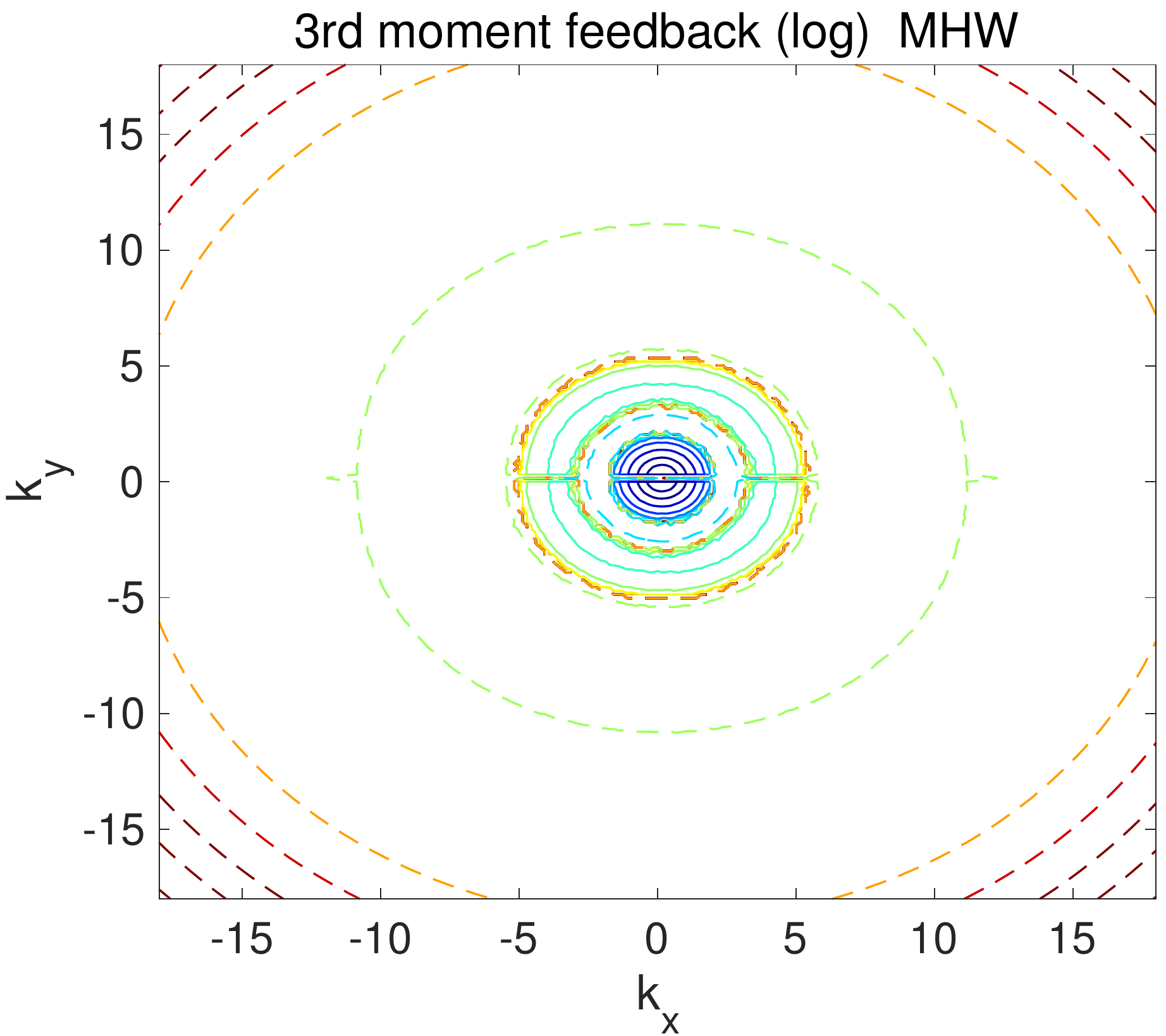}\includegraphics[scale=0.29]{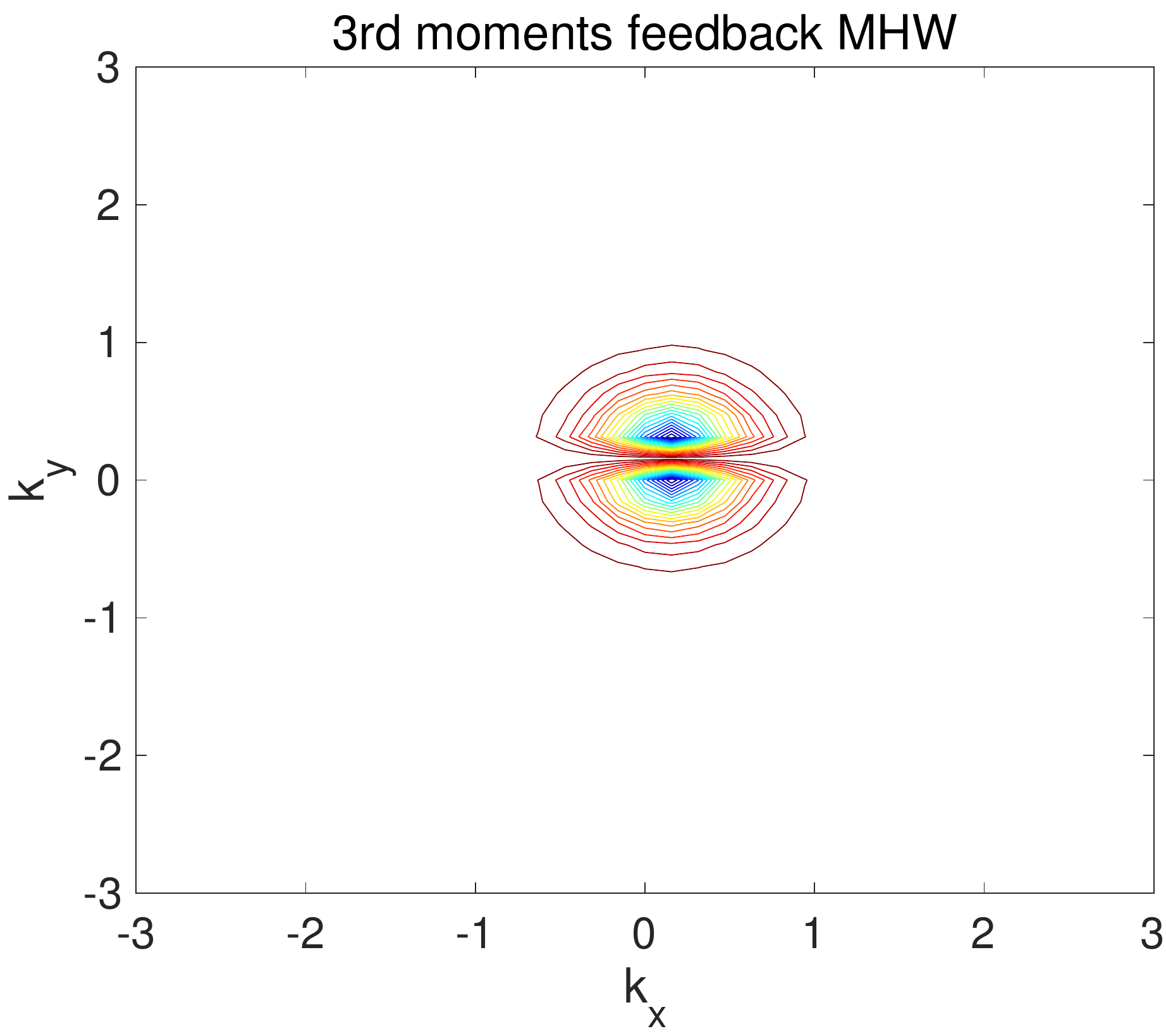}

}

\caption{Equilibrium second-order moments and third-order moment feedbacks
in the Fourier domain for the bHW and mHW models with $\alpha=0.01,\kappa=0.5$.
The first column shows the logarithm  of the variance $\tilde{k}^{2}\left\langle |\hat{\varphi}_{k}|^{2}\right\rangle $ of each spectral mode. The second column plots the
logarithm of the third-order moment feedbacks. Negative values, which correspond to effective
damping, are plotted with solid lines and positive values, which correspond to an injection of energy, are plotted with dashed lines. In the third column, the contours of
the third-order moment feedbacks are plotted without taking the logarithm,
thus emphasizing the modes with the strongest third-order moments. Blue colors are
for the largest negative values, red colors for the largest positive values.\label{fig:Second-moment}}

\end{figure}

To numerically investigate the differences and similarities between the mHW and bHW models from the point of view of the statistical moments, we choose a regime with small $\alpha$, namely 
$\alpha=0.01$, which corresponds to a physical situation with strong mean flow -- drift wave interactions. Figure \ref{fig:Second-moment} presents the equilibrium statistics of the bHW and mHW models for both the second-order variances and the third-order moment feedbacks
in the spectral domain. We first look at the first column, which shows contours of the logarithm of the second-order variance of each energy mode $k^{2}\left|\hat{\varphi}_{k}\right|^{2}$. The
stronger variability for the zonal modes ($k_{y}=0$) in the bHW model is clearly visible. This is consistent with the observation we made in the previous section that even in the small $\alpha$ regime, the bHW model generates energetic zonal flows which are responsible for most of the flow variability. In contrast, we note that the contour plot for the mHW model is radially symmetric in the Fourier domain, indicating the absence of anisotropy in the flow statistics.

We next turn to the third-order moment feedbacks shown in the second and third columns of Figure \ref{fig:Second-moment}. In the middle column, the logarithm of the magnitude of the third-order moments is plotted in order to amplify the small values, and we use dashed lines to represent positive values and solid lines for negative values. In the last column, we plot the contours of the third-order moments themselves (i.e. not their logarithm). The red colors correspond to positive values, and the blue colors to negative values. We first observe that both the bHW and the mHW models show strong non-zero contributions from the third-order moments, particularly at the largest scales in the system. Comparing the results of the linear stability analysis we present in Figure \ref{fig:Growth-rate-HW} of Appendix \ref{sec:Linear-Instability} with these figures, we see that the linearly unstable modes are subject to negative third-order moment feedbacks which act as an effective damping mechanism against the growth in energy. At the same time, the linearly stable modes are subject to positive third-order moment feedbacks. This is characteristic of a nonlinear transfer of energy from the unstable modes, whose energy grows in time, to the stable modes, which then dissipate the energy. Note that the third-order moment feedback conserves energy as a whole, so it cannot act as a source or sink of energy, and instead redistributes energy at different scales. Finally, one observes much stronger third-order moment feedbacks for the zonal modes in the bHW model than in the mHW model, which we would expect given our previous results: once again, even in the regime $\alpha\ll 1$, zonal modes play a central role in the dynamics in the bHW model, while they are essentially absent in the mHW model. These results also highlight the need for accurate approximations of the third-order moment feedback terms when applying model reduction strategies to investigate the statistical properties of a system and for uncertainty quantification. In the present case, the differences between the bHW and mHW models, localized in the vicinity of $k_{y}=0$, lead to major differences in the observed dynamics.

\subsection{Numerical convergence to the modified Hasegawa-Mima model as $\alpha\rightarrow\infty$}

\begin{figure}
\subfloat[bHW model]{\includegraphics[scale=0.3]{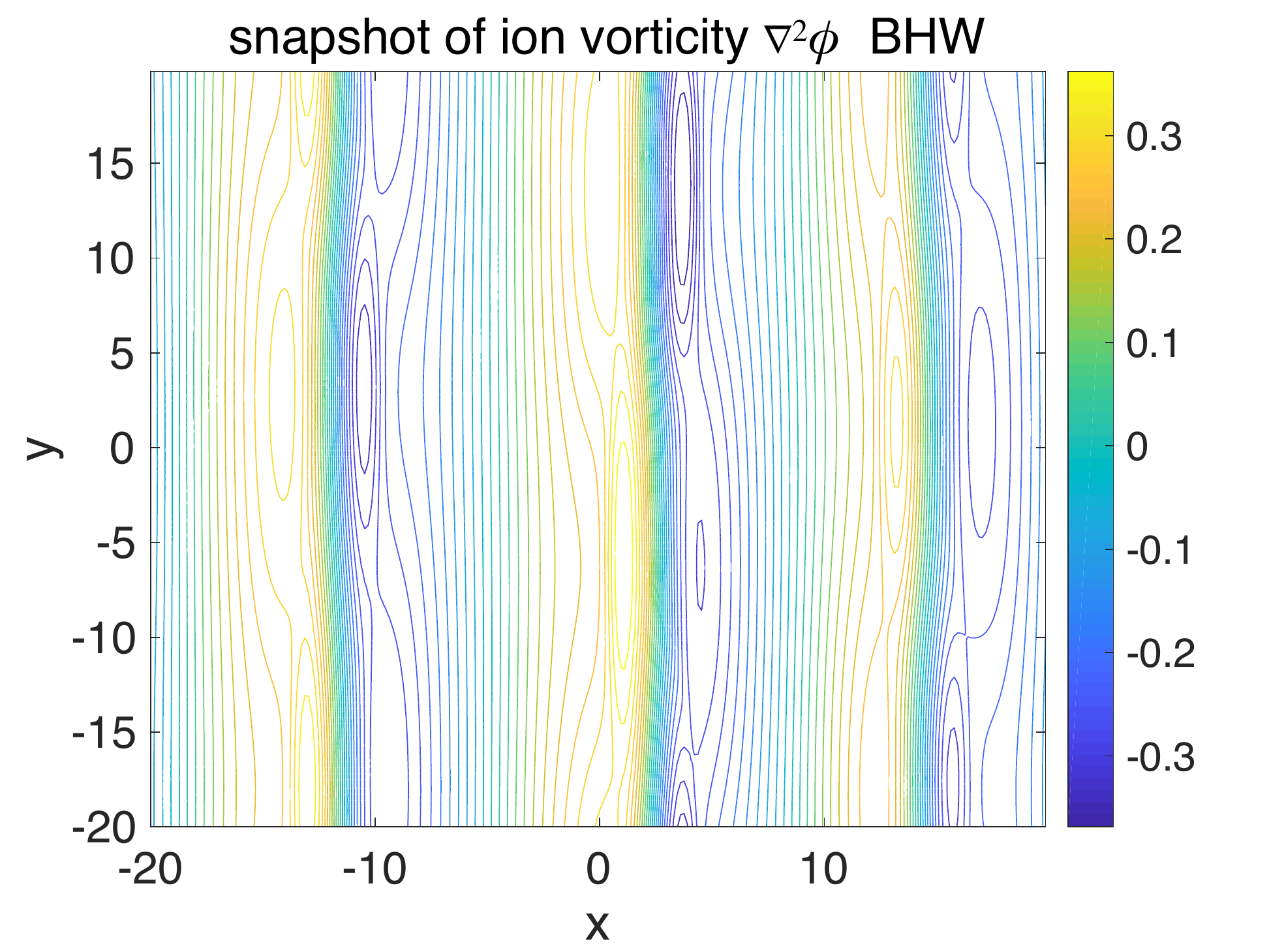}\includegraphics[scale=0.3]{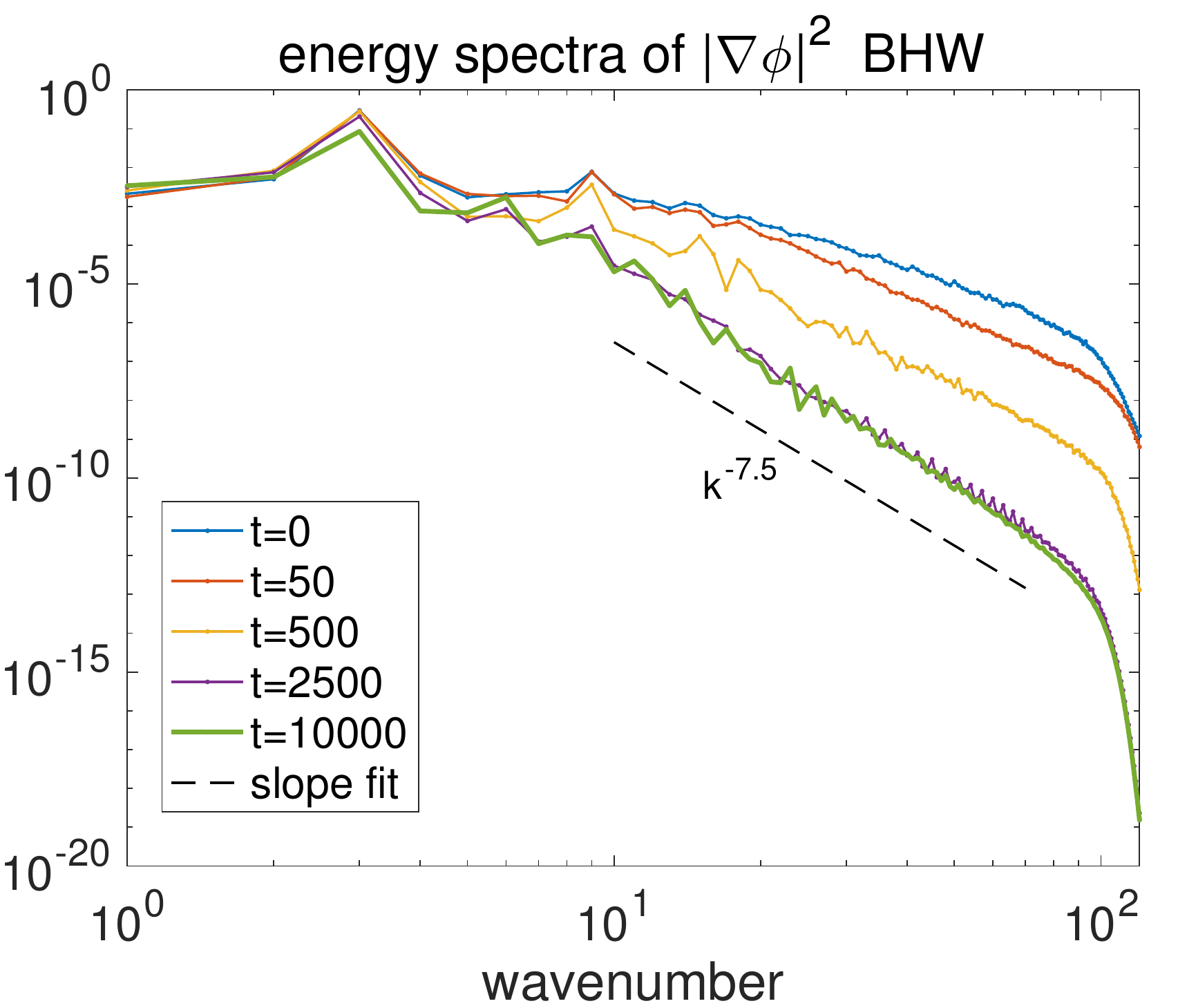}

}\enskip{}\subfloat{\includegraphics[scale=0.3]{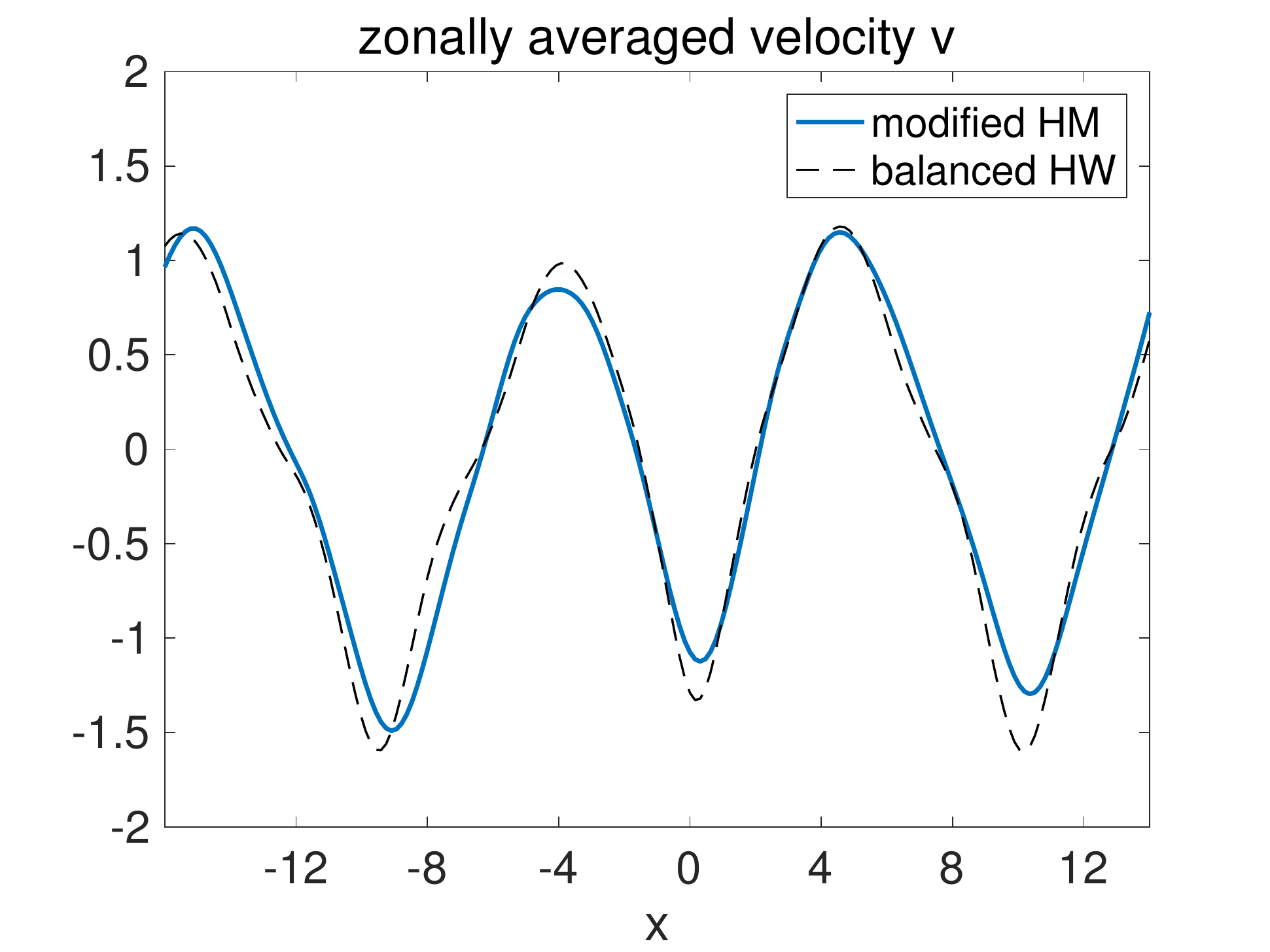}}
\addtocounter{subfigure}{-1}

\subfloat[mHW model]{\includegraphics[scale=0.3]{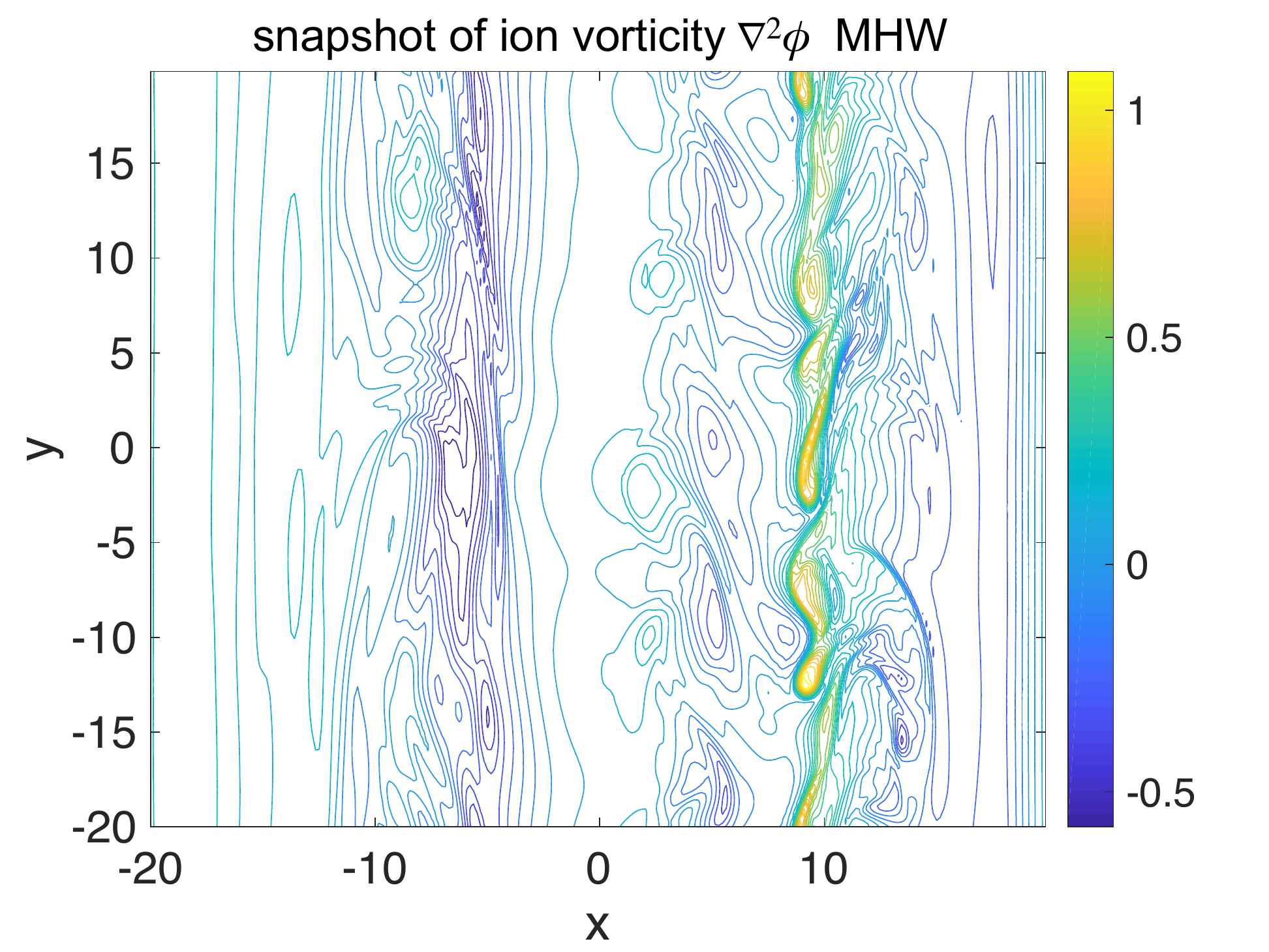}\includegraphics[scale=0.3]{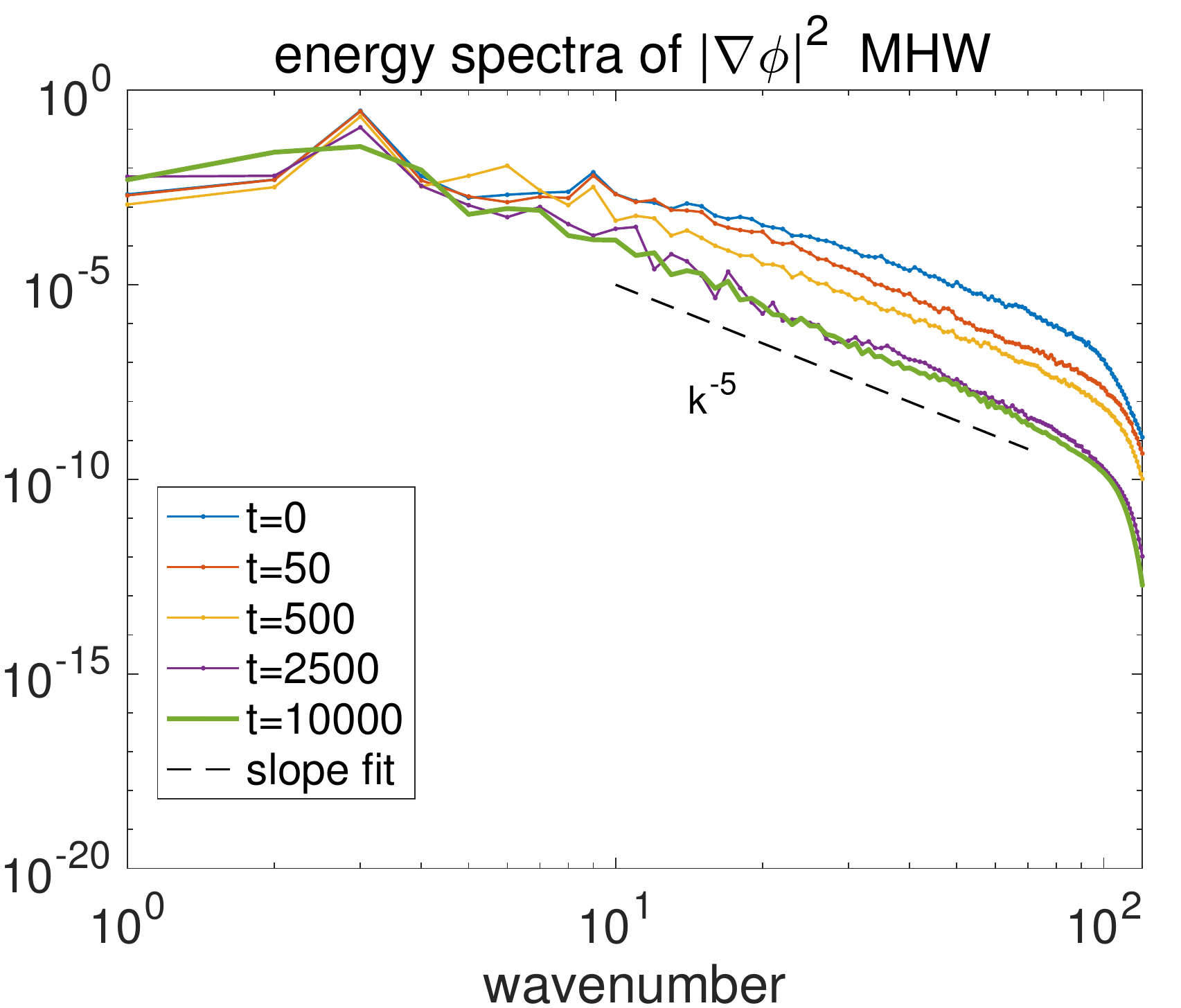}

}\enskip{}\subfloat[mHM model]{\includegraphics[scale=0.3]{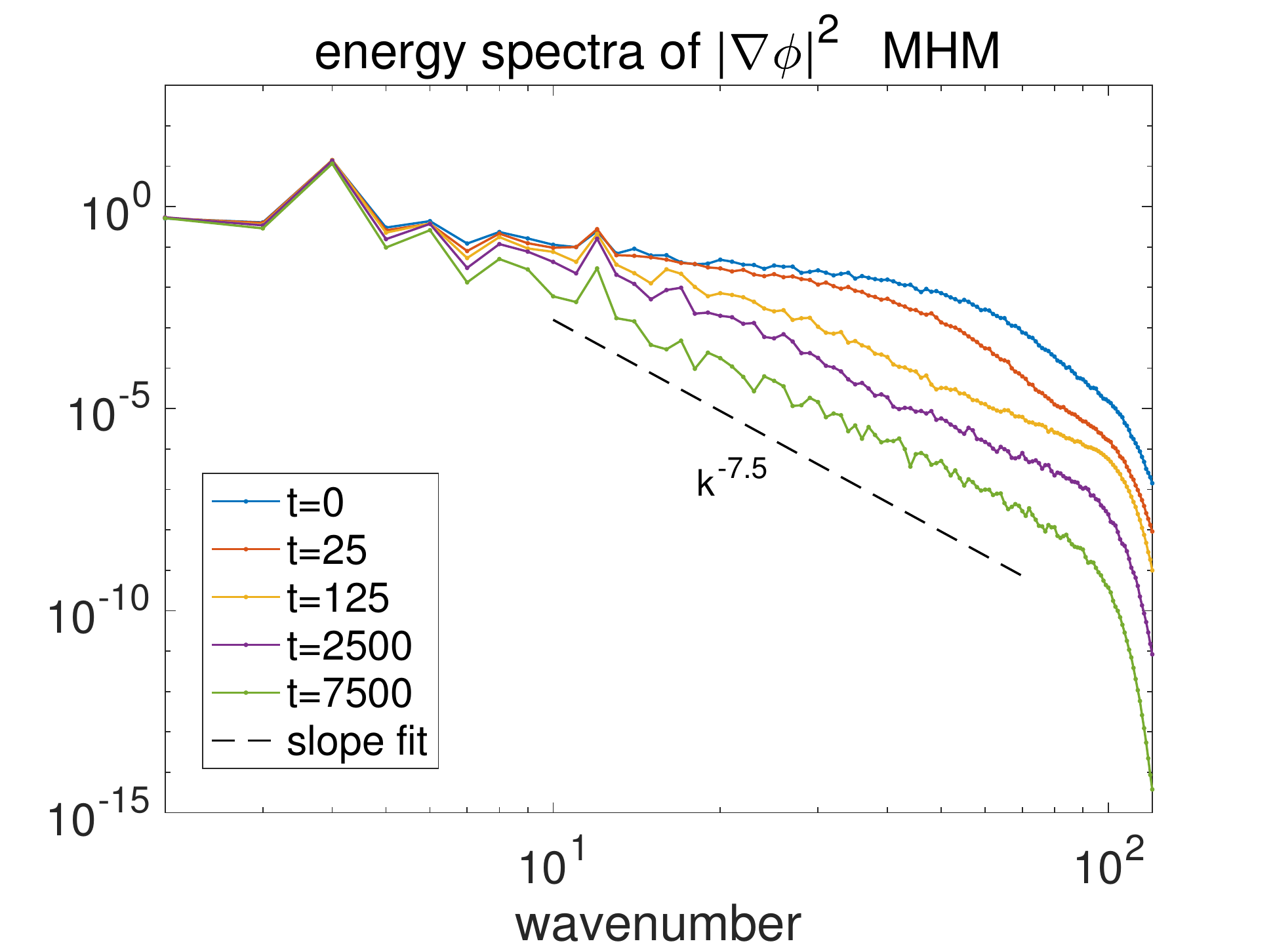}

}

\caption{(Left column) Contour plot of the ion vorticity at the final time of the simulation for the bHW model (top) and the mHW model (bottom). (Middle column) Kinetic energy spectra at different simulation times for the bHW model (top) and the mHW model (bottom). (Third column) Profile of the zonally averaged velocity in the $y$-direction for the mHM model in solid blue and the bHW model in dashed black (top) and kinetic energy spectra at different simulation times for the mHM model (bottom). The final state of the bHW simulation was chosen as the initial state for the mHM simulation, with random noise added at small scales. For all these results, we used $\alpha=0.5$ and the values used throughout Section \ref{sec:Features-of-BHW} for the other parameters. \label{fig:Comparison-decay}}
\end{figure}

Both the bHW and the mHW models converge to an HM-like model without instability in the limit of large $\alpha$, as we confirmed numerically in section \ref{sub:Comparison-in-transition}. We showed in section \ref{subsec:cons} that the bHW
model converges to the mHM equation (\ref{eq:HM_nondim}) (with an additional dissipation term) in the limit $\alpha\rightarrow\infty$, with the balanced potential vorticity converging to its limiting form $q^{\mathrm{b}}=\nabla^{2}\varphi-\tilde{n}\rightarrow\nabla^{2}\varphi-\tilde{\varphi}$.
We also explained in section \ref{subsec:fbpres} that the mHW equations do not guarantee exact convergence to the mHM model due to the incomplete treatment of the zonal mean density $\overline{n}$. In this section, we verify our reasoning with numerical simulations. Specifically, we consider a case in which $\alpha$ is relatively large, $\alpha=5$, which is a regime where the linear growth rates are small, so the energy in the small-scale modes is mostly dissipated, and a dominant zonal mode with multiple jets sets in, as is clearly visible in the first column of Figure \ref{fig:Comparison-decay}, which shows contour plots of the ion vorticity $\zeta$ at the final time of the simulation, in the top left corner for the bHW model, and in the bottom left corner for the mHW model. To obtain a more quantitative idea of the selective energy decay in each spectral mode, we plot the kinetic energy $\tilde{k}^{2}|\hat{\varphi}_{\mathbf{k}}|^{2}$ of each mode at different simulation times in the second column of Figure \ref{fig:Comparison-decay}, with the bHW results at the top and the mHW results at the bottom as before. The energy decay in the small scale modes can be observed for both models at
early times, and for both models the decay saturates at around $t=2500$. At the final time, only a dominant zonal mode with three zonal jets remains in the flow field, as can be seen in the contour plots in the first column of Figure \ref{fig:Comparison-decay}. However, the decay rate in the mHW is much slower, with considerable amount of kinetic energy maintained in the small scales, while the small scale kinetic energy in the bHW model decays quickly to negligible amounts after a short transient state. This fundamental difference between the mHW and bHW models is clearly apparent in the contour plots of the vorticity field in the first column. In the bHW simulation in the upper left corner, the vorticity field is largely dominated by the large scale zonal jet structure, with very limited smaller scale features. On the other hand, the mHW simulation shows many persistent small scale vortices in the vorticity field. 

To further verify the convergence to the mHM model, we numerically solve the mHM equations with dissipation and random noise perturbations added at small scales, starting from an initial state given by the final state of the bHW simulation with parameters $\kappa=0.5,\alpha=5$. The results are shown in the right column of Figure \ref{fig:Comparison-decay}. Since the mHM model does not contain any instability, the fluctuations get dissipated while the zonal jet structure is maintained in time. The top right figure shows the final and converged zonally averaged velocity $\overline{v}=\partial_x\overline{\varphi}$ for both the bHW and mHM model 
simulations. The zonal mean flow clearly converges to the same saturated limit. We have verified that this is true whatever the initial state is chosen for the mHM simulation. Although not shown in the figure, the contour plot for the mHM ion vorticity has strong similarities with the bHW figure, with the same dominant zonal jets, and very few small scale structures, unlike the mHW results.

This concludes our numerical proof of the convergence of the bHW model to the mHM model in the adiabatic limit, and of the incomplete convergence of the mHW model in that same limit.

\section{Direct Numerical Simulations of the bHW and mHW Models\label{sec:Direct-Numerical-Simulations}}

In this section, we illustrate the key results discussed in Section \ref{sec:Features-of-BHW} with plots of the ion vorticity and mean flow obtained with direct numerical simulations for different values of the adiabaticity parameter $\alpha$. We also study the separate roles of the dissipation terms $\mu\Delta\zeta$, $D\Delta n$ and $C\omega_{*}\varphi$ by considering different values for the parameters $\mu$, $D$ and $C$ than what we had in the previous section.

\subsection{Weak and strong zonal regimes}

\begin{figure}
\subfloat[bHW model]{\includegraphics[scale=0.28]{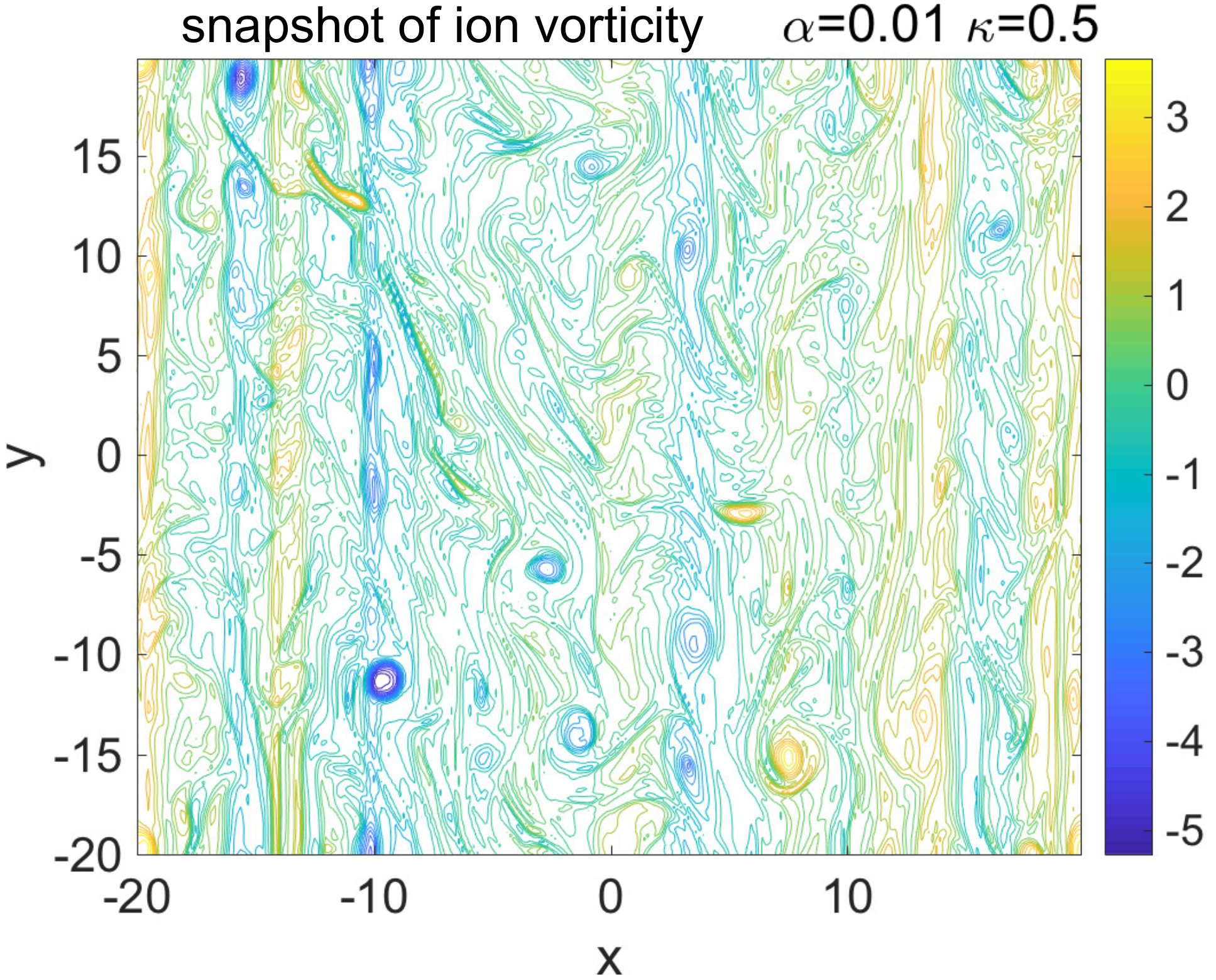}\includegraphics[scale=0.28]{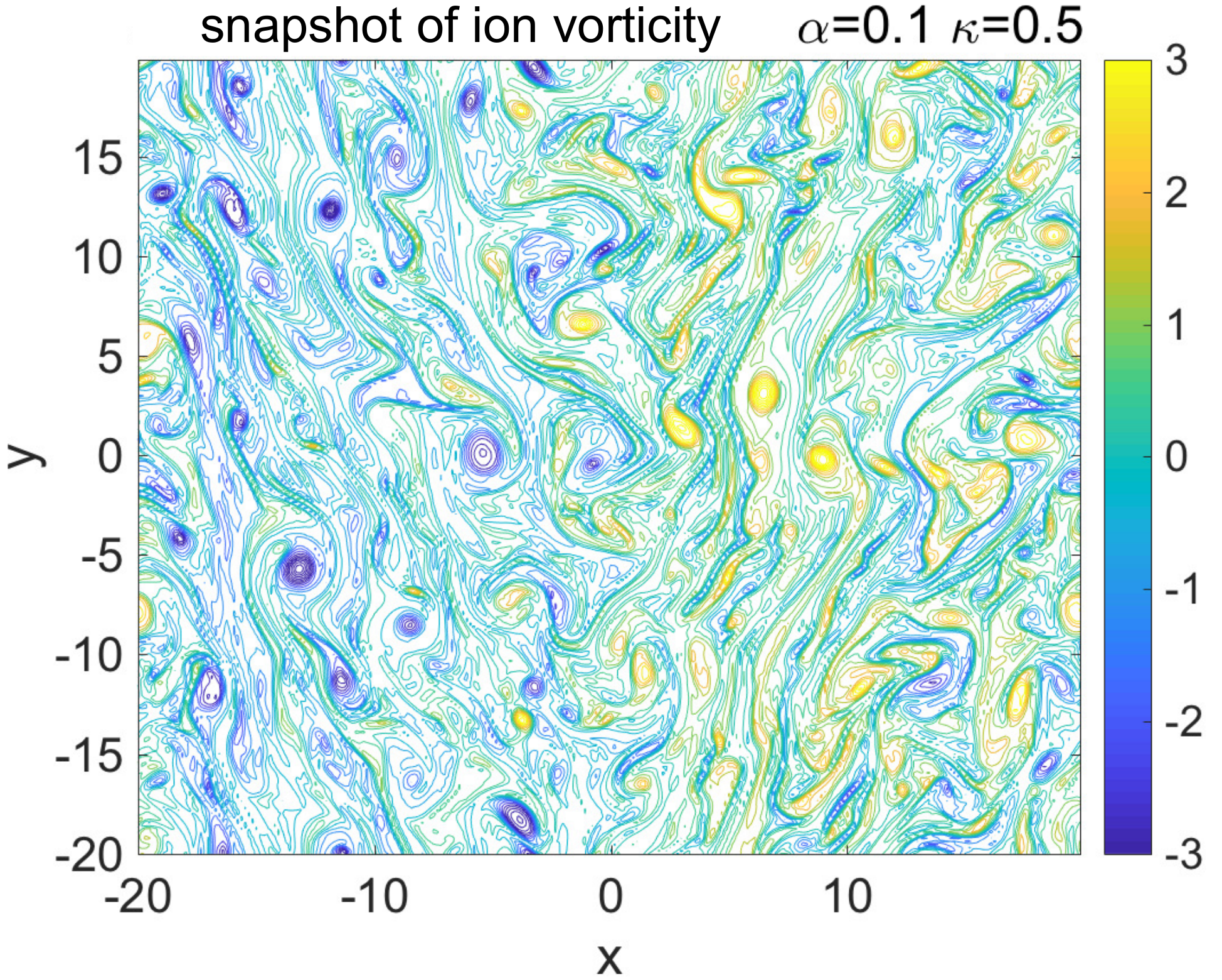}\includegraphics[scale=0.28]{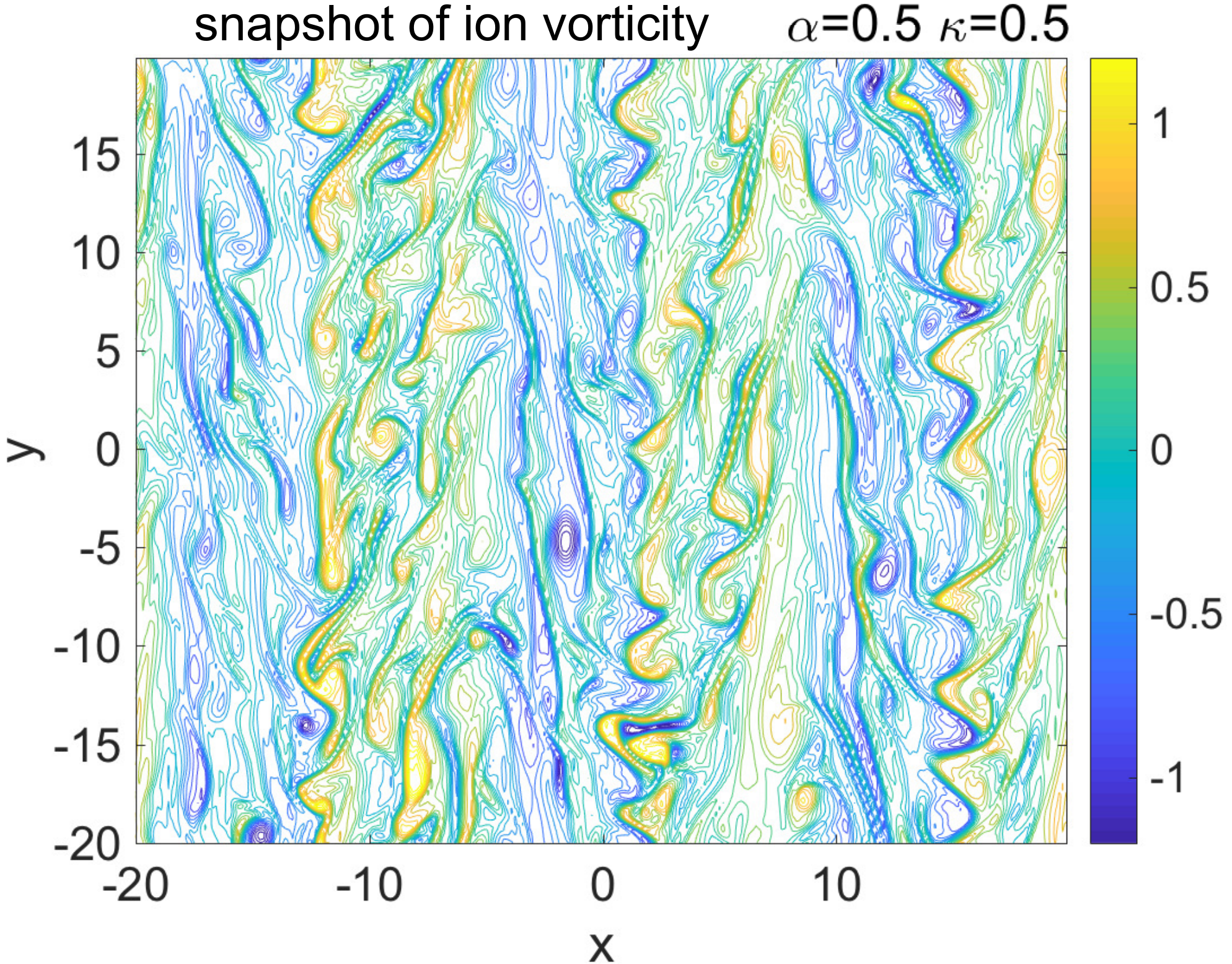}
}

\subfloat[mHW model]{\includegraphics[scale=0.28]{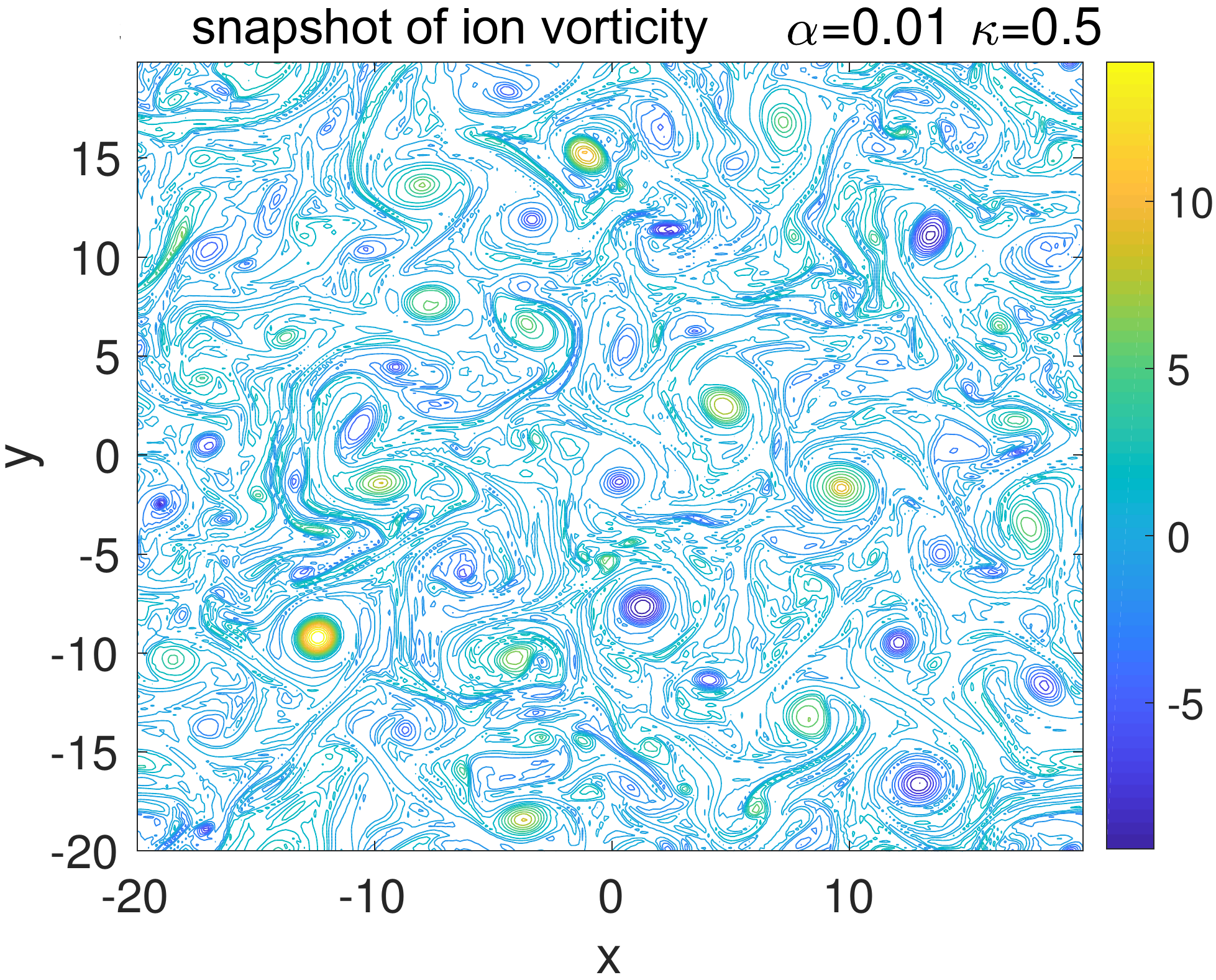}\includegraphics[scale=0.28]{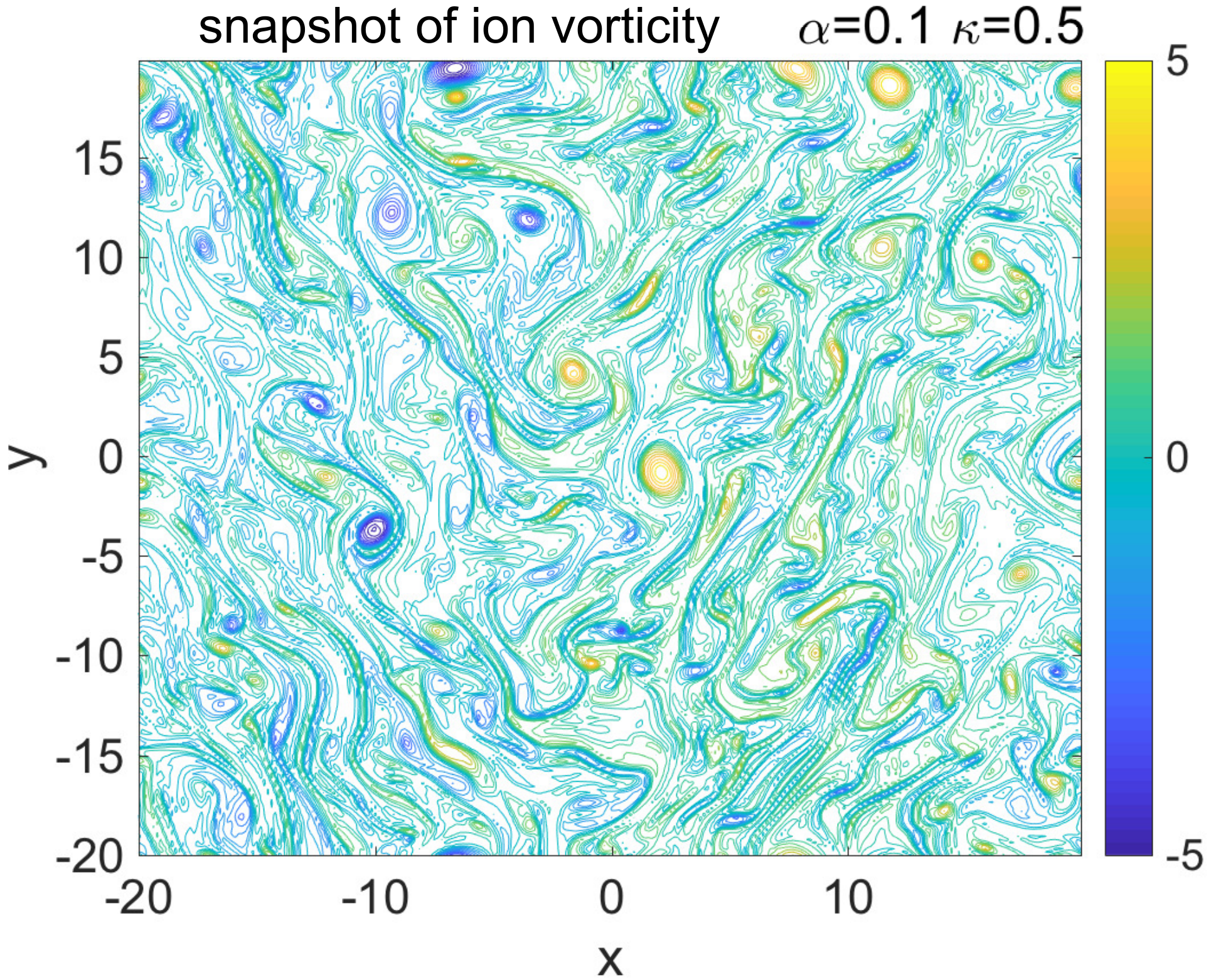}
\includegraphics[scale=0.28]{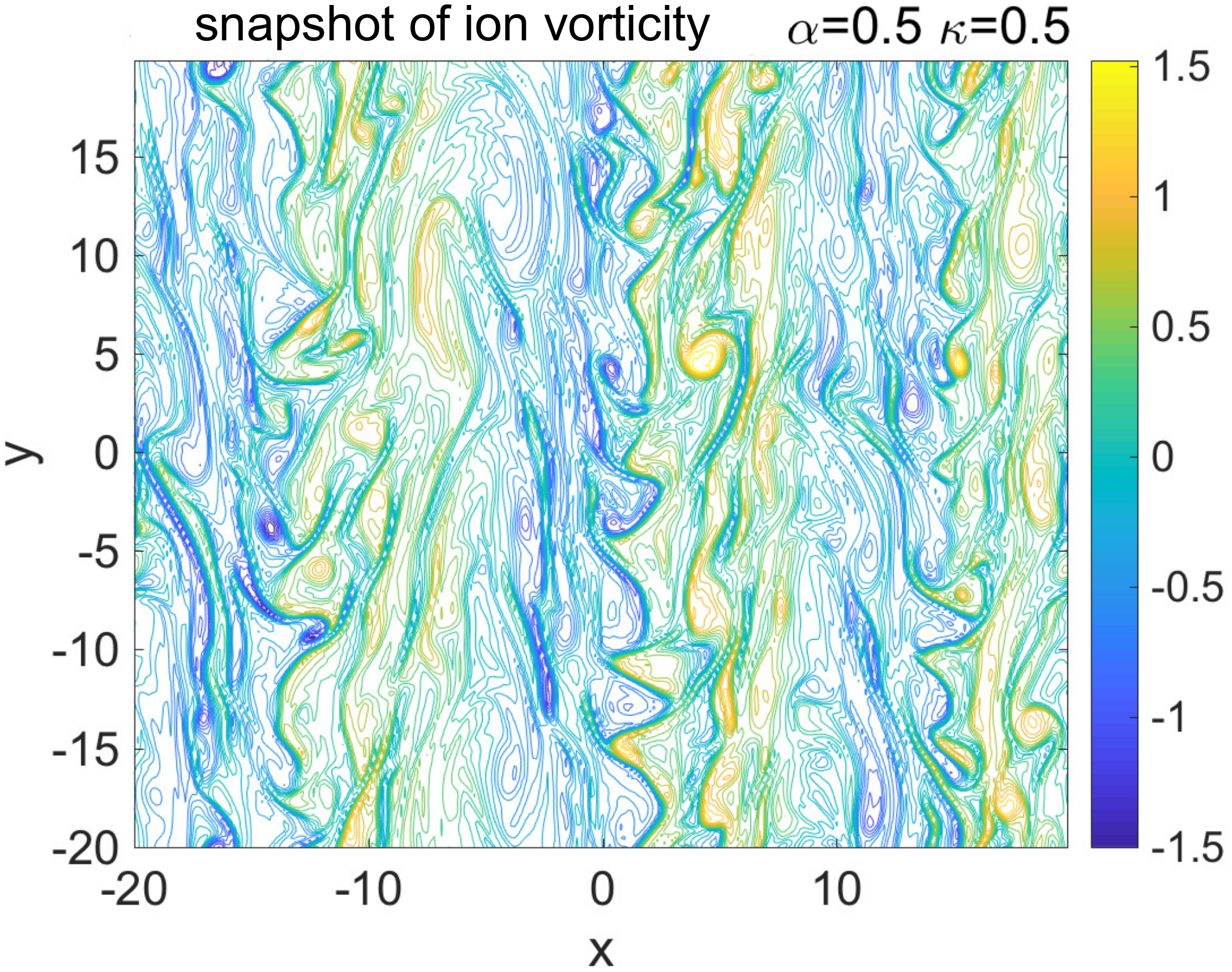}
}

\caption{Snapshots of the ion vorticity $\zeta=\Delta\varphi$
at the final time of the simulation for the bHW model (top row) and the mHW model (bottom row). For each figure $\kappa=0.5$, and each column corresponds to a different value of $\alpha$: $\alpha=0.01,0.1,0.5$ from left to right. Notice the clear zonally elongated structures obtained in the bHW model for $\alpha=0.01$ in comparison with the homogeneous field
in the mHW model.\label{fig:Snapshots}}
\end{figure}

In section \ref{sec:Features-of-BHW}, we highlighted the critical role of the adiabaticity parameter $\alpha$ in determining the turbulent flow regime in the HW models, with a transition to strong zonal jets as $\alpha$ increases. This transition is clearly visible in Figure \ref{fig:Snapshots}, which shows contours of the ion vorticity $\zeta$ for the bHW model in the first row, and for the mHW model in the second row. These figures are obtained for the parameters $\mu=D=5\times10^{-4}$ and $C=0$, and for the final time of the simulation, which is much later than the time when statistical equilibrium has been reached. The three contour plots in each row correspond to three different values of the adiabaticity parameter $\alpha$. From left to right, the values are $\alpha=0.01$, $\alpha=0.1$, and $\alpha=0.5$. We see that for $\alpha=0.01$, many small-scale vortices are visible on top of the background mean vorticity. As the parameter value increases to $\alpha=0.1$, a stronger single
jet pattern begins to form, while still coexisting with many smaller scale fluctuations. Finally for $\alpha=0.5$ the vorticity field becomes fully
dominated by several strong zonal jets. These figures offer an explicit illustration of the statistical results given in Figure \ref{fig:Total-statistical-energy}.

Figure \ref{fig:Snapshots} also confirms another key observation regarding differences between the bHW and mHW models. The bHW model maintains jet structures for a wide range of values of the adiabaticity parameter $\alpha$, throughout the transition from a regime with dominant zonal jets (large $\alpha$) to the strong drift wave turbulence regime (small $\alpha$). For small values of $\alpha$, the jets are more turbulent and shift in time, but the anisotropic zonal dynamics persists. This is in stark contrast to the mHW model, which loses the jets as $\alpha\rightarrow0$, as a regime with fully homogeneous turbulence and strong vortices sets in. Because of the persistence of the zonal jets, the particle flux is always smaller in the bHW model than in the mHW model for small values of $\alpha$. Physically, the unbalanced density flux in the mHW model is responsible for the highly turbulent vorticity and strong particle transport in the limit of small $\alpha$.

\begin{figure}
\includegraphics[scale=0.45]{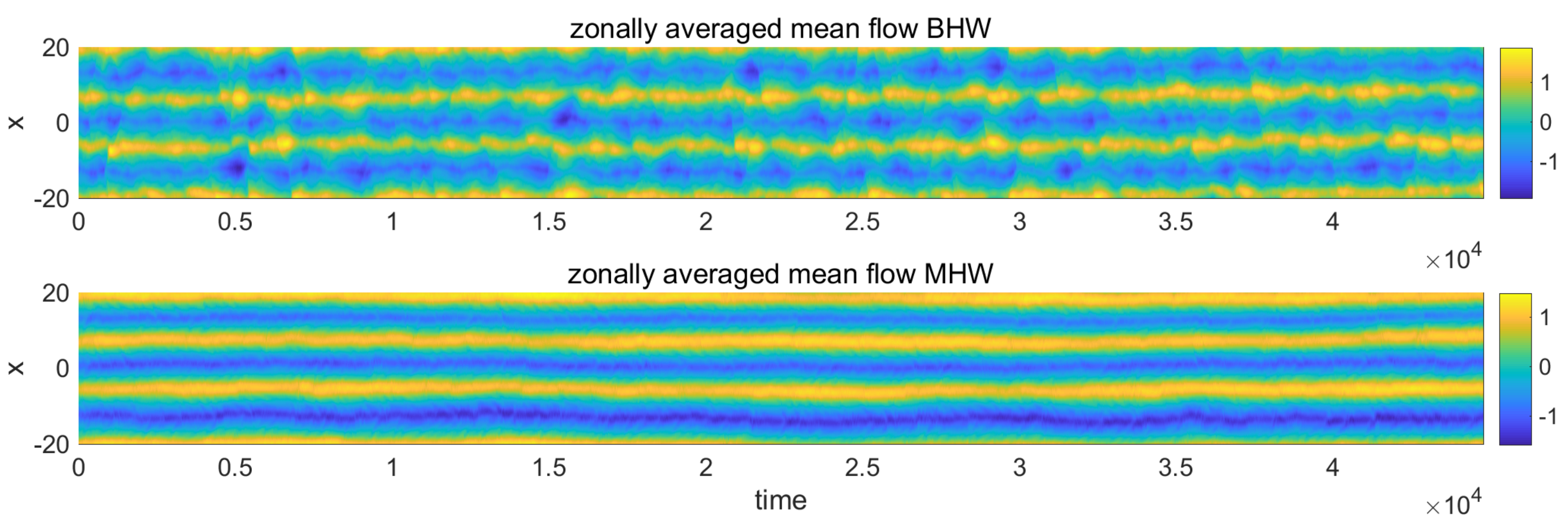}
\caption{Time-series of the zonally averaged mean flow $\overline{v}=\overline{\varphi}_{x}$
for the bHW model (top) and the mHW model (bottom) in the zonal jet dominated
regime $\alpha=0.5,\kappa=0.5$.\label{fig:Time-series-of-jets}}
\end{figure}

In the large $\alpha$ regime, we highlight another critical difference between the mHW and the bHW models, namely the fact that in the bHW model, both the zonal mean flow and the fluctuations have a much larger variability than in the mHW model. To demonstrate this, in Figure \ref{fig:Time-series-of-jets} we plot the time series of the zonal
mean velocity field $\overline{v}=\partial\overline{\varphi}/\partial x$ for $\alpha=0.5$, which corresponds to a strong zonal jet regime. The jets generated in the bHW model have large amplitude variations in time, whereas the zonal velocity in the mHW model is mostly steady in time with an almost constant jet amplitude.

\subsection{Role of dissipation terms\label{sub:The-effects-inhomo}}

Thus far, we have always turned off ion Landau damping by setting $C=0$, and set the dissipation coefficients $D$ and $\mu$ to be equal, with $\mu=D=5\times10^{-4}$. In this last section, we take a closer look at the right-hand side of Eq. (\ref{eq:plasma_balc1}) by dissociating $\mu$ and $D$ and considering finite values for $C$. Specifically, we study the consequences of increasing the values of $\mu$ to $\mu=2\times10^{-3}$ while keeping the other parameters to their original values, and of increasing $D$ to $D=2\times10^{-3}$ while keeping the other parameters to their original values. In addition, we test a small value for $C$, namely $C=0.01$. Since Landau damping mostly acts on the largest
scales, we focus in this section on the regime corresponding to $\alpha=0.5,\kappa=0.5$, where there exist strong but fluctuating large-scale zonal mean modes.

\begin{figure}
\subfloat{\includegraphics[scale=0.26]{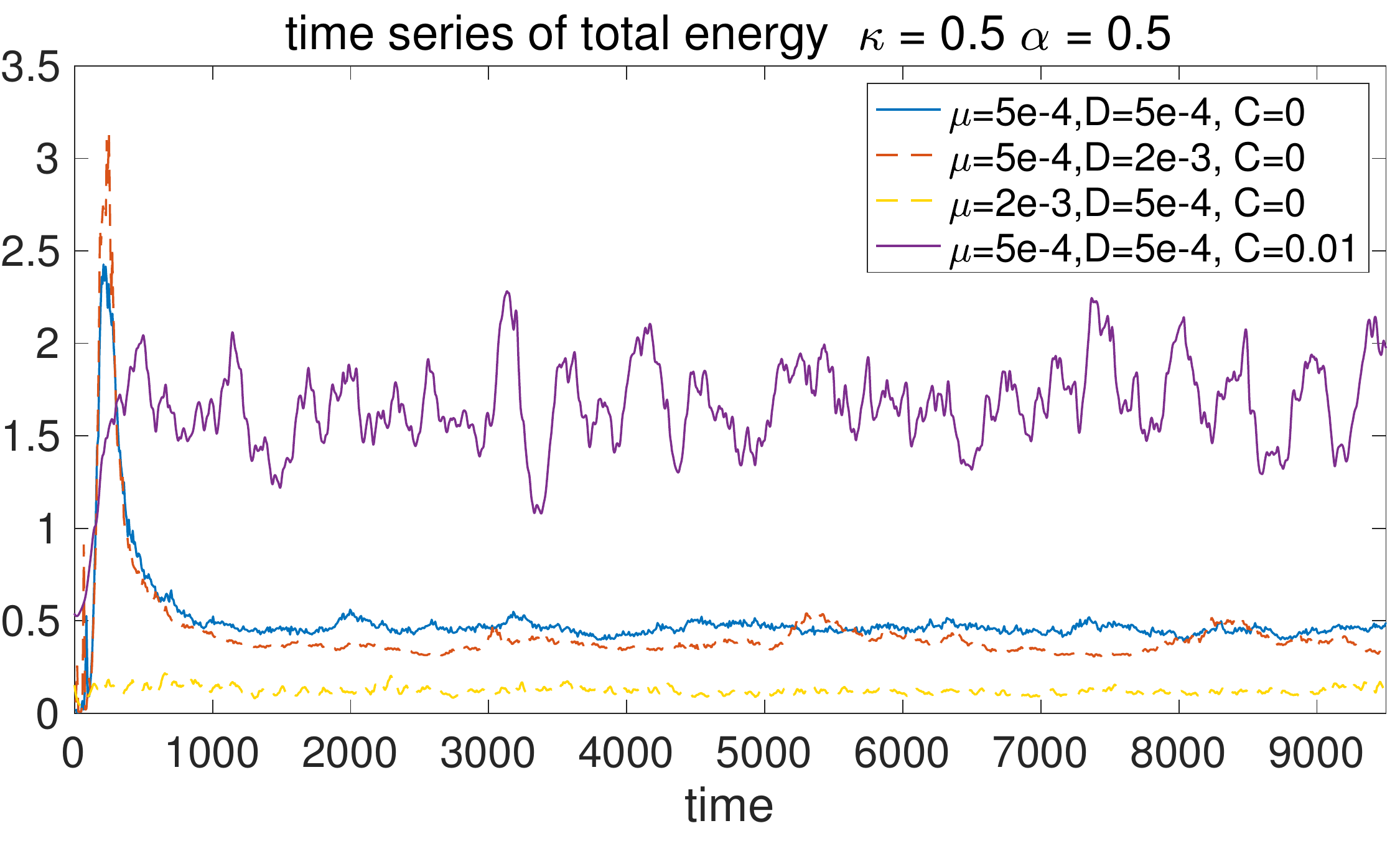}\includegraphics[scale=0.26]{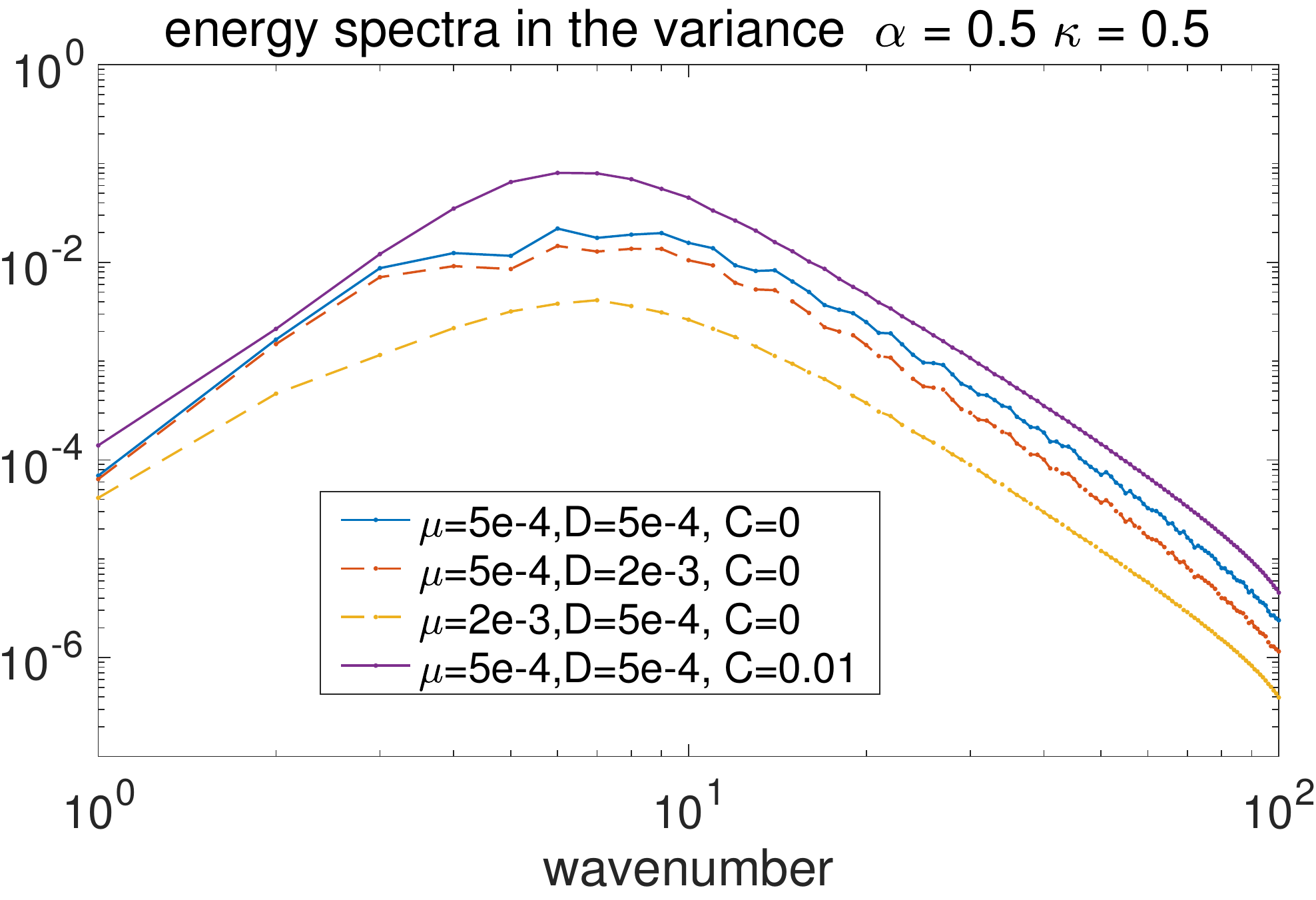}\includegraphics[scale=0.26]{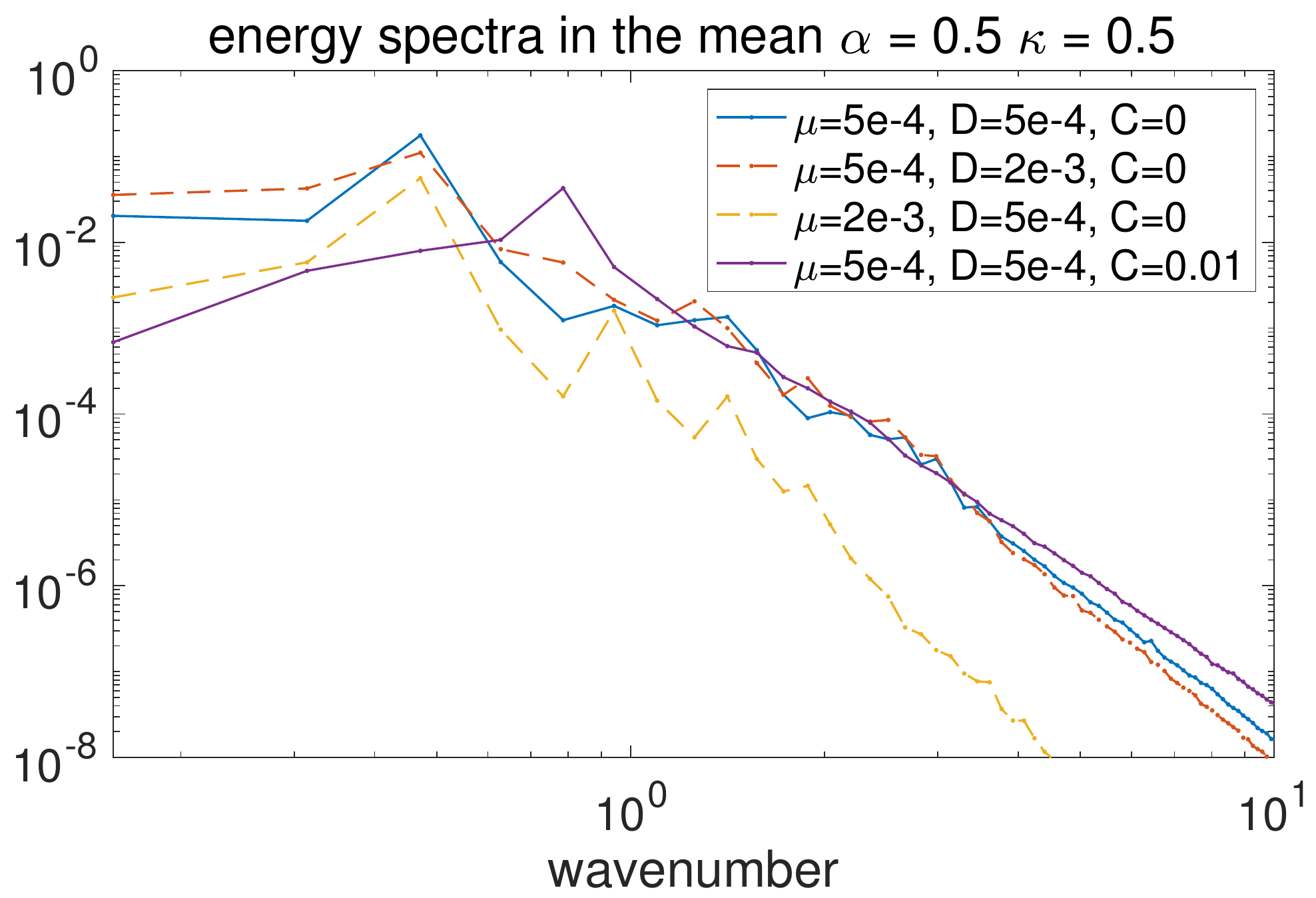}
}

\subfloat{\includegraphics[scale=0.37]{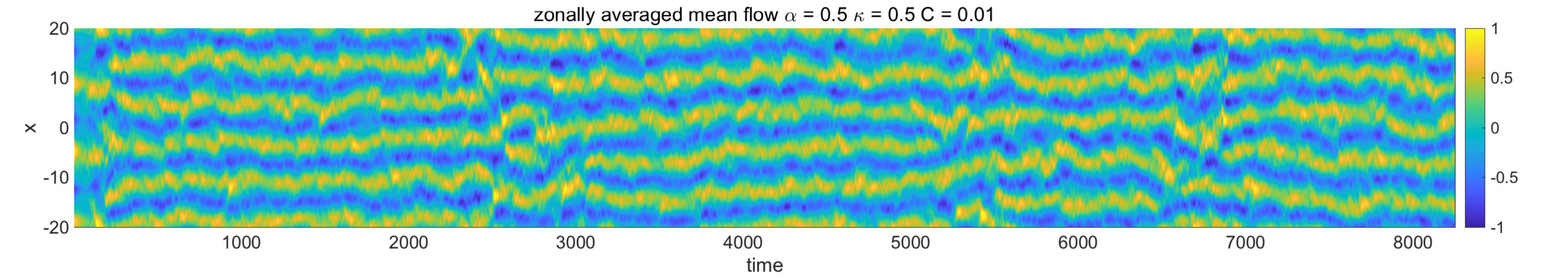}
}

\caption{(Top row) Time-series of the total energy and time-averaged energy spectra for the bHW model, for different values of the coefficients $\mu$, $D$, and $C$. (Bottom row) Time-series of the zonally averaged mean flow $\overline{v}=\overline{\varphi}_{x}$ for the bHW model when ion Landau damping is included, with $\mu=D=5\times 10^{-4}$ and $C=0.01$.  Notice how the jets are more turbulent when $C\neq 0$, with stronger small-scale fluctuations. For all these results, $\alpha$ and $\kappa$ are fixed, with $\alpha=0.5,\kappa=0.5$.\label{fig:spectra-damping}}
\end{figure}

Figure \ref{fig:spectra-damping} provides a summary of our main results. The top left panel shows time series of the total energy in the system for the different values of $\mu$, $D$ and $C$ considered in this section. The blue red, and yellow curves correspond to cases in which there is no Landau damping, and we see that increasing $D$ by a factor of 4, to $D=2\times10^{-3}$, only decreases the total energy by a small amount. As expected from our linear stability analysis in Appendix \ref{sec:Linear-Instability}, increasing $\mu$ by a factor of 4, to $\mu=2\times10^{-3}$ has a slightly stronger effect, decreasing the total energy to a noticeably lower value, but it does not significantly change the nature of the dynamics. In contrast, ion Landau damping greatly changes the energy time series, even for small values of $C$, as shown with the purple curve. In this regime with strong zonal jets, Landau damping increases the total energy and leads to much larger fluctuations. This is because our naive model for Landau damping damps the large scales, but increases the energy at the intermediate and small scales, as we now explain with a spectral viewpoint.

To better understand the results found for the time series of the total energy, we show the energy spectra of the variance $\langle E_{\mathrm{tot}}'^2\rangle_{\mathrm{eq}}$ and the statistical mean $\langle E_{\mathrm{tot}}\rangle_{\mathrm{eq}}$ in the top center and top right panels of Figure \ref{fig:spectra-damping} respectively. These figures confirm that increasing $D$ has a small impact on small scale modes, and does not significantly modify the largest zonal mean modes of the flow either. The impact of a larger value of $\mu$ is more noticeable, but approximately uniform across all scales. On the other hand, setting the Landau damping coefficient $C$ to a finite value has a strong effect in changing the zonal flow profile $\overline{v}$. Landau damping effectively removes the energy of the zonal modes at the largest scales, and drives the original zonal flow with 3 jets to a configuration with 5 dominant jets in the flow. At the same time, it increases the variance at all scales, and particularly strongly at intermediate scales. We illustrate this effect by showing the time-series of the zonal mean flow $\overline{v}=\partial_x\overline{\varphi}$ with Landau damping in the second row of Figure \ref{fig:spectra-damping}. Comparing this figure with the analogous Figure \ref{fig:Time-series-of-jets} for the same regime in the absence of Landau damping, we see that the jets are more turbulent when Landau damping is included, meandering in time with stronger fluctuations. Moreover, instead of a persistent 3-jet structure as in the $C=0$ case, the flow shows a dominant 5-jet structure, with sudden merging events leading to a 4-jet flow, before the 5 jets reemerge after a short period of time.

\section{Summary and Discussion}\label{sec:conclusion}

In this article, we presented a new two-field fluid model for the study of the drift wave -- zonal flow dynamics in magnetically confined plasmas. Our model is inspired by the rich tradition of Hasegawa-Wakatani (HW) models \cite{wakatani1984collisional,numata2007bifurcation,pushkarev2013}, which are the simplest known two-field models for magnetized plasmas which contain a drift-wave instability and can generate zonal flows through nonlinear drift wave interactions. The main novelty of our balanced HW (bHW) model as compared to previous HW models is its improved treatment of the electron dynamics parallel to the magnetic field lines, which guarantees a balanced electron density response \cite{dewar2007zonal,dorland1993gyrofluid} in the limit in which electrons are adiabatic. To achieve this, the bHW model does not solve for the ion vorticity, but instead for a well constructed potential vorticity defined by $q^{\mathrm{b}}=\nabla^2\varphi-\tilde{n}$, which we call the balanced potential vorticity and does crucially not include the zonal mean density $\overline{n}$. 

Our new bHW model inherits the desirable features of former HW models \textit{as well as} those of the Hasegawa-Mima model with a modified adiabatic electron response, which we have called mHM \cite{dewar2007zonal}. Like other HW models, the bHW model contains a drift wave instability, and like the mHW model, the bHW model has the desired Galilean invariance along a magnetic flux surface; on the other hand, like the mHM model, the improved electron response in the bHW model leads to enhanced zonal flows and strong fluctuations. Mathematically, the bHW model has the satisfying property of converging to the mHM model in the proper adiabatic limit, which is not the case of previous HW models.

We relied on direct numerical simulations to investigate the main features of the bHW model, with numerical simulations of the recent modified HW model (mHW) \cite{numata2007bifurcation} as a point of comparison. We paid close attention to the transition from a strongly drift wave turbulent regime to a much more organized regime with strong zonal structures, which is associated with a decrease of the plasma resistivity, allowing electrons to become more and more adiabatic. We observed that the bHW model has stronger fluctuations than the mHW model in the turbulent regime and throughout the transition to the regime with strong zonal jets. Remarkably, however, the bHW model maintains mainly zonal dynamics even at large plasma resistivity, while in contrast the drift wave turbulent regime in the mHW model is dominated by small scale dynamics. In other words, in the bHW model, the zonal flows are more robust, in the sense that they are observed for any value of the plasma resistivity, but they have a larger variability than in the mHW model in the regime in which they are seen in both models. We also used direct numerical simulations to confirm the convergence to the mHM model in the zero resistivity limit, and to study the role of the dissipation terms introduced in our model. We noticed that the simple model for Landau damping introduced by Wakatani and Hasegawa \cite{wakatani1984collisional} leads to more turbulent dynamics, with zonal structures which are fluctuating more, and more energy and variability at intermediate and small scales. The significant effect of the Landau damping term on the bHW dynamics provides yet again a clear justification for the development and implementation of more accurate fluid approximations for this effect, as was first done in the pioneering work of Hammett, Perkins, and Dorland \cite{hammettperkins,dorland1993gyrofluid}.

This article is meant to be the first in a series of articles in which we study the interplay between drift waves and zonal flows using the new bHW model presented here. In a second article \cite{qmc2018}, more focused on the analysis of detailed numerical simulations, we investigate the effect of the shape and size of the computational domain on the properties of the drift wave turbulence and zonal structures. In a subsequent article, we will treat the bHW model and the drift wave - zonal flow dynamics it describes as a test bed to explore the applicability of statistical model reduction strategies \cite{majda2018strategies,qi2016low} for the efficient description of turbulent magnetized plasmas. The latter is the topic of ongoing research, with progress to be reported in the near future.

%

\section*{Acknowledgements}
A.J.C. would like to thank Jeffrey B. Parker for the suggestions he made during an insightful conversation which led to this work. A.J.M. is partially supported by the Office of Naval Research through MURI N00014-16-1-2161 and DARPA through
W911NF-15-1-0636. D. Q. is supported as a postdoctoral fellow on the
second grant. A.J.C is partially supported by the US
Department of Energy, Office of Science, Fusion Energy
Sciences under Award Numbers DE-FG02-86ER53223 and
DE-SC0012398.

\appendix 
\numberwithin{equation}{section}

\section{Linear drift wave instability in the flux-balanced Hasegawa-Wakatani model\label{sec:Linear-Instability}}

In this appendix, we present a detailed linear stability analysis for the bHW model given by Eq. (\ref{plasma_balanced}). For this analysis, we assume a zero background mean flow, so the linearized equations for the perturbations
$\left(\varphi_{1},q_{1}^{\mathrm{b}},n_{1}\right)$ to the background state are given by
\[
\begin{aligned}\partial_{t}q^{\mathrm{b}}_{1}-\kappa\partial_{y}\tilde{\varphi}_{1} & =\left[\left(\mu-D\right)\Delta^{2}+C\omega_{*}\right]\varphi_{1}+D\Delta q_{1}^{\mathrm{b}},\\
\partial_{t} n_{1}+\kappa\partial_{y}\tilde{\varphi}_{1} & =\alpha\left(\tilde{\varphi}_{1}-\tilde{n}_{1}\right)+D\Delta n_{1}.
\end{aligned}
\]
The dispersion relation for the bHW model is found by considering a single mode solution for the perturbations, i.e.
\[
\varphi_{1}=\hat{\varphi}_{1}\exp\left(i\left(\mathbf{k\cdot x}-\omega t\right)\right),\quad n_{1}=\hat{n}_{1}\exp\left(i\left(\mathbf{k\cdot x}-\omega t\right)\right),\quad\zeta_{1}=-k^{2}\hat{\varphi}_{1}\exp\left(i\left(\mathbf{k\cdot x}-\omega t\right)\right),
\]
where we assume $k_{y}\neq 0$. Substituting the above single mode solution in the linearized equations, we find the following equations for $\hat{\varphi}_{1}$ and $\hat{n}_{1}$:
\begin{equation}
\begin{aligned}i\omega k^{2}\hat{\varphi}_{1} & =\alpha\left(\hat{\varphi}_{1}-\hat{n}_{1}\right)+\mu k^{4}\hat{\varphi}_{1}+C\frac{\kappa k_{y}}{1+k^{2}} \hat{\varphi}_{1},\\
-i\omega\hat{n}_{1} & =\alpha\left(\hat{\varphi}_{1}-\hat{n}_{1}\right)-i\kappa k_{y}\hat{\varphi}_{1}-Dk^{2}\hat{n}_{1},
\end{aligned}
\label{eq:linear_dispersion}
\end{equation}
We first derive and analyze the dispersion relation for the situation without dissipation $D=\mu=C=0$. We will then discuss the modifications due to the various dissipation effects.

\subsection{Linear stability analysis without dissipation}

From the coupled equations (\ref{eq:linear_dispersion})
in the absence of dissipation effects, $\mu=D=C=0$, it is straightforward to derive the dispersion relation
\begin{equation}
\omega^{2}+ib\left(\omega-\omega_{*}\right)=0,\label{eq:dispersion1}
\end{equation}
where the drift wave frequency $\omega_{*}$ and
the wavenumber dependent coefficient $b$ are given by
\[
\omega_{*}\left(k\right)=\kappa k_{y}/\left(1+k^{2}\right),\quad b\left(k\right)=\alpha\left(1+k^{-2}\right).
\]
We immediately see that if $\alpha=0$, $b=0$ and $\omega=0$, so all the modes are marginally stable. Furthermore, when $\alpha\neq 0$ but $\kappa=0$, $\omega=0$ and $\omega=-ib$ are solutions, so there are only damped and marginally stable modes. For the general case with $\alpha\neq0,\kappa\neq0$, we write the solution to equation (\ref{eq:dispersion1}) as $\omega =\omega_{r}+i\omega_{i}$, with
\begin{equation}
\omega_{r}=\pm\frac{b}{2}\left(1+16\gamma^{2}\right)^{1/4}\cos\frac{\theta}{2},\;\omega_{i}=-\frac{b}{2}\pm\frac{b}{2}\left(1+16\gamma^{2}\right)^{1/4}\sin\frac{\theta}{2},\theta=\mbox{Arg} (-1+i4\gamma),\;\gamma=\frac{\omega_{*}}{b}.\label{eq:soln1}
\end{equation}
Here, $\mbox{Arg}$ stands for the principal value argument. Thus, for the case $\alpha\ne0,\kappa\neq0$, linear stability is determined by comparing the term $\left(1+16\gamma^{2}\right)\sin^{4}\frac{\theta}{2}$ with 1. $\theta=\mbox{Arg} (-1+i4\gamma)$ implies that $\cos^{2}\theta=\left(1+16\gamma^{2}\right)^{-1}$. Hence,
\begin{equation}
\begin{aligned}\left(1+16\gamma^{2}\right)\sin^{4}\frac{\theta}{2} & =\frac{1}{4}\left(1+\left(1+16\gamma^{2}\right)^{-1/2}\right)^{2}\left(1+16\gamma^{2}\right)\\
 & =\frac{1}{4}\left(\sqrt{1+16\gamma^{2}}+1\right)^{2}>1\:\mathrm{for\:any}\:\gamma\neq0.
\end{aligned}
\label{eq:bnd_coeff}
\end{equation}
We conclude that for all modes with $k_{y}\neq0$, there is a solution with $\omega_{i}>0$. In other words, in the absence of dissipation, all modes with $k_{y}\neq0$ are unstable.

Keeping in mind the complete equations including dissipation, it is valuable to construct a boundary curve outside of which the growth rate can be regarded as small compared to dissipation mechanisms. We know that the system is marginally stable if $\gamma=\omega_{*}/b=0$. The growth rate of linear instability is small if $\omega_{*}\left(k\right)\ll b\left(k\right)$, which we write as
\[
\frac{\omega_{*}}{b}=\epsilon,\quad\epsilon\ll1.
\]
The above relation can be rewritten as
\[
\left(k+k^{-1}\right)^{2}=\frac{k_{y}}{\epsilon}\frac{\kappa }{\alpha}.
\]
In the asymptotic regime $\epsilon\ll 1$, the above equation has one solution for $k\ll 1$ and one solution for $k\gg 1$. We discard the former since for our finite size periodic box, the smallest wavenumber has magnitude $2\pi/L$. We thus have, for $\epsilon\ll 1$,
\[
k+\frac{1}{k}=\left(\frac{k_{y}}{\epsilon}\right)^{1/2}\left(\frac{\kappa}{\alpha}\right)^{1/2}.
\]
The curve separating slowly growing modes from faster growing modes for our drift wave instability is therefore given by the following Cartesian equation in wavenumber space:
\begin{equation}
k_{x}^{2}+\left(k_{y}-\frac{1}{2\epsilon}\frac{\kappa}{\alpha}\right)^{2}=\frac{1}{4\epsilon^{2}}\left(\frac{\kappa}{\alpha}\right)^{2}.\label{eq:meta_stable}
\end{equation}
It describes a circle in the spectral domain, centered in $\left(0,\kappa/2\epsilon\alpha\right)$ and
with radius $\kappa/2\epsilon\alpha$. Inside
the circle the modes have large linear growth rates,
and outside the circle the modes are quasi-stable. We note that our analysis
shows that the range of wavenumbers which correspond to very unstable modes only depends on the ratio $\kappa/\alpha$.

Now, returning to the general situation with arbitrary values for $\omega_{*}/b$, we can combine our expression for the growth rate in Eq. (\ref{eq:soln1}) and
the relation (\ref{eq:bnd_coeff}) to write 
\[
\omega_{i}=\frac{b}{2}\left[-1+\frac{1}{\sqrt{2}}\left(\sqrt{1+16\gamma^{2}}+1\right)^{1/2}\right].
\]
The magnitude of the term in the square parentheses is determined, for a given wave vector, by the magnitude of the ratio $r=\kappa/\alpha$. We can derive insightful explicit formulae for the wavenumber dependence of $\omega_{i}$ for the two asymptotic regimes $r\ll1$ and $r\gg1$.

For $r\ll 1$, $\gamma\ll 1$ and we find
\[
\omega_{i}\sim\alpha\frac{(k^2+1)}{k^2}\gamma^{2}=\frac{\kappa^2}{\alpha}\frac{k_{y}^2 k^2}{(1+k^2)^3};
\]
For fixed values of $\kappa$ and $\alpha$, we can look for the mode with the largest growth rate. Given the symmetry of the linearized equations with respect to the wavenumber $k_{x}$, we look for a mode with $k_{x}=0$ and $k_{y}$ finite. In that case,
\[
\omega_{i}\sim\frac{\kappa^2}{\alpha}\frac{k_{y}^4}{(1+k_{y}^2)^3};
\]
The maximum growth rate $\omega_{i}^{\max}=\frac{4}{27}(\kappa^2/\alpha)$ is reached for $k_{y}^{\max}=\sqrt{2}$, corresponding to $\gamma^{\max}=\frac{2\sqrt{2}}{9} (\kappa/\alpha)$ and $b^{\max}=3\alpha/2$.

For $r\gg 1$, $\gamma\gg1$ and we find
\[
\omega_{i}\sim\alpha\frac{(k^2+1)}{2 k^2}\sqrt{2\gamma}=\sqrt{\frac{\alpha\kappa}{2 k_{y}}}.
\]
In this regime, the largest growth rate is obtained for the smallest finite $k_{y}$.

\begin{figure}
\includegraphics[scale=0.25]{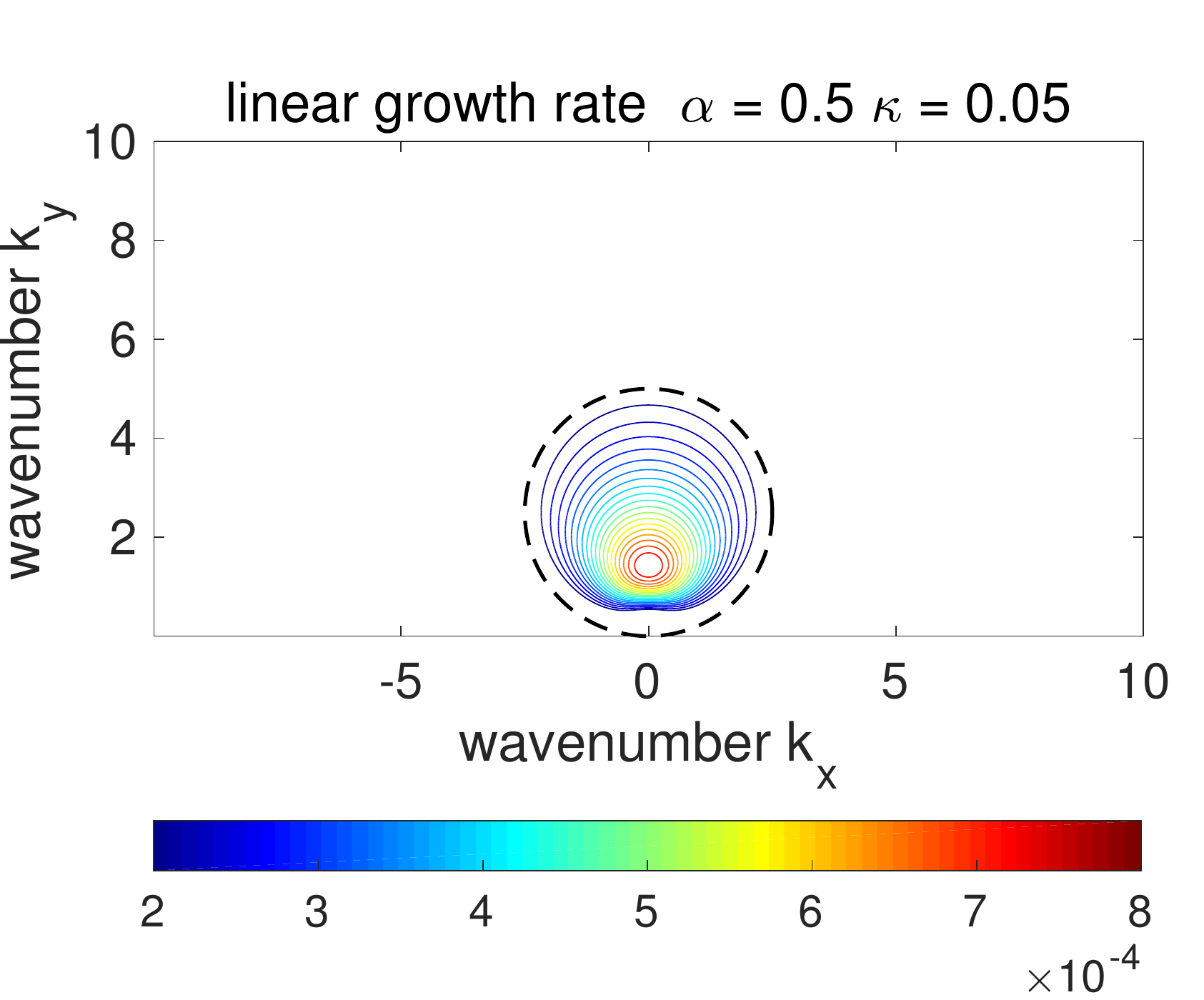}\includegraphics[scale=0.25]{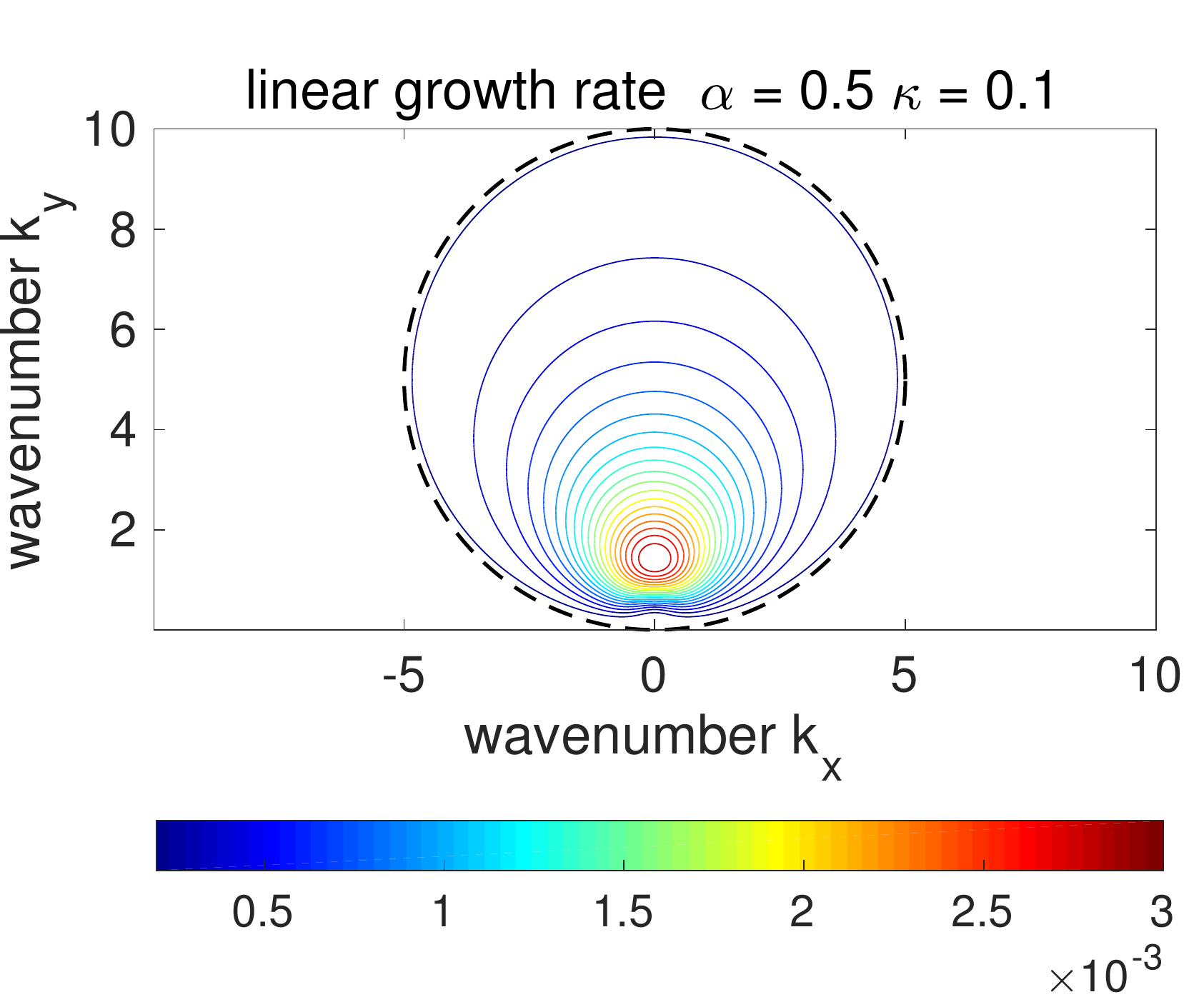}\includegraphics[scale=0.25]{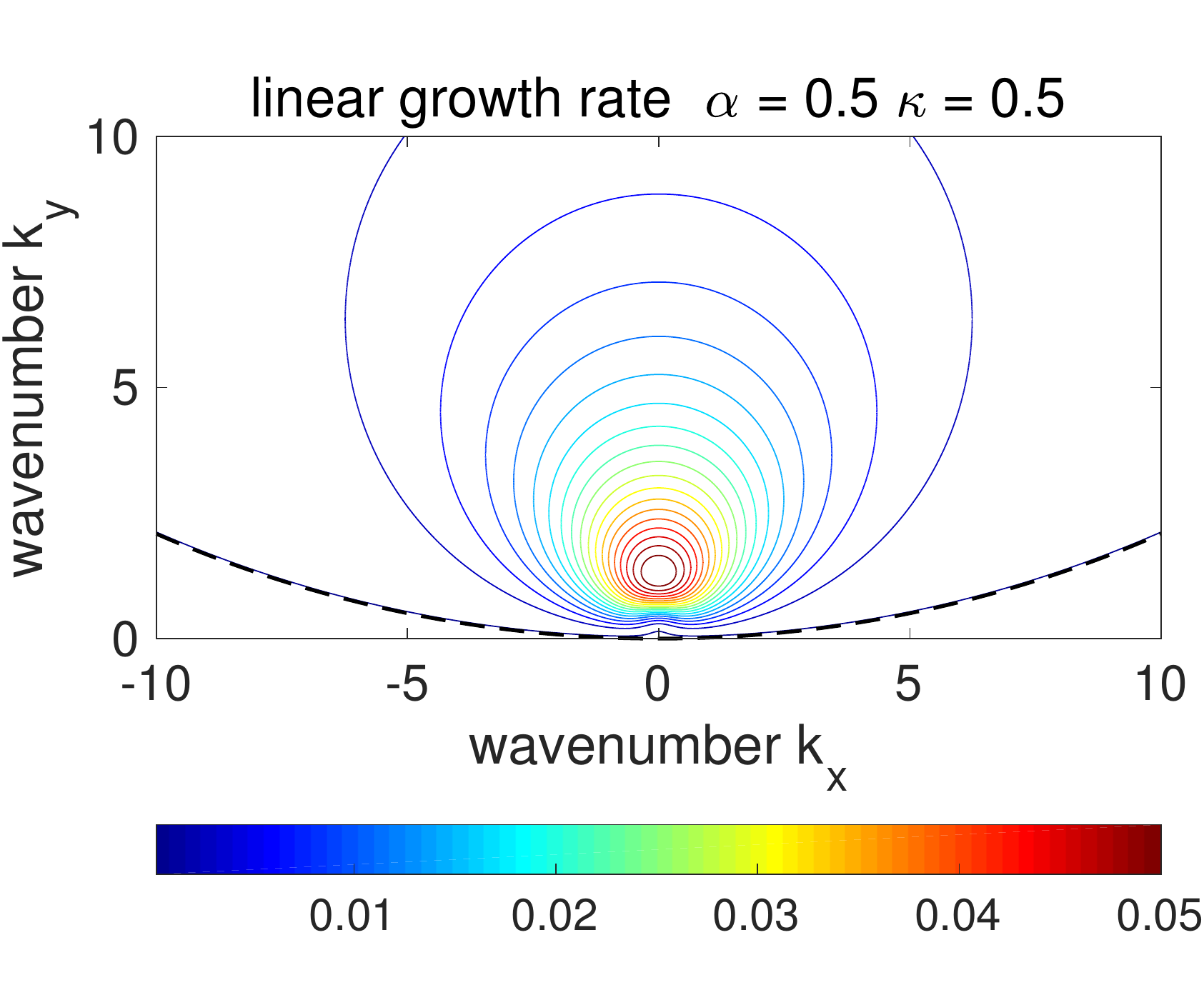}\includegraphics[scale=0.25]{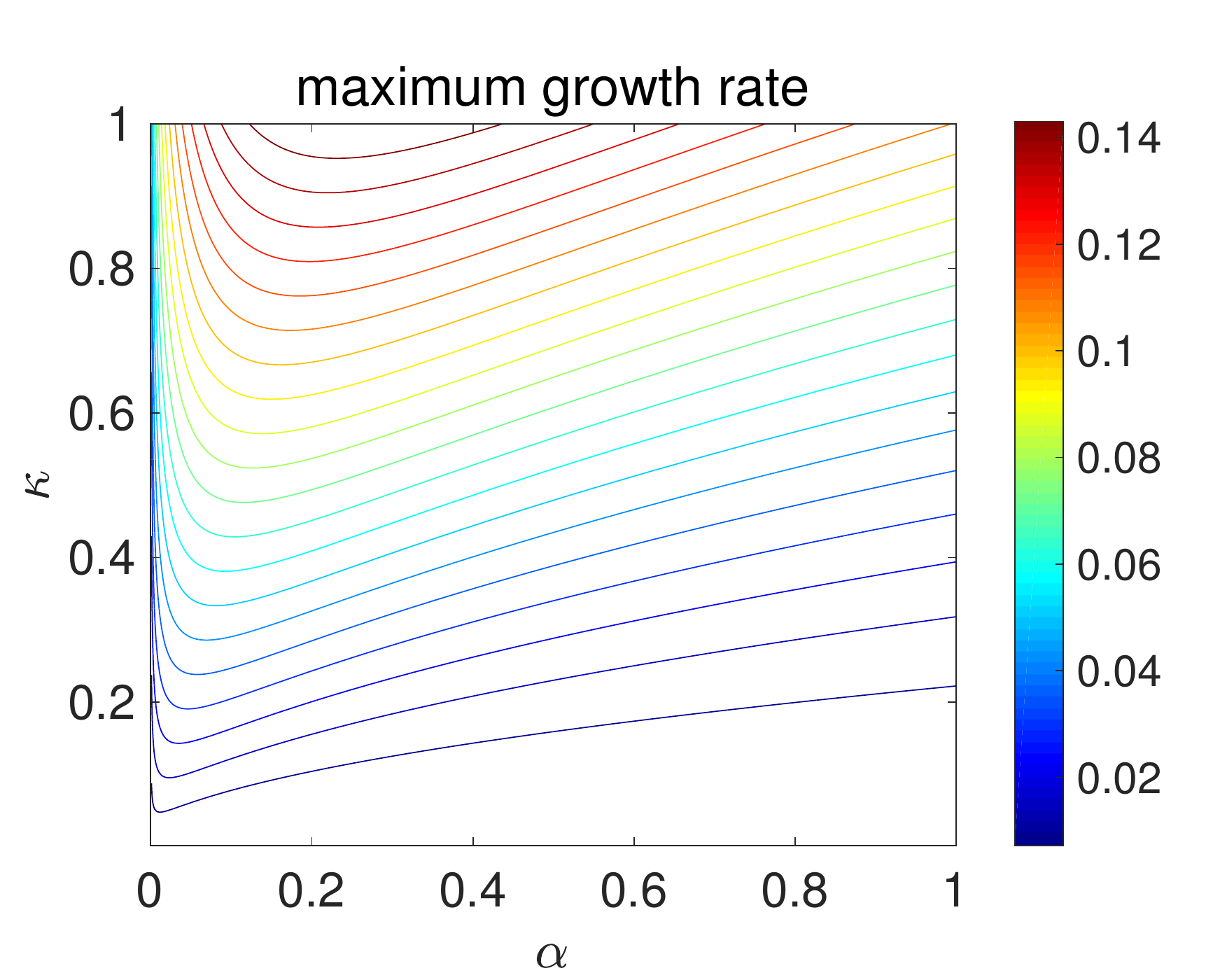}
\caption{Linear growth rates from the solution of (\ref{eq:soln1}) with different
ratios $\kappa/\alpha$.
The black dashed line corresponds to the circle of meta-stability given by (\ref{eq:meta_stable}),
with $\epsilon=2\times10^{-2}$. The fast growing modes are located inside the circle. 
The figure on the right shows the maximum linear growth $\omega_{i}^{\max}$
rate as a function of the model parameters $\alpha$ and $\kappa$.\label{fig:Growth-rate-HW}}
\end{figure}

Our linear stability analysis is confirmed by our numerical results. The first two plots on the left in Figure \ref{fig:Growth-rate-HW} are contour plots of the growth rate $\omega_{i}$ in Fourier space, calculated from Eq. (\ref{eq:soln1}). For these plots, we chose $\epsilon=2\times10^{-2}$ and did not plot contours corresponding to slow growing modes, outside of the circle of meta-stability given by Eq. (\ref{eq:meta_stable}) and marked by the dashed black lines. The size of the region with strong growing modes increases with the ratio $\kappa/\alpha$, as predicted. The axial symmetry with respect to the line $k_{x}=0$ is clearly visible, and we verify that the maximum growth rate is reached for $k_{x}=0$ and $k_{y}\sim\sqrt{2}$ for $\kappa/\alpha\ll 1$. The rightmost plot in Figure \ref{fig:Growth-rate-HW} shows contours of the maximum growth rate in $\kappa$-$\alpha$ space. The dependence on $\kappa$ and $\alpha$ in the regimes $r\ll 1$ and $r \gg 1$ agree with our analytic expressions derived for these asymptotic regimes. 

\subsection{Modifications due to dissipative effects}

\begin{figure}
\includegraphics[scale=0.31]{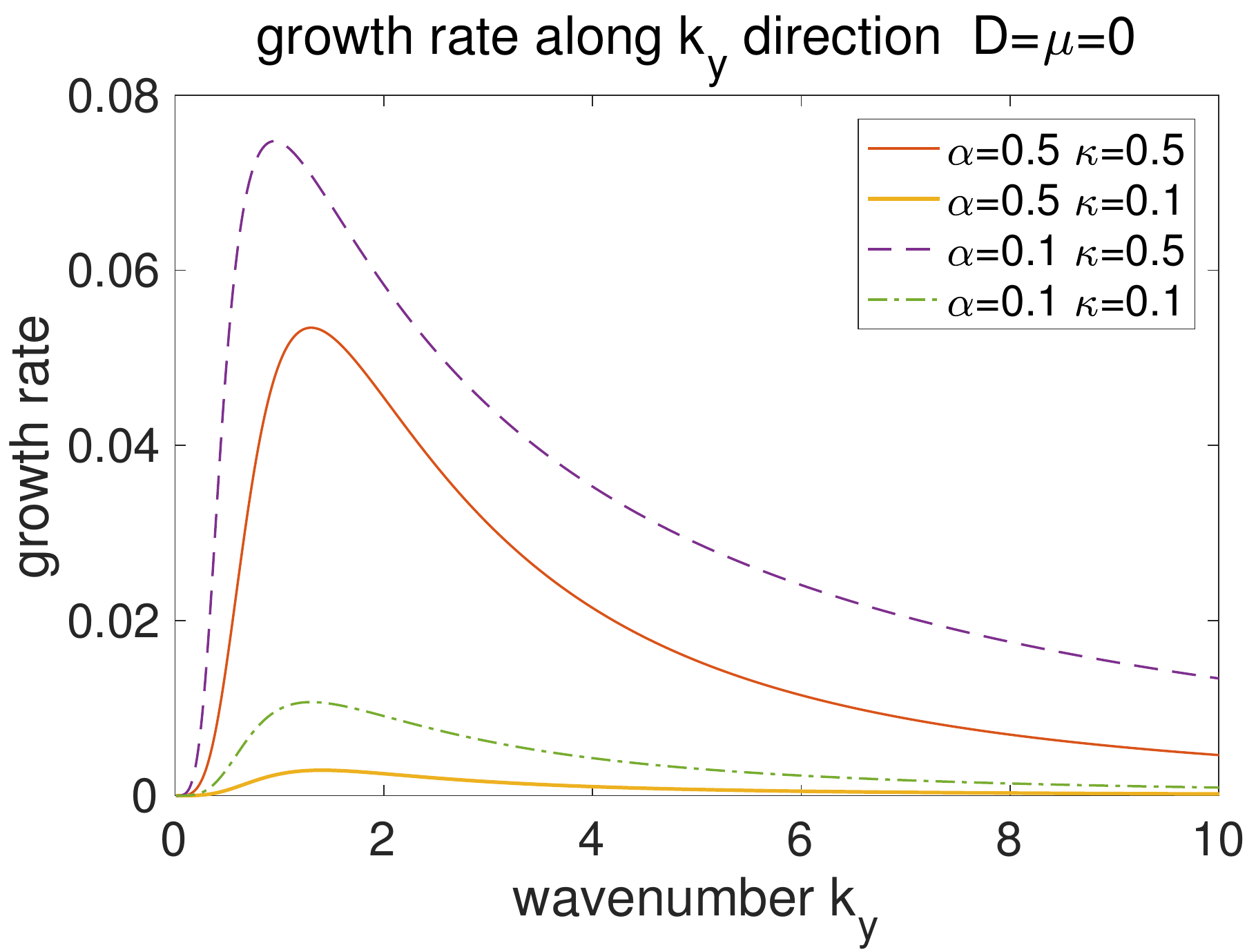}\includegraphics[scale=0.31]{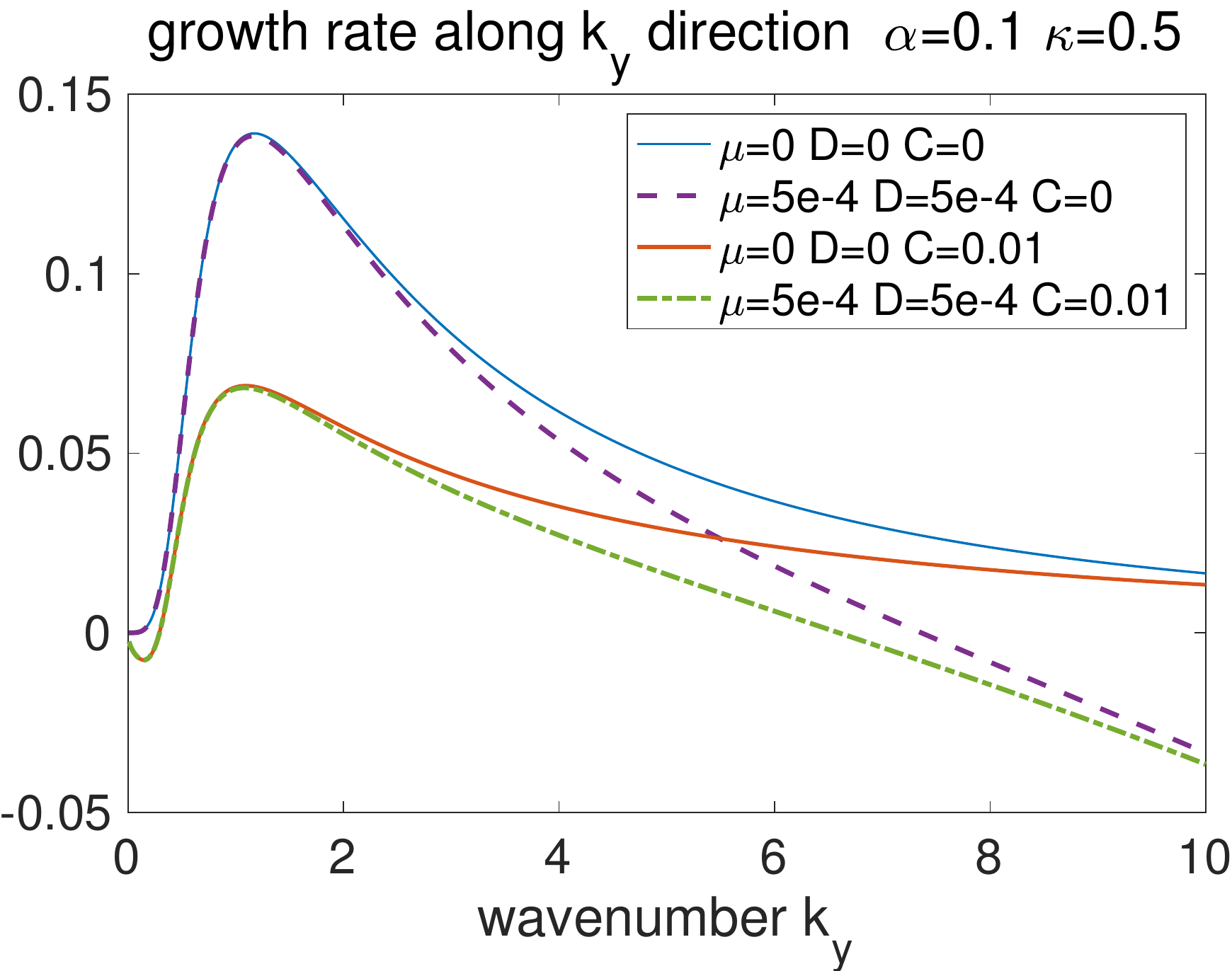}\includegraphics[scale=0.30]{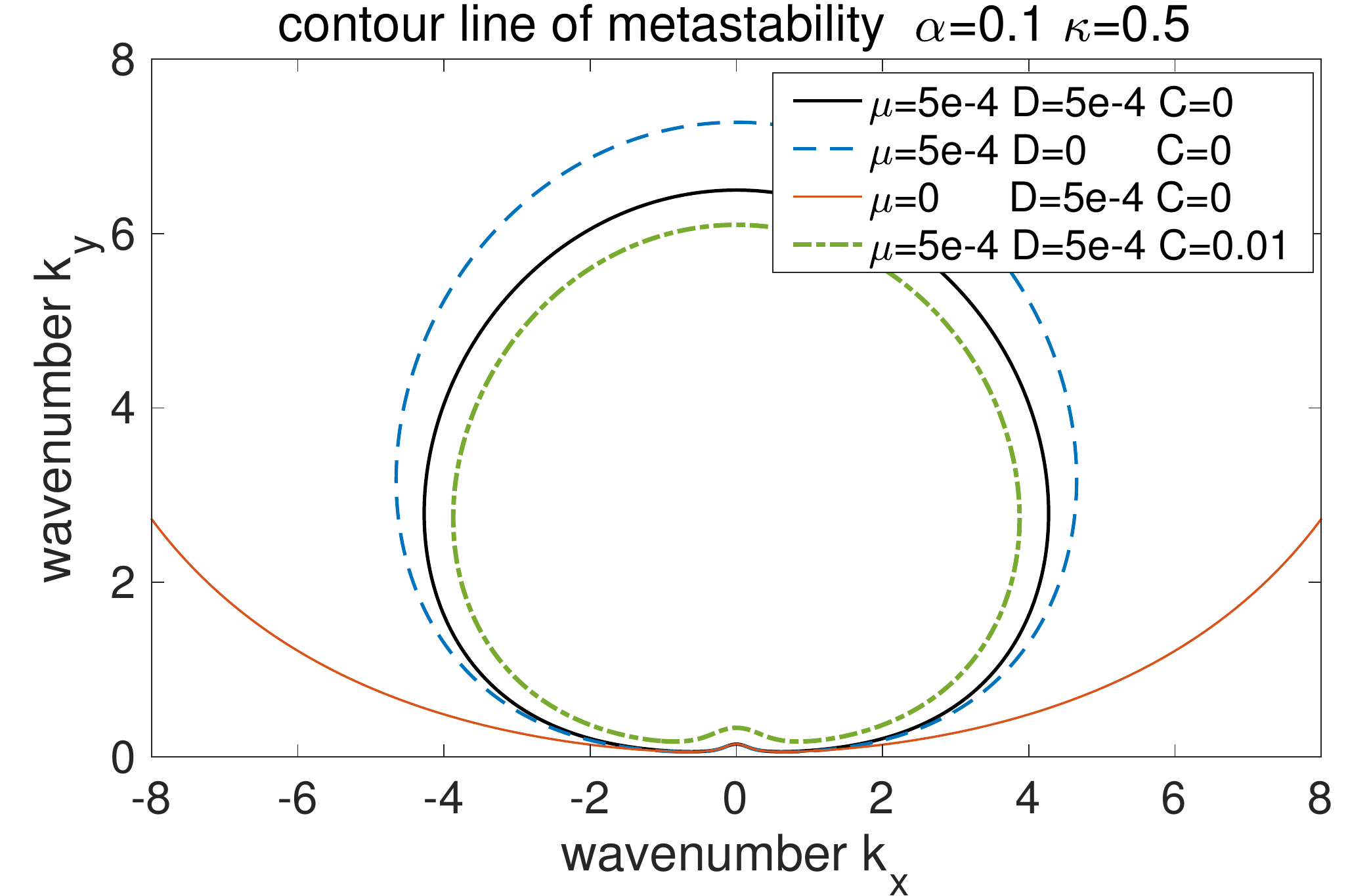}

\caption{Comparison of growth rates with changing parameter values for $\alpha,\kappa$, $\mu$, $D$, and $C$. The left and middle plots show the linear
growth rates as a function of $k_{y}$, for $k_{x}=0$ and different values of $\alpha,\kappa$, $\mu$, $D$, and $C$. The plot on the right shows the contours of meta-stability with small growth rate $\omega_{i}=1\times10^{-3}$ as the coefficients of the dissipation terms are changed.\label{fig:Growth-rate-comp}}
\end{figure}

We now consider the effects of the dissipation terms, namely $\mu\Delta\zeta+C\omega_{*}\varphi$
for the vorticity equation, and $D\Delta n$ for the density equation.
The linearized equation (\ref{eq:linear_dispersion}) can be written as a $2\times2$
system of equations for each wavenumber, given by
\begin{equation}
\begin{bmatrix}i\omega-\alpha k^{-2}-\left(\mu k^{2}+C\omega_{*}(k)k^{-2}\right) & \alpha k^{-2}\\
-\alpha+i\kappa k_{y} & -i\omega+\alpha+Dk^{2}
\end{bmatrix}\begin{bmatrix}\hat{\varphi}\\
\hat{n}
\end{bmatrix}=0.\label{eq:linearized}
\end{equation}
Nontrivial solutions to the system exist provided the following dispersion relation has solutions:
\begin{equation}
\omega^{2}+i\left(b+L_{1}\right)\omega-ib\omega_{*}(k)-\alpha L_{2}-D\left(\mu k^{4}+C\omega_{*}(k)\right)=0,\label{eq:dispersion2}
\end{equation}
where 
\[
L_{1}\left(k\right)=Dk^{2}+\mu k^{2}+C\omega_{*}(k)k^{-2},\quad L_{2}\left(k\right)=\left(D+\mu k^{2}+C\omega_{*}(k)k^{-2}\right).
\]
Because of the complexity of the dispersion relation (\ref{eq:dispersion2}), we will only solve it numerically. To do so, we choose the same numerical values for the parameters as we used in the numerical simulations
discussed in the main text. Specifically, if $\mu\neq 0$, then $\mu=5\times10^{-4}$, if $D\neq 0$, then $D=5\times10^{-4}$, and if $C\neq 0$, then $C=0.01$. Our results comparing linear growth rates with and without dissipation terms are shown in Figure \ref{fig:Growth-rate-comp}.
The first two plots on the left show the dependence of the linear growth rates on $k_{y}$ along the line $k_{x}=0$ in $k$-space, which is the line along which the maximum growth rate is reached. The situation without dissipation is considered in the first plot on the left, for different values of $\kappa$ and $\alpha$. The curves confirm our analytic results in the previous section, with a maximum reached for $k_{y}\sim\sqrt{2}$ for $r\ll 1$ and moving to the left as $r$ increases. In the second plot, we analyze the effect of including dissipation terms, for $\alpha=0.1$ and $\kappa=0.5$. We see that as expected, our simple model for Landau damping stabilizes the large scale modes, but has a negligible effect on small scale modes. In contrast, the terms $\mu\Delta\zeta$ and $D\Delta n$ only stabilize the small scale modes, and have little effect on the large scale modes. In order to determine the relative importance of these two terms, we turn to the plot on the right of Figure \ref{fig:Growth-rate-comp}, which shows the contour of meta-stability, set here at $\omega_{i}=1\times10^{-3}$, for different values of the parameters $\mu$, $D$, and $C$, and $\alpha=0.1$ and $\kappa=0.5$. We see that the term $\mu\Delta \zeta$ has a much stronger damping effect, since setting $\mu$ to zero leads to a much increased region of strong instability. We also have the confirmation that the model Landau damping term stabilizes the large scale modes.

\section{Summary of similarities and differences between the MHW and BHW models\label{sec:comparison-table}}
In this appendix, we present a summarized comparison between the modified Hasegawa-Wakatani model and the new flux-balanced Hasegawa-Wakatani model we introduced in this article.

\begin{minipage}[t]{0.47\columnwidth}%
\begin{center}
\textbf{modified Hasegawa-Wakatani (mHW) Model}
\par\end{center}

Governing equations for the vorticity $\zeta=\nabla^{2}\varphi$
and the density $n$
\[
\begin{aligned}\frac{\partial\zeta}{\partial t}+J\left(\varphi,\zeta\right)\qquad\quad\;\: & =\alpha\left(\tilde{\varphi}-\tilde{n}\right)\;+D\Delta\zeta,\\
\frac{\partial n}{\partial t}+J\left(\varphi,n\right)+\kappa\frac{\partial\tilde{\varphi}}{\partial y} & =\alpha\left(\tilde{\varphi}-\tilde{n}\right)\;+D\Delta n.
\end{aligned}
\]

Auxiliary equation for the balanced potential vorticity $q^{\mathrm{b}}=\nabla^{2}\varphi-\tilde{n}$, with $\tilde{u}=-\frac{\partial\tilde{\varphi}}{\partial y}$  
\[
\frac{\partial q^{\mathrm{b}}}{\partial t}+J\left(\varphi,q^{\mathrm{b}}\right)+\frac{\partial\left(\overline{\tilde{u}\tilde{n}}\right)}{\partial x}-\left(\kappa-\frac{\partial\overline{n}}{\partial x}\right)\frac{\partial\tilde{\varphi}}{\partial y}=D\Delta q^{\mathrm{b}}.
\]
\
\
\

Zonal mean and fluctuation equations for the vorticity $\zeta=\overline{\zeta}+\tilde{\zeta}$
\[
\begin{aligned} &\frac{\partial\overline{\zeta}}{\partial t}+\frac{\partial\left(\overline{\tilde{u}\tilde{\zeta}}\right)}{\partial x} =D\frac{\partial^2\overline{\zeta}}{\partial x^2},\\
&\frac{\partial\tilde{\zeta}}{\partial t}+J\left(\varphi,\tilde{\zeta}\right)+\left[\frac{\partial\overline{\zeta}}{\partial x}\tilde{u}-\frac{\partial}{\partial x}\left(\overline{\tilde{u}\tilde{\zeta}}\right)\right] =\alpha\left(\tilde{\varphi}-\tilde{n}\right)\\
&\qquad\qquad\qquad\qquad\qquad\qquad\qquad\qquad+D\Delta\tilde{\zeta}.
\end{aligned}
\]
\
\
\

Conservation laws for the energy $E$ and the enstrophy $W$
\[
\begin{aligned}& E=\frac{1}{2}\int_{\Omega}\left(\left|\nabla\varphi\right|^{2}+n^{2}\right)dV,\quad W=\frac{1}{2}\int_{\Omega}\left(\zeta-n\right)^{2}dV\\
& \frac{dE}{dt} =\kappa\int_{\Omega}\tilde{n}\tilde{u}dV-\alpha\int_{\Omega}\left(\tilde{\varphi}-\tilde{n}\right)^{2}dV-D_{E},\\
& \frac{dW}{dt} = \kappa\int_{\Omega}\tilde{n}\tilde{u}dV-D_{W}.
\end{aligned}
\]
\end{minipage}\hspace{2em}%
\begin{minipage}[t]{0.5\columnwidth}%
\begin{center}
\textbf{flux-balanced Hasegawa-Wakatani (bHW) Model}
\par\end{center}

Governing equations for the balanced potential vorticity $q^{\mathrm{b}}=\nabla^{2}\varphi-\tilde{n}$
and the density $n$
\[
\begin{aligned}\frac{\partial q^{\mathrm{b}}}{\partial t}+J\left(\varphi,q^{\mathrm{b}}\right)-\kappa\frac{\partial\tilde{\varphi}}{\partial y} & =\left[\left(\mu-D\right)\Delta^{2}+C\omega_{*}\right]\varphi+D\Delta q^{\mathrm{b}},\\
\frac{\partial n}{\partial t}+J\left(\varphi,n\right)+\kappa\frac{\partial\tilde{\varphi}}{\partial y} & =\alpha\left(\tilde{\varphi}-\tilde{n}\right)+D\Delta n.
\end{aligned}
\]
Auxiliary equation for the vorticity $\zeta=\nabla^{2}\varphi$, with $\tilde{u}=-\frac{\partial\tilde{\varphi}}{\partial y}$

\[
\frac{\partial\zeta}{\partial t}+J\left(\varphi,\zeta\right)+\left[
\frac{\partial\overline{n}}{\partial x}\tilde{u}-\frac{\partial\left(\overline{\tilde{u}\tilde{n}}\right)}{\partial x}\right]=\alpha\left(\tilde{\varphi}-\tilde{n}\right)+\mu\Delta\zeta+C\omega_{*}\varphi.
\]

Zonal mean and fluctuation equations for the vorticity $\zeta=\overline{\zeta}+\tilde{\zeta}$
\[
\begin{aligned} &\frac{\partial\overline{\zeta}}{\partial t}+\frac{\partial\left(\overline{\tilde{u}\tilde{\zeta}}\right)}{\partial x}-\frac{\partial\left(\overline{\tilde{u}\tilde{n}}\right)}{\partial x} =\mu\Delta\overline{\zeta}+C\omega_{*}\overline{\varphi}\\
&\frac{\partial\tilde{\zeta}}{\partial t}+J\left(\varphi,\tilde{\zeta}\right)+\left[\frac{\partial\zeta}{\partial x}\tilde{u}-\frac{\partial}{\partial x}\left(\overline{\tilde{u}\tilde{\zeta}}\right)\right]+\frac{\partial\overline{n}}{\partial x}\tilde{u} =\alpha\left(\tilde{\varphi}-\tilde{n}\right)\\
&\qquad\qquad\qquad\qquad\qquad\qquad\qquad\qquad\qquad+\mu\Delta\tilde{\zeta}+C\omega_{*}\tilde{\varphi}
\end{aligned}
\]

Conservation laws for the energy $E$ and the enstrophy $W$, with $\overline{v}=\overline{\varphi_{x}}$
\[
\begin{aligned}& E=\frac{1}{2}\int_{\Omega}\left(\left|\nabla\varphi\right|^{2}+n^{2}\right)dV, \quad W=\frac{1}{2}\int_{\Omega}\left(\zeta-\tilde{n}\right)^{2}dV \\
& \frac{dE}{dt} = \kappa\int_{\Omega}\tilde{n}\tilde{u}\left(1+\kappa^{-1}\overline{v}\right)dV-\alpha\int_{\Omega}\left(\tilde{n}-\tilde{\varphi}\right)^{2}dV-D_{E},\\
& \frac{dW}{dt} = \kappa\int_{\Omega}\tilde{n}\tilde{u}dV-D_{W}.
\end{aligned}
\]
\end{minipage}%

\bigskip
\bibliographystyle{plain}
\bibliography{./ref1}

\end{document}